\documentclass[nofootinbib,amsmath,prd,aps,superscriptaddress,tightenlines,12pt]{revtex4}
\usepackage[hypertex]{hyperref}
\usepackage{graphicx}
\usepackage{epstopdf}
\usepackage{bm}
\usepackage{epsfig}
\usepackage{graphics}
\usepackage{xspace}
\def\OMIT#1{}

\newcommand{\nn}{\nonumber}

\newcommand{\bn}{{\bar n}}
\newcommand{\bea}{\begin{eqnarray}}
\newcommand{\eea}{\end{eqnarray}}

\newcommand{\shat}{\hat{s}}

\newcommand{\be}{\begin{equation}}
\newcommand{\ee}{\end{equation}}

\DeclareMathOperator{\Tr}{Tr}

\begin{document}
\setlength\baselineskip{17pt}

%%%%%%%%%%%%%%%%%%%%%%%%%%%%%%%%%%%%%%%%%%
%Define Title, Author, Address, Preprint#

%\preprint{ \vbox{  \hbox{CALT-68-}
% \hbox{hep-ph/xxxxxxx}

\title{\bf Factorization and Resummation of Higgs Boson Differential Distributions in Soft-Collinear Effective Theory}

%\vspace*{1cm}

\author{Sonny Mantry}
\email[]{mantry147@gmail.com}
\author{Frank Petriello}
\email[]{frankjp@physics.wisc.edu}
\affiliation{University of Wisconsin, Madison, WI, 53706}

%\date{\today\\ \vspace{1cm} }

%%%%%%%%%%%%%%%%%%%%%%%%%%%%%%%%%%%%%%%%%%
%Create the title page

\newpage
\begin{abstract}
  \vspace*{0.3cm}
 We derive a factorization theorem for the Higgs boson transverse momentum  ($p_T$) and rapidity ($Y$) distributions at hadron colliders, using the Soft Collinear Effective Theory (SCET), for  $m_h\gg p_T\gg \Lambda_{QCD}$  where  $m_h$  denotes  the Higgs mass. In addition to the factorization of the various scales involved, the perturbative physics at the $p_T$ scale is further factorized into two collinear impact-parameter Beam Functions (iBFs) and an inverse Soft Function (iSF).  These newly defined functions are of a universal nature for the study of differential distributions at hadron colliders.  The additional factorization of the $p_T$-scale physics simplifies the implementation of higher order radiative corrections in $\alpha_s(p_T)$.  We derive formulas for factorization in both momentum and impact parameter space and discuss the relationship between them. 
 Large logarithms of the relevant scales in the problem are summed using the renormalization group equations of the effective theories.  
 Power corrections to the factorization theorem in $p_T/m_h$ and $\Lambda_{QCD}/p_T$ can be systematically derived.  We perform multiple consistency checks on our factorization theorem including a comparison with known fixed order QCD results. We compare the SCET factorization theorem with the Collins-Soper-Sterman approach to low-$p_T$ resummation.
 \end{abstract}

\maketitle

%%%%%%%%%%%%%%%%%%%%%%%%%%%%%%%%%%%%%%%%%%
\newpage
\tableofcontents
\section{Introduction}

The Higgs boson is the last undiscovered particle of the Standard Model (SM), and its discovery is a major goal of both the Tevatron and the Large Hadron Collider (LHC) physics programs.   If a scalar particle is discovered at either collider, the measurement of its properties will be crucial to determine whether the particle found is the Standard Model Higgs boson, or whether it hints at physics beyond the SM.  The theoretical community has devoted significant effort to understanding precisely the production cross section and decay widths of the SM Higgs particle in order to facilitate such studies, as reviewed recently in Ref.~\cite{Boughezal:2009fw}.  The dominant production mode at both the Tevatron and the LHC is the partonic mechanism $gg\to H$ proceeding through a top-quark loop~\cite{Ellis:1975ap,Georgi:1977gs,Wilczek:1977zn,Shifman:1979eb,Rizzo:1979mf}.  The perturbative QCD corrections are known through next-to-leading order in full QCD~\cite{Djouadi:1991tka,Spira:1995rr}, while the corrections in the $m_t \to \infty$ limit 
are known through next-to-leading order~\cite{Dawson:1990zj} and next-to-next-to-leading order~\cite{Harlander:2002wh,Anastasiou:2002yz,Ravindran:2003um}.  Resummation of logarithmically-enhanced threshold corrections to the cross section has been studied~\cite{Kramer:1996iq,Catani:2003zt,Ahrens:2008qu,Ahrens:2008nc}.  The inclusion of such theoretical calculations is crucial for experimental searches for the Higgs boson, as they increase the predicted cross section in the SM by a factor of two at the LHC and by more than a factor of three at the Tevatron.

The study of differential distributions of the Higgs boson is also needed in experimental analyses.  For example, for a SM Higgs in the mass range 
130 GeV $ \leq m_h \leq$ 160 GeV, one of the most promising discovery modes is through the partonic process $gg \to  h\to W^+W^-\to \ell^+\nu \ell^- \bar{\nu}$.  Since the final state contains two neutrinos, reconstruction of the Higgs mass peak is not possible.  An understanding of the kinematic distributions for both signal and backgrounds is needed in this search channel.  The NNLO differential distributions in the $m_t \to \infty$ effective theory were obtained in Refs.~\cite{Anastasiou:2004xq,Anastasiou:2005qj,Catani:2007vq,Catani:2008me}, and detailed studies of the effects of experimental cuts on Higgs boson cross sections have been performed~\cite{Davatz:2006ut,Anastasiou:2007mz,Anastasiou:2008ik}.  However, a large reducible background comes from $pp \to t \bar{t}\to bW^+ \bar{b}W^-\to \ell^+\nu \ell^-\bar{v} + \text{jets}$. Such backgrounds are brought under control with a series of cuts which include a jet veto so that any process involving a jet with high transverse momentum, taken to be roughly $p_T > 20$ GeV~\cite{Dittmar:1996ss,Ball:2007zza,Aad:2009wy}, is rejected. Such a cut selects Higgs boson with primarily low transverse momentum, and therefore 
a proper implementation of such jet vetoes requires a good understanding of the Higgs differential distributions at low $p_T$ where resummation of 
large $p_T / m_h$ logarithms is necessary.  The study of the low-transverse momentum region in hadronic collisions has been under investigation since the early days of QCD~\cite{Dokshitzer:1978yd,Parisi:1979se,Curci:1979bg}.  It has been studied for the Higgs boson following the seminal analysis of Collins, Soper, and Sterman (CSS)~\cite{Collins:1981uk,Collins:1984kg} in several works~\cite{Kauffman:1991jt,Yuan:1991we,Berger:2002ut,Bozzi:2003jy,Bozzi:2005wk}.

\begin{figure}
\includegraphics[width=4.5in, height=3.0in]{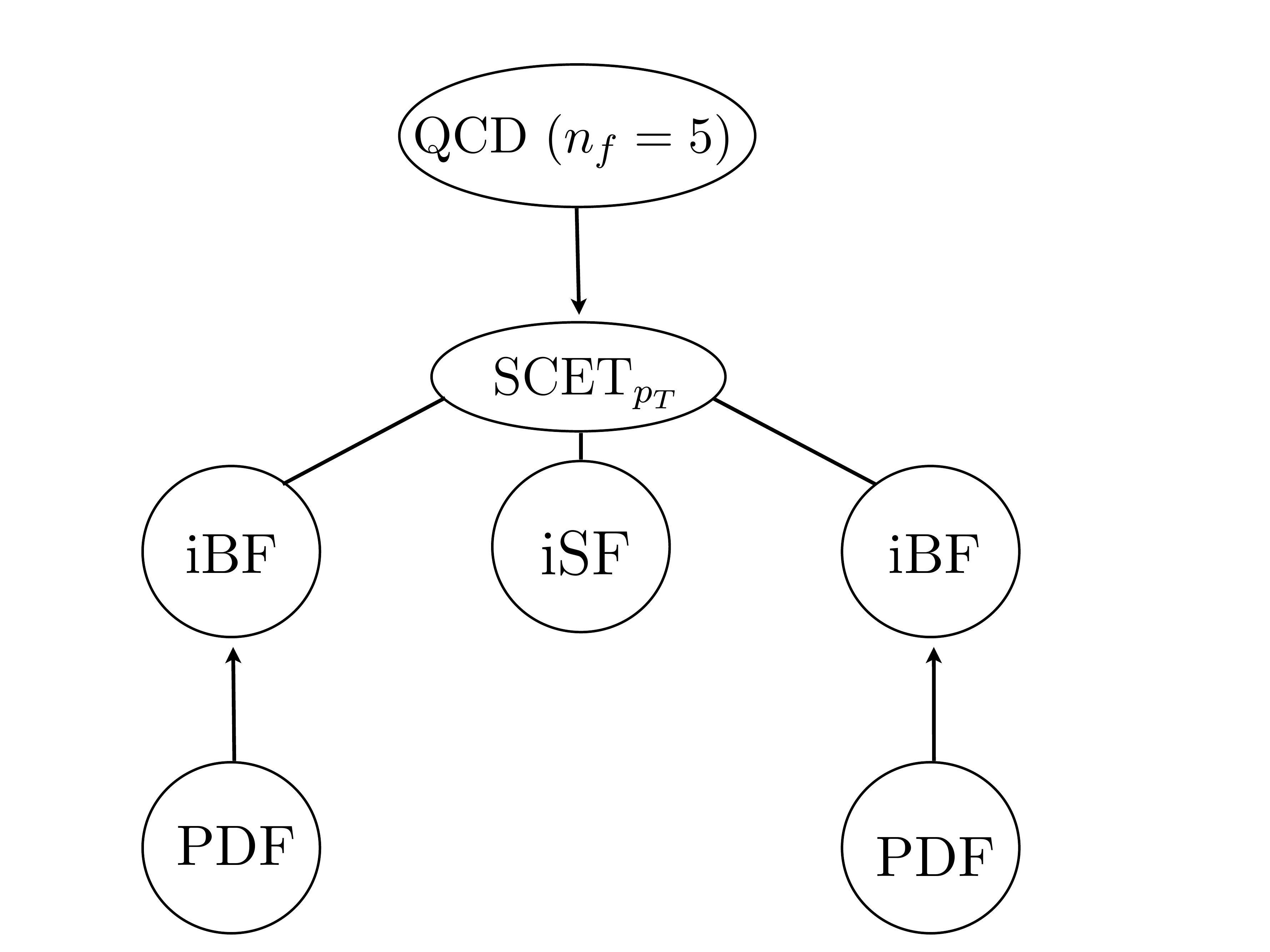}
\caption{Structure of the factorization theorem. The QCD $(n_f=5)$ theory obtained after integrating out the top is matched onto SCET$_{p_T}$ at the $\mu \sim m_h$ scale followed by RG running down to the $\mu \sim $ p$_T$ scale  summing logarithms in $m_h/p_T$ in the process. Using the soft-collinear decoupling property of the leading order SCET$_{p_T}$ Lagrangian, the cross-section is factorized into a n-collinear iBF(Impact-parameter Beam Function), a $\bn$-collinear iBF, and an iSF(Inverse Soft Function). The iBFs are then matched onto the standard QCD PDFs at the $\mu \sim p_T$ scale and the logarithms of $\Lambda_{QCD}/p_T$ are summed via the DGLAP equations which determine the PDFs at the $\mu \sim p_T$ scale.   }
\label{cartoon}
\end{figure}

The purpose of this paper is to derive a factorization theorem for the Higgs transverse momentum $p_T$ and rapidity $Y$ distribution, in the region $\hat{Q} \sim m_h \gg p_T\gg \Lambda_{QCD}$, using the Soft Collinear Effective Theory(SCET)~\cite{Bauer:2000yr, Bauer:2001yt, Bauer:2002nz}. Here $\hat{Q}$ and $m_h$ denote the partonic center of mass energy and the Higgs mass respectively.  Although we focus on Higgs production, our methods and results can be immediately generalized to the differential distributions of any one or more color neutral particles. The factorization theorem we derive has the form
\bea
\label{intro-fac}
\frac{d^2\sigma}{d\text{p}_T^2 \>dY} &=& \frac{\pi^2}{4(N_c^2-1)^2 Q^2 }     \int_0^1\frac{dx_1}{x_1}\int _0^1\frac{dx_2}{x_2} \int_{x_1}^1\frac{dx'_1}{x'_1}\int _{x_2}^1 \frac{dx'_2}{x'_2}\nn \\
&\times&  H(x_1, x_2,\mu_Q;\mu_T){\cal G}^{ij}(x_1,x_1',x_2,x_2',p_T,Y,\mu_T)f_{i/P}(x_1' ,\mu_T) f_{j/P}(x_2' ,\mu_T),
\eea
%,
where $Q$ is the hadronic center of mass energy, $H$ is the hard Wilson coefficient arising matching QCD onto SCET, and $f_{i/P}$ is the standard parton distribution function (PDF) for taking a parton of species $i$ from the proton.  ${\cal G}^{ij}$ is a perturbative coefficient at the $p_T$ scale that has the form
\bea
\label{intro-G}
{\cal G}^{ij}(x_1,x_1',x_2,x_2',p_T,Y,\mu_T)&=&  \int dt_n^+ \int  dt_\bn^-\> \int \frac{d^2b_\perp  }{(2\pi)^2}\> J_0 (b_\perp \>p_T)\;g^\perp_{\alpha \sigma} g^\perp_{\beta \omega}\nn \\
&\times&  \> {\cal I}_{n;g,i}^{\alpha\beta } (\frac{x_1}{x_1'}, t_n^+,b_\perp,\mu_T)\> {\cal I}_{\bn;g,j}^{\sigma \omega} (\frac{x_2}{x_2'}, t_\bn^-,b_\perp,\mu_T) \\
&\times&  {\cal S}^{-1}(x_1 Q-e^{Y}\sqrt{\text{p}_T^2+m_h^2}-\frac{t_\bn^-}{Q}, x_2 Q-e^{-Y}\sqrt{\text{p}_T^2+m_h^2}- \frac{t_n^+}{Q},b_\perp,\mu_T), \nn 
\eea
which is a convolution over the collinear functions ${\cal I}_{n,\bn;g,i}^{\alpha \beta}$ and the Inverse Soft Function (iSF) ${\cal S}^{-1}$.
Logarithms of $m_h/p_T$ are summed by the Renormalization Group (RG) equations in SCET and are encoded in $H(x_1, x_2,\mu_Q;\mu_T)$ which is the hard coefficient evolved from the renormalization scale $\mu_Q \sim m_h$ down to $\mu_T \sim p_T$. The logarithms of $\Lambda_{QCD}/p_T$ are summed via the standard DGLAP evolution  of the PDFs and are encoded in the PDFs evaluated at $\mu_T \sim p_T$.
The factorization formula in Eq.(\ref{intro-fac}) is derived by matching QCD onto a sequence of effective field theories EFT:
\bea
\text{QCD} (n_f=6) \to \text{QCD} (n_f=5) \to \text{SCET} _{p_T} \to \text{SCET}_{\Lambda_{QCD}},
\eea
which is shown graphically in Fig.~\ref{cartoon}.  The first step $\text{QCD} (n_f=6) \to \text{QCD} (n_f=5)$ denotes the usual procedure of integrating out the top quark to get an effective coupling of the Higgs to gluons. The Higgs production mechanism then proceeds via this effective coupling. The hard scale $\hat{Q}\sim m_h$ is then integrated out by matching onto SCET$_{p_T}$, which describes the dynamics of soft and collinear modes with transverse momenta of order $p_T$. The factorization theorem in SCET$_{p_T}$ takes the form
\bea
\label{fac-intro-pT}
\frac{d^2\sigma}{dp_T^2 \>dY} &=& \frac{\pi}{4(N_c^2-1)^2 } \int dp_h^+ dp_h^- \int d^2k_h^\perp    \int \frac{d^2b_\perp}{(2\pi)^2}  e^{-i \vec{k}_h^\perp \cdot \vec{b}_\perp}  \nn \\
%%%%
&\times & \delta \left [  p_h^- -e^Y\sqrt{p_T^2+m_h^2}\right ] \delta \left [  p_h^+-e^{-Y}\sqrt{p_T^2+m_h^2}\right ]\delta \left [p_h^+p_h^- - \vec{k}_{h\perp}^2 - m_h^2 \right ]  \nn \\
&\times&\int_0^1 dx_1 \int_0^1 dx_2 \int  dt_n^+ \int dt_\bn^- H(x_1x_2Q^2,\mu_Q;\mu_T) \>\tilde{B}_n^{\alpha \beta}(x_1,t_n^+,b_\perp,\mu_T)\> \tilde{B}_{\bn \alpha \beta}(x_2,t_\bn^-,b_\perp,\mu_T)\nn \\
&\times&{\cal S}^{-1}(x_1 Q-p_h^- - \frac{t_\bn^-}{Q}, x_2 Q-p_h^+ - \frac{t_n^+}{Q},b_\perp,\mu_T), \nn \\
\eea
where the collinear functions $\tilde{B}_{n,\bn}^{\alpha \beta}$ are the Impact-parameter Beam Functions(iBFs). The iBFs $\tilde{B}_{n,\bn}^{\alpha \beta}$  are extensions of the beam functions that appear in \cite{Fleming:2006cd, Stewart:2009yx} and reduce to them for $b_\perp=0$ after contraction of the transverse indices $\alpha$ and $\beta$. The beam functions of  
Ref.~\cite{Stewart:2009yx} were shown to have wide applicability to the analysis of observables at the LHC.  The iBFs are proton matrix elements evaluated at the scale $\mu_T\sim p_T$. The iBFs are matched onto the standard QCD PDFs by performing an OPE in $\Lambda_{QCD}/p_T$ and the logarithms of $\Lambda_{QCD}/p_T$ are summed via the standard DGLAP equations used to evaluate the PDFs at the scale $\mu_T \sim p_T$. This is shown schematically in Fig.~\ref{cartoon} and gives  the final form of the factorization theorem shown in Eqs.(\ref{intro-fac}) and (\ref{intro-G}) where the collinear functions ${\cal I}^{\alpha \beta}_{n,\bn;g,i}$ are just the iBF to PDF matching coefficients. 

While the factorization and resummation of transverse-momentum distributions has been studied extensively in the QCD literature, and SCET analyses \cite{Gao:2005iu, Idilbi:2005er} have been performed in the past, 
our analysis contains several interesting differences that we believe are worth further investigating.  A summary of the main points of this paper is given below:
\begin{enumerate}

\item We derive a factorization theorem for the Higgs transverse momentum and rapidity distributions using effective field theory methods. A clear separation of  the dynamics associated with the  scales $\hat{Q}\sim m_h\gg p_T\gg \Lambda_{QCD}$ into perturbative Wilson coefficients and standard QCD PDFs is achieved. Large logarithms of ratios of the relevant scales are summed using RG equations in the effective theories. Power corrections in $p_T/m_h$ and $\Lambda_{QCD}/p_T$ can be systematically derived by going to higher orders in the power counting of the effective theories.

\item In addition to the factorization of the scales $m_h\gg p_T\gg \Lambda_{QCD}$, the perturbative physics of the $p_T$ scale is further factorized into an iSF ${\cal S}^{-1}$  and two distinct collinear functions ${\cal I}_{n;gi}^{\alpha \beta}$ and ${\cal I}_{\bn;gi}^{\alpha \beta}$. This additional factorization simplifies the structure of higher order perturbative corrections at the $p_T$ scale.  They can now be obtained through higher order computations of the simpler perturbative functions ${\cal S}^{-1},{\cal I}_{n;gi}^{\alpha \beta}$ and ${\cal I}_{\bn;gi}^{\alpha \beta}$.  

\item  The factorization in SCET naturally occurs in terms of \textit{purely collinear} PDFs and soft functions. The purely collinear PDFs differ from the standard QCD PDFs by soft subtraction terms required to avoid double counting soft emissions that are already contained in the soft function. However, the equivalence of soft zero-bin and soft subtractions allow us to rewrite the factorization in terms of the standard QCD PDFs.

\item We give expressions for the factorization formula in both impact-parameter space (b-space) and momentum space and discuss how they are related. The factorization can be formulated entirely in momentum space; see Eqs.(\ref{intro-fac}), (\ref{momspace-1}), and (\ref{ftI}). However, the matching coefficients  ${\cal I}_{n;gi}^{\alpha \beta}$ and ${\cal I}_{\bn;gi}^{\alpha \beta}$ are obtained by matching the iBFs in b-space onto the standard QCD PDFs. 

\item The iBFs $\tilde{B}_{n,\bn}^{\alpha \beta}(x,t,b_\perp,\mu)$ that appear in the SCET$_{p_T}$ factorization theorem in Eq.(\ref{fac-intro-pT})  are extensions of the beam functions introduced recently in ~\cite{Fleming:2006cd, Stewart:2009yx}.  The additional functional dependence on $b_{\perp}$ found here is required to facilitate resummation of the low $p_T$ region, and our functions reduce to those studied previously for $b_{\perp}=0$.  

\item The iBFs are similar to transverse-momentum dependent PDFs studied previously in the literature~\cite{Ji:2002aa,Belitsky:2002sm,Collins:2003fm,Idilbi:2008vm}, except for an additional dependence on a residual light-cone momentum component.  The iBFs are defined in covariant or non-singular gauges and are invariant under non-singular gauge transformations. In singular gauges, like the light cone gauge where the gauge potential is non-vanishing at infinity, the iBFs must be modified~\cite{Ji:2002aa,Belitsky:2002sm,Collins:2003fm,Idilbi:2008vm} with a transverse gauge link to be gauge invariant  under the singular gauge transformations. However, the convolutions over perturbative quantities $\tilde{B}_{n}^{\alpha \beta}\otimes  \tilde{B}_{\bn \alpha \beta} \otimes {\cal S}^{-1}$ is fully gauge invariant since it is obtained by applying a soft-collinear decoupling transformation to a manifestly gauge invariant operator in SCET$_{p_T}$.  Correspondingly, after matching the iBFs onto gauge invariant PDFs, the quantity ${\cal G}^{ij}={\cal I}_{n;g,i}^{\alpha \beta}\otimes  {\cal I}_{\bn;g,j \alpha \beta} \otimes {\cal S}^{-1}$ is also gauge invariant. As a result, even though the  iBFs are defined in covariant gauges, the factorization theorem is completely gauge invariant as expected.  The additional functional dependence on the residual light-cone momentum avoids singularities that occur when this variable is inclusively integrated over, as it is when considering the usual transverse-momentum dependent PDFs.

\item The iBFs come with perpendicular indices $\alpha , \beta$ which are contracted between the $n$-collinear and the $\bn$-collinear iBFs.  This structure has not been discussed in previous SCET analyses.  This non-trivial index structure encodes dot products between the momenta of final state gluons emitted in different directions, and is needed to separately calculate contributions from $n$ and $\bn$ gluons.

\item The Landau poles associated with impact-parameter integrations in the CSS formulation can be avoided in the effective field theory approach employed here. As already mentioned, the factorization theorem can be formulated entirely in momentum space with no reference to any impact-parameter or impact-parameter integrations. The perturbative function ${\cal G}^{ij}$ and the PDFs in Eq.(\ref{intro-fac}) are naturally evaluated at the renormalization scale $\mu \sim p_T$ so that one never encounters the Landau pole associated with the limit $\mu \to 0$.

\item The iBFs are not specific to Higgs production, but have universal applicability to the study of differential distributions in gluon-initiated processes at colliders.  Analogous functions exist for quark-initiated processes, and can be derived following the approach discussed in this work.  

\item In the non-perturbative region $p_T\sim \Lambda_{QCD}$, where the effective field  theory power counting in $\Lambda_{QCD}/p_T$ breaks down,  a phenomenological model  can be introduced for the quantity ${\cal G}^{ij}$  analogous to what is done in the standard approach.

\item We perform several consistency checks on the factorization theorem including a comparison with known fixed order QCD results.  We ensure that the renormalization scale dependence properly cancels among the various factored objects  in the factorization formula, which is a non-trivial check on the structure of the factorization theorem.

\end{enumerate}
There have also been many other recent~\cite{Bauer:2008jx, Bauer:2009km, Chiu:2009yz, Chiu:2009mg, Chiu:2009ft, Trott:2006bk, Cheung:2009sg, Stewart:2009yx, Becher:2009cu, Becher:2009qa, Becher:2009th,Trott:2006bk} developments in the application of SCET to describe other processes at hadron colliders and it is a promising avenue to pursue in the LHC era. 

Our paper is organized is follows.  In Section~\ref{sec:higgsqcd} we review the calculation of Higgs production through the gluon-fusion mechanism in perturbative QCD.  We derive our factorization formula  in Section~\ref{sec:eft}.   We discuss the factorization formula and  compare  to other results in the literature in Section~\ref{sec:eftsum}.  In Section~\ref{fixedorder} we calculate the various quantities in fixed-order perturbation theory, while Section~\ref{running} is devoted to resummation of logarithms. We perform a series of consistency checks on our factorization formula including a comparison with fixed-order perturbative QCD  in Section~\ref{sec:consistency}.  Finally, we conclude in Section~\ref{sec:conc}.

%%%%%%%%%%%%%%%%%%%%%%
\section{Higgs production in QCD}
\label{sec:higgsqcd}

We begin by reviewing the gluon-initiated production of a Higgs boson in QCD.  The coupling of the Higgs boson to gluons arises primarily from a top-quark loop.  For $m_h < 2 m_t$, we can integrate out the top quark to derive an effective coupling of the Higgs boson to gluons~\cite{Ellis:1975ap,Shifman:1979eb,Inami:1982xt,Dawson:1990zj,Djouadi:1991tka}.  The effective Lagrangian is given by
\begin{eqnarray}
\label{Lmt}
\mathcal{L}_{m_t} = C_{GGh} \,  \frac{h}{v} \, G^a_{\mu \, \nu} \, G^{\mu \, \nu}_a, \quad \quad C_{GGh}  = \frac{\alpha_s}{12\pi}\left\{1+\frac{11}{4} \frac{\alpha_s}{\pi} +\mathcal{O}(\alpha_s^2) \right\},
\end{eqnarray}
where $C_{GGh}$, the Wilson coefficient in the $\rm \overline{MS}$ scheme, is known through $\mathcal{O}(\alpha_s^4)$~\cite{Kramer:1996iq,Chetyrkin:1997iv,Chetyrkin:1997un,Chetyrkin:2005ia,Schroder:2005hy}.  Calculations of the total cross section at higher orders in QCD perturbation theory using this effective Lagrangian have been shown to reproduce the result of the full theory to percent-level accuracy when $m_h < 2 m_t$ if the effective-theory cross section is normalized by the full top-quark mass-dependent leading-order cross section~\cite{Spira:1995rr,Harlander:2009mq}.  Additional corrections to the transverse momentum spectrum of $\mathcal{O}(p_T^2 /m_t^2)$ are also present in the full theory.  These affect the low-$p_T$ region at the percent level, and have recently been studied in Refs.~\cite{Keung:2009bs,Anastasiou:2009kn}.  An effective field theory approach for studying such corrections at higher orders has recently been developed~\cite{Neill:2009tn,Neill:2009mz}.  The scale $\mu$ at which the coupling constant is evaluated should be chosen as $\mu \approx m_t$ to minimize logarithms that appear in the $\mathcal{O}(\alpha_s^3)$ expression for $C_{GGh}$.  For notational ease we define
\bea
g^2 \, c =  - 4 \, C_{GGh}
\eea
and express later results using $c$.

\begin{figure}
\includegraphics[]{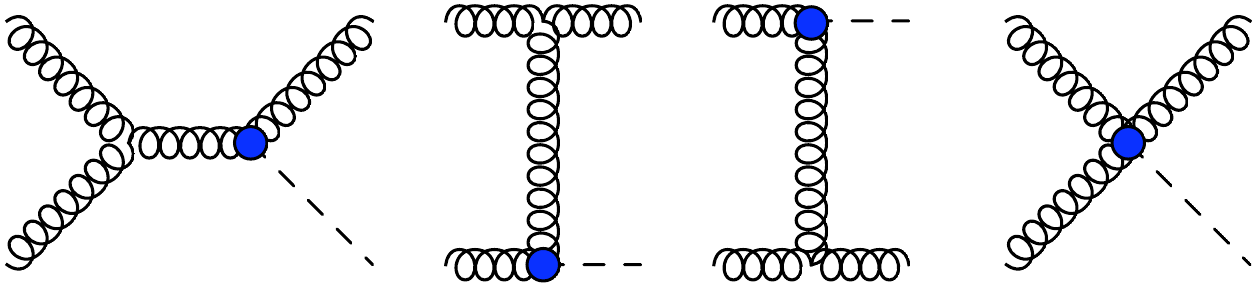}
\caption{The leading order diagrams in QCD that contribute to Higgs production with non-zero $p_T$. }
\label{QCDdiagram}
\end{figure}

The Higgs boson must recoil against at least one parton in order to have non-zero transverse momentum.  At leading-order in perturbative QCD, three partonic processes contribute to Higgs production at non-zero $p_T$: $gg \to hg$, $q(\bar{q})g \to hq(\bar{q})$, and $q\bar{q} \to hg$.  We focus here on the dominant process $gg \to hg$.  The other partonic production channels are suppressed by smaller PDF luminosities, and can be straightforwardly included in the analysis if desired.  The diagrams contributing to this process are shown in Fig.~\ref{QCDdiagram}.  We denote the incoming proton momenta as $p_1, p_2$, the outgoing parton momentum as $p_3$, and the outgoing Higgs momentum as $p_h$.  We define the light-cone vectors $n$ and $\bn$ through
\bea
n^\mu = (1,0,0,1) , \qquad \bn ^\mu = (1,0,0,-1).
\eea
The momenta of the initial hadrons and the Higgs boson are given by 
\begin{eqnarray}
p_1^{\mu} &=& \frac{Q}{2}n^{\mu}, \;\;\; p_2^{\mu} = \frac{Q}{2} \bn^{\mu}, \nonumber \\
p_h^{\mu} &=& \left( \sqrt{p_T^2+m_h^2}\,{\rm cosh} \,Y, \vec{p}_T,  \sqrt{p_T^2+m_h^2}\,{\rm sinh} \,Y \right),
\end{eqnarray}
where $Y= 1/2 \, {\rm ln} (\bn \cdot p_h / n \cdot p_h)$ is the rapidity of the Higgs boson.  The partons entering the hard-scattering process have momenta and virtualities given by 
\begin{equation}
\hat{p}_1 = x_1 p_1+ \mathcal{O}(\Lambda_{QCD}), \;\;\;\hat{p}_2 = x_2 p_2+ \mathcal{O}(\Lambda_{QCD}), \;\;\; \hat{p}_{1,2}^2 \sim \mathcal{O}(\Lambda_{QCD}^2),
\end{equation}
where $x_{1,2}$ are the usual Bjorken momentum fractions. 

Several distinct kinematic regions contribute to production of a Higgs boson at low transverse momentum.  In the first, the Higgs recoils against a $n$-collinear gluon with the following momentum in light-cone coordinates:
\begin{equation}
p_3 \sim (n\cdot p_h , \bn \cdot p_h , p_{h\perp}) \sim m_h(p_T^2/m_h^2, 1, p_T/m_h).
\end{equation}
This corresponds to a Higgs boson produced at high rapidity and low transverse momentum.  A similar region exists with the Higgs recoiling against an $\bn$-collinear gluon with scaling $p_3 \sim m_h(1, p_T^2/m_h^2 ,p_T/m_h).$  Finally, the Higgs boson may recoil against a gluon with an ``soft" momentum $p_3 \sim (p_T, p_T,p_T)$.  This corresponds to the production of a Higgs at central rapidity and low transverse momentum.  Since the final-state gluons in the low-momentum region are restricted to be either soft or collinear to the initial partons, large logarithms of the form ${\rm ln} \, (m_h / p_T)$ appear in the perturbative expansion and must be resummed to all orders.  The $n$-collinear, $\bn$-collinear and soft gluon modes will be used to construct the effective theory that facilitates this resummation.

For future use we reproduce here the differential and total cross sections for the gluon-gluon scattering process.  The hadronic cross section can be expressed as a convolution of parton distribution functions and partonic cross sections:
\begin{equation}
\sigma_{PP\to h} = \int dx_1 dx_2 f_{g/P}(x_1,\mu) f_{g/P}(x_1,\mu) \hat{\sigma}_{gg\to h}(\hat{s},\hat{t},\hat{u},\mu),
\label{totcross}
\end{equation}
where $\hat{s},\hat{t},$ and $\hat{u}$ are the usual partonic Mandelstam variables.  For production of the Higgs with non-zero $p_T$, the differential partonic cross section is given by~\cite{Ellis:1987xu}
\begin{equation}
\frac{d \hat{\sigma}}{d\hat{t}} = \frac{\pi}{384 v^2} \left(\frac{\alpha_s}{\pi}\right)^3 \left\{ \frac{m_h^8+\hat{s}^4+\hat{t}^4+\hat{u}^4}{\hat{s}\hat{t}\hat{u}} \right\}.
\label{ptspectrum}
\end{equation}
The total partonic cross section for $gg\to h$ through next-to-leading order in QCD perturbation theory is~\cite{Dawson:1990zj,Djouadi:1991tka}
\begin{eqnarray}
\hat{\sigma} &=& \frac{\pi}{576 v^2} \left( \frac{\alpha_s}{\pi}\right)^2 \left\{ \delta(1-z) + \frac{\alpha_s}{\pi} \left[ \delta(1-z) \left( \pi^2+\frac{11}{2}\right) - \frac{11}{2}(1-z)^3  \right.\right. \nonumber \\
	&+&\left.6 \left(1+z^4+(1-z)^4\right) \left( \frac{\text{ln}(1-z)}{1-z}\right)_{+}  \right\},
\label{qcdcross}
\end{eqnarray}
where $z= m_h^2/\shat$.
This result assumes the scale choice $\mu^2=\hat{s}$.  The dependence of the partonic cross section on the renormalization and factorization scales can be restored by using the known renormalization group running of the cross section.  The result is presented in Ref.~\cite{Dawson:1990zj}.

%%%%%%%%%%%%%%%%%%
\section{EFT framework}
\label{sec:eft}

We derive a factorization theorem via a sequence of effective theories
\bea
\label{eft-seq}
\text{QCD} (n_f=6) \to \text{QCD} (n_f=5) \to \text{SCET}_{p_T} \to \text{SCET}_{\Lambda_{QCD}},
\eea
which factorize the physics associated with the different scales $Q\sim m_h \gg p_T \gg \Lambda_{QCD}$ into calculable perturbative functions and standard QCD PDFs.  As we are assuming that the mass of the Higgs is sufficiently small ($m_h < 2 \, m_t$), we can integrate out the top quark in the matching step $\text{QCD} (n_f=6) \to \text{QCD} (n_f=5) $ to obtain an effective coupling of the Higgs boson to gluons.  The cross sections obtained using this effective theory were described in Sec.~\ref{sec:higgsqcd}.  To derive a renormalization group equation allowing resummation of large logarithms $\text{ln}\, (m_h/p_T)$ that appear at low transverse momenta, the matching to $\text{SCET}_{p_T}$ is required.  The soft-collinear decoupling property of the leading order SCET$_{p_T}$ Lagrangian also leads to a factorization of the soft and collinear sectors, which simplifies calculations of the cross section in the low $p_T$ region.  Finally, the matching to 
$\text{SCET}_{\Lambda_{QCD}}$ expresses the cross section in terms of the standard parton distribution functions.  We describe in this section the details of each stage in the matching in 
$ \text{QCD} (n_f=5) \to \text{SCET}_{p_T} \to \text{SCET}_{\Lambda_{QCD}}$.

\subsection{ QCD to SCET$_{p_T}$}

As already mentioned, the perturbative expansion in QCD for the transverse momentum spectrum of the Higgs contains logarithms of $m_h/p_T$. In the low transverse momentum region $\Lambda_{QCD} \ll p_T\ll m_h$, these logarithms become large and must be resummed to all orders in perturbation theory. In the effective theory formulation, this is done by matching QCD onto the effective theory SCET$_{p_T}$, which describes the dynamics of the degrees of freedom recoiling against the Higgs, and solving the RG equations of the effective theory operators. The effective theory SCET$_{p_T}$ is formulated in terms of collinear and soft modes with momentum scalings
\bea
p_n \sim m_h (\eta^2, 1, \eta), \qquad p_\bn \sim m_h (1,\eta^2,\eta), \qquad p_{s}\sim m_h(\eta,\eta,\eta), \qquad \eta \sim\frac{p_T}{m_h}, \nn \\
\eea
where $p_n, p_\bn$, and $p_{s}$ denote typical momenta for the n-collinear, $\bn$-collinear and soft modes respectively.  The effective theory has a well defined power counting in the parameter $\eta$ and has distinct quark and  gluon fields for each of these modes.  We note that ultrasoft gluons 
with scaling $p_{us} \sim m_h(\eta^2,\eta^2,\eta^2)$ would not produce a Higgs boson with sufficient transverse momentum, and therefore do not 
need to be included as separate modes in the effective theory.  The gluon fields $A_{n,\tilde{p}_n}^\mu(x)$, $A_{\bn,\tilde{p}_\bn}^\mu(x)$, and $A_{s,\tilde{q}}^\mu(x)$ destroy n-collinear, $\bn$-collinear, and soft gluons respectively. The presence of distinct collinear and soft gluons requires the theory to be invariant under collinear and soft gauge transformations~\cite{Bauer:2001yt, Bauer:2001ct}. The momenta of the effective theory modes are separated into label $\tilde{p}$ and residual $k$ parts
\bea
\label{SCETpTmodes}
p^\mu = \tilde{p}^\mu + k^\mu, \qquad \tilde{p}^\mu \sim m_h(1, \eta) , \qquad k^\mu \sim m_h\eta^2.
\eea
Derivative operators are similarly separated into label and residual operators so that, for example,  a derivative acting on the n-collinear field takes the form
\bea
i\partial ^\mu \to \frac{n^\mu}{2} \bar{\mathcal{P}} + \mathcal{P}_\perp^\mu + i \partial^\mu ,
\eea
such that the label operators act on the label momentum subscripts
\bea
 \bar{\mathcal{P}_n}A_{n,\tilde{p}_n}^\mu (x)= \bn \cdot \tilde{p} A_{n,\tilde{p}_n}^\mu(x) , \qquad  \mathcal{P}_\perp^\nu A_{n,\tilde{p}_n}^\mu(x)  = \tilde{p}_\perp^\nu A_{n,\tilde{p}_n}^\mu(x),
\eea
and the residual derivative operator acts on the residual co-ordinate dependence $x^\mu$.  We note that such a field with label momenta can be written explicitly as a Fourier transform of a standard quantum field.  As an example, a field with no dependence on residual coordinates can be expressed as 
\begin{equation}
X_{\tilde{p}_n}(0) = \int \frac{d y}{4\pi} \,e^{-i y \tilde{p}_n /2} \,X(y).
\end{equation}

As already discussed, after integrating out the top quark,  the $gg\to h$ process is mediated by the effective QCD operator
\bea
\label{QCDop}
O_{QCD} = g^2 \> h\> \text{Tr} \Big [ G_{\mu \nu} G^{\mu \nu}\Big ] = -\frac{4v}{c} {\cal L}_{m_t},
\eea
where ${\cal L}_{m_t}$ is given in Eq.~(\ref{Lmt}).
In SCET$_{p_T}$, the leading order operator that mediates this process is~\cite{Gao:2005iu, Ahrens:2008qu}
\bea
\label{vertexop}
{\cal O}(\omega_1, \omega_2) = g_{\mu \nu}h \> T\{ \text{Tr} \Big [ S_n (g B_{n\perp}^\mu)_{\omega_1}S_n^\dagger S_\bn (g B_{\bn \perp}^\nu)_{\omega_2} S_\bn^\dagger\Big ] \},
\eea
where we have suppressed the perpendicular labels on the soft and collinear fields, the sum of which are constrained to be the negative of the Higgs transverse momentum, and we use the standard notation
\bea
(X)_{\omega} \equiv X \delta (\omega - \bar{{\cal P}}^\dagger), 
\eea
and   the big brackets in Eq.~(\ref{vertexop}) denote the fact that any label operators appearing inside do not act outside the brackets.  The $B$ field is defined as~\cite{Marcantonini:2008qn}
\bea
g_s B_{n\perp}^\mu \equiv \left[ \frac{1}{\bar{\mathcal{P}_n}} [i \, \bar{n} \cdot \mathcal{D}_n, i \mathcal{D}_n^{\perp \, \mu} ] \right],
\eea
and the soft Wilson line $S_n$ in position space is given by
\bea
S_{n} (x^\mu) = P {\rm exp} \left(i g_s \int_{\infty}^0 ds \, n \cdot A^a_{s}(x^\mu + s n^{\mu}) \right),
\eea
with an analogous expression for $S_\bn$.
The covariant derivatives are dressed with momentum space Wilson lines so that
\bea
\mathcal{D}_n^{\mu} \equiv W_n^\dagger \, D_n^\mu \, W_n , \qquad W_n(x) = \Big [ \sum_{\text{perms}} \text{Exp} \big (\frac{-g}{\bar{\mathcal{P}}} \bn \cdot A_{n,q}\big )\Big ],
\eea
and $D_n^\mu$ are the usual covariant derivatives
\bea
i \bn \cdot D_n = \bar{\mathcal{P}_n} + g\bn \cdot A_{n,\tilde{p}}, \qquad iD^{\perp \mu}_n = \mathcal{P}_{n\perp}^\mu + g A_{n,\tilde{p}}^{\perp \mu}, \qquad in\cdot D_n = i n\cdot \partial + g n\cdot A_{n,p}.
\eea

We match the QCD operator onto the the effective SCET$_{p_T}$ operator via
\bea
\label{QCDtoSCETmatch}
 O_{QCD}= \int d\omega_1 \int d\omega_2  \>C(\omega_1, \omega_2 ) \>{\cal O}(\omega_1, \omega_2), 
\eea
and determine the Wilson coefficient by computing the $gg\to h$ process on the LHS in QCD and on the right side in SCET$_{p_T}$.
At tree level the Wilson coefficient is
\bea
\label{tree-wil}
C^{(0)}(\omega_1, \omega_2) = \frac{c\> \omega_1\omega_2}{v}.
\eea
The process $gg\to gh$ where the transverse momentum of the Higgs is of order $m_h \eta \sim p_T$ is mediated through an extra soft or collinear emission from the Wilson lines in ${\cal O}(\omega_1, \omega_2)$. 

We note that one can instead implement a two-step matching procedure by first matching onto an intermediate SCET with scaling parameter $\xi \sim \sqrt{p_T/m_h}$.  The relevant modes in this 
theory are collinear and ultrasoft gluons with momenta $p_n \sim m_h (\xi^2,1,\xi)$, $p_{\bn} \sim m_h (1,\xi^2,\xi)$, and $p_{us} \sim m_h (\xi^2, \xi^2 \xi^2)$.  The leading operator in this theory takes the 
schematic form $\mathcal{O} \sim \text{Tr}\left[ Y_n B_{n\perp} Y_n^{\dagger} Y_{\bn} B_{\bn\perp} Y_{\bn}^{\dagger} \right]$, where $Y_{n,\bn}$ are ultrasoft Wilson lines built up from eikonal 
interactions with the collinear gluons.   The ultrasoft gluons of the intermediate theory just become the soft modes of $\text{SCET}_{p_T}$, and 
we replace $Y_{n,\bn} \to S_{n,\bn}$ to obtain Eq.~(\ref{vertexop}).  The Wilson coefficient from matching these operators is unity to all orders as also seen in~\cite{Gao:2005iu}. A more detailed
explanation of this point can also be found in \cite{Bauer:2003mga}. 

For convenience, in deriving the factorization theorem, we will work with soft fields  that are in position space and collinear fields that are in position space conjugate to the transverse label momentum so that no label momenta of the transverse momentum appear in the soft and collinear fields. This amounts to not separating from momentum components of order $m_h \eta$ a residual part of order $m_h\eta^2$. The separation into label and residual components is employed only for the hard light cone collinear momentum components of order $m_h$.

\subsection{Factorization in SCET$_{p_T}$: iBFs and the soft function}
It is easier to work in terms of the hadronic Mandelstam variables $u$ and $t$ instead of $p_T$ and $Y$, which correspond to the Higgs transverse momentum and rapidity respectively. These two sets of variables are related by
\bea
\label{ut}
u = (p_2 -p_h)^2 = m^2 - Q\> \sqrt{p_T^2 + m^2}\> e^Y, \nn \\
t= (p_1 -p_h)^2 = m^2 - Q \>\sqrt{p_T^2 + m^2} \>e^{-Y}.
\eea
The transformation between these sets of variables has a rather simple Jacobian given by
\bea
du dt = Q^2\> dp_T^2 dY.
\eea
Thus, a restriction on the $u$ and $t$ Mandelstam variables is equivalent to a restriction on the p$_T$ and $Y$ of the Higgs.
The double differential cross-section in the Mandelstam variables can be written in SCET as
\bea
\label{fac-A}
\frac{d^2\sigma}{du \>dt} &=& \frac{1}{2Q^2} \Big [\frac{1}{4} \Big ]\int \frac{d^2 p_{h_\perp}}{(2\pi)^2} \int \frac{dn\cdot p_h d\bn\cdot p_h}{2(2\pi)^2} (2\pi)\theta (n\cdot p_h + \bn \cdot p_h)\delta (n\cdot p_h \bn \cdot p_h - \vec{p}_{h_\perp}^{\> 2} -m_h^2 ) \nn \\
%%%
&\times& \delta (u-(p_2-p_h)^2) \delta (t-(p_1 -p_h)^2)  
\sum_{\text{initial pols.}} \sum_{X} \big |C(\omega_1,\omega_2) \otimes \> \langle h X_n X_\bn X_{s} |  {\cal O}(\omega_1,\omega_2) | pp \rangle \>\big |^2 \nn \\
&\times& (2\pi)^4\delta^{(4)} (p_1+p_2-P_{X_n}-P_{X_\bn}-P_{X_s}-p_h) ,\nn \\
\eea
where ${\cal O}$ and $C$ denote the SCET$_{p_T}$ operator and the matching coefficient respectively. The $\otimes$ symbol denotes a convolution in the label momenta $\omega_{1,2}$ as in Eq.~(\ref{QCDtoSCETmatch}). Note that the constraint delta functions $\delta (u-(p_2-p_h)^2) $ and $\delta (t-(p_1 -p_h)^2) $ restrict the final states to those that satisfy $u = (p_2 -p_h)^2$ and $t= (p_1 -p_h)^2 $, or equivalently pick out the states with the corresponding values of p$_T$ and $Y$. The states $X_n, X_\bn, X_{s}$ correspond to final state particles with the n-collinear, $\bn$-collinear and soft momentum scaling respectively. It is only the states with such momentum scalings that will have a non-zero overlap with the SCET$_{p_T}$ operator ${\cal O}(\omega_1,\omega_2)$. The overall factor of $1/4$ in square brackets in Eq.~(\ref{fac-A}) is from the average over the initial proton spins.

Using the fact that the soft and collinear modes are  decoupled in the leading order SCET$_{p_T}$ Lagrangian we arrive at the factorization formula
\bea
\label{fac-first}
\frac{d^2\sigma}{du \>dt} &=& \frac{(2\pi)}{(N_c^2-1)^2 8 Q^2} \int dn\cdot p_h \int d\bn\cdot p_h \int d^2k_h^\perp \int dk_n^+ d^2k_n^\perp \int dk_\bn^- d^2k_\bn^\perp \int d^4k_s \nn \\
%%%
&\times& \int \frac{dx^- d^2x_\perp}{2(2\pi)^3}\int \frac{dy^- d^2y_\perp}{2(2\pi)^3}\int \frac{d^4z}{(2\pi)^4} e^{\frac{i}{2}k_n^+x^- - i\vec{k}_n^\perp \cdot x_\perp}e^{\frac{i}{2}k_\bn^-y^+ - i\vec{k}_\bn^\perp \cdot y_\perp} e^{i k_s \cdot z} \nn \\
%%%%%%%%%%%%
&\times& \delta \left ( u - m_h^2 +Q \bn \cdot p_h\right ) \delta \left ( t - m_h^2 +Q n \cdot p_h\right ) \delta \left ( \bn \cdot p_h n \cdot p_h - \vec{k}_{h\perp}^{\>2} - m_h^2\right ) \nn \\
%%%%
&\times&\int d\omega_1 d\omega_2 |C(\omega_1,\omega_2,\mu)|^2 J_n^{\alpha \beta}(\omega_1,x^-,x_\perp,\mu)\> J_{\bn \alpha \beta}(\omega_2,y^+,y_\perp,\mu) \>S(z,\mu) \nn \\
%%%%
&\times& \delta \left ( \omega_1 - \bn \cdot p_h - k_\bn^- - k_s^- \right ) \delta (\omega_2 - p_h^+ - k_n^+ -k_s^+) \delta^{(2)} (k_s^\perp + k_n^\perp + k_\bn^\perp + k_h^\perp), \nn \\
\eea
where the jet and soft functions are defined as
\bea
 J_n^{\alpha \beta}(\omega_1,x^-,x_\perp,\mu) &=&\sum_{\text{initial pols.}}\langle p_1 |  \big [ g B_{1n \perp \beta}^A  (x^-,x_\perp)\delta(\bar{{\cal P}} -\omega_1 )g B_{1n \perp \alpha}^A  (0)  \big ] | p_1 \rangle \nn \\
 %%%
J_\bn^{\alpha \beta}(\omega_1,y^+,y_\perp,\mu) &=&\sum_{\text{initial pols.}}\langle p_2 |  \big [ g B_{1n \perp \beta}^A  (y^+,y_\perp)\delta(\bar{{\cal P}} -\omega_2 )g B_{1n \perp \alpha}^A  (0)  \big ] | p_2 \rangle \nn \\
%%%%
S(z,\mu) &=& \langle 0 | \bar{T} \left[\text{Tr} \left ( S_\bn T^D S_\bn^\dagger S_n T^C S_n^\dagger \right ) (z)\right] T \left[ \text{Tr} \left ( S_n T^C S_n^\dagger S_\bn T^D S_\bn^\dagger \right ) (0)\right] | 0 \rangle .\nn \\
\label{jetsoftdef}
\eea
$T$ is the time-ordering symbol, and $\bar{T}$ denotes anti-time ordering.  Details of the derivation of this formula  are given in appendix~\ref{fac-derv}.
The above factorization theorem can be brought into  a more concise form involving a simpler convolution structure so that
\bea
\label{bspacefac}
\frac{d^2\sigma}{du \>dt} &=& \frac{(2\pi)}{(N_c^2-1)^2 8 Q^2} \int dp_h^+ dp_h^- \int d^2k_h^\perp \int d\omega_1 d\omega_2   \int \frac{db^+ db^- d^2b_\perp}{4(2\pi)^4}e^{\frac{i}{2} ( \omega_1 -p_h^-)b^+} e^{\frac{i}{2}( \omega_2 -p_h^+)b^-} \nn \\
&\times&  e^{-i \vec{k}_h^\perp \cdot \vec{b}_\perp}   \delta \left [  u - m_h^2 +Q  p_h^- \right ] \delta \left [  t - m_h^2 +Q p_h^+\right ]\delta \left [p_h^+p_h^- - \vec{k}_{h\perp}^2 - m_h^2 \right ]\nn \\
%%%%
&\times&|C(\omega_1, \omega_2,\mu)|^2 J_n^{\alpha \beta}(\omega_1,b^-,b_\perp,\mu)\> J_{\bn \alpha \beta}(\omega_2,b^+,b_\perp,\mu) \>S(b^+,b^-,b_\perp,\mu). \nn \\
\eea
We recast this factorization theorem in terms of jet and soft functions that have momentum space light cone coordinates as
\bea
\label{kpkmspacefac}
\frac{d^2\sigma}{du \>dt} &=& \frac{(2\pi)}{(N_c^2-1)^2 8Q^2} \int dp_h^+ dp_h^- \int d^2k_h^\perp    \int \frac{d^2b_\perp}{(2\pi)^2}  e^{-i \vec{k}_h^\perp \cdot \vec{b}_\perp}  \nn \\
%%%%
&\times & \delta \left [  u - m_h^2 +Q p_h^-\right ] \delta \left [  t - m_h^2 +Q  p_h^+\right ]\delta \left [p_h^+p_h^- - \vec{k}_{h\perp}^2 - m_h^2 \right ] \int d\omega_1 d\omega_2|C(\omega_1, \omega_2,\mu)|^2  \nn \\
&\times&\int dk_n^+ dk_\bn^-\>B_n^{\alpha \beta}(\omega_1,k_n^+,b_\perp,\mu)\> B_{\bn \alpha \beta}(\omega_2,k_\bn^-,b_\perp,\mu) \>{\cal S}(\omega_1-p_h^- - k_\bn^-, \omega_2-p_h^+ - k_n^+,b_\perp,\mu), \nn \\
\eea
where we have defined the hybrid-fourier space jet and soft functions as
\bea
 B_n^{\alpha \beta}(\omega_1,k_n^+,b_\perp,\mu) &=& \int \frac{db^-}{4\pi} e^{\frac{i}{2} k_n^+ b^-} J_n ^{\alpha \beta}(\omega_1,b^-,b_\perp,\mu), \nn \\
  B_\bn^{\alpha \beta}(\omega_2,k_\bn^-,b_\perp) &=& \int \frac{db^+}{4\pi} e^{\frac{i}{2} k_\bn^- b^+} J_n ^{\alpha \beta}(\omega_2,b^+,b_\perp,\mu), \nn \\
  {\cal S}(\tilde{\omega}_1, \tilde{\omega}_2,b_\perp,\mu)&=&  \int \frac{db^+ db^-}{16\pi^2} e^{\frac{i}{2} \tilde{\omega}_1 b^+} e^{\frac{i}{2} \tilde{\omega}_2 b^-}S(b^+,b^-,b_\perp,\mu).\nn \\
\label{beamsoftdef}
\eea
The hybrid jet functions $ B_{n,\bn}^{\alpha \beta}(\omega,k^\pm,b_\perp,\mu)$ are similar to the functions that appeared  in \cite{Fleming:2006cd} and more recently in \cite{Stewart:2009yx}, but differ because of their dependence on the impact parameter $b_\perp$ and because their perpendicular indices $\alpha,\beta$ are not contracted with each other. We will refer to these functions as impact-parameter Beam Functions (iBFs) in analogy to the Beam Functions of \cite{Stewart:2009yx} which have $b_\perp=0$.  These iBFs are implicitly defined with a zero-bin~\cite{Manohar:2006nz} subtraction in order to avoid double counting the soft region already present in the soft function $ {\cal S}(\tilde{\omega}_1, \tilde{\omega}_2,b_\perp,\mu)$. For clarity we will refer to the zero-bin subtracted iBF as a \textit{purely collinear} iBF. We will refer to the iBF defined without a zero-bin subtraction as  the \textit{naive} iBF or simply the iBF ~\footnote{Note that the iBF is still always implicitly defined with an ``ultra-soft" zero-bin subtraction~\cite{Stewart:2009yx}  to avoid double-counting  between the collinear and the ultra-soft modes with scaling $ p_{us}\sim M(\eta^2,\eta^2,\eta^2)$. In  the paper, the phrase ``zero-bin subtraction" refers to the ``soft" zero-bin subtraction, to avoid double-counting between the collinear and soft modes, unless otherwise specified. The ultra-soft and soft zero-bin subtractions are distinct since the soft modes are constrained by $p_T$ while the ultra-soft modes are unaffected.  The soft zero-bin subtraction in the beam functions
can be recast in terms of the two iBFs and an inverse soft function (iSF) as shown in the next section. Now the iBF is defined without the soft zero-bin subtraction which is instead taken into account by the presence of the  iSF.  However, there is still a double-counting left over between the collinear and the ultra-soft modes (which are unaffected by $p_T$ constraints) which is avoided by the ultra-soft zero-bin subtraction in the iBF. This ultra-soft zero-bin subtraction in the iBF should  be  understood to be implicitly part of its definition. Thus, at $b_\perp=0$, the iBF reduces exactly to the beam functions in \cite{Stewart:2009yx}. In pure dimensional regularization, the ultrasoft zero-bin subtraction is scaleless and vanishes. We thank I.W. Stewart and F. Tackmann for detailed discussions which led to an understanding of this point while at the Aspen Center for Physics summer workshop on Forefront QCD and LHC Discoveries, 2010. We also thank A. Jain, M. Procura, and W. Waalewijn for an inquiry which prompted us to  explain  this point here explicitly. } when the context is clear. These purely collinear iBFs that appear in the factorization theorem are in general gauge dependent quantities. This is seen from their dependence on the impact parameter $b_\perp\neq 0$ which leads to a spatial separation between the fields in the matrix element not connected by a Wilson line.  However, this additional gauge link can be placed at infinity along the light-cone, and it does not contribute in covariant gauges where the gauge potential vanishes at infinity.  The iBF is thus well-defined in covariant gauges.  This is similar to what occurs for transverse-momentum dependent PDFs in QCD~\cite{Ji:2002aa,Belitsky:2002sm,Collins:2003fm,Idilbi:2008vm}.  In light-cone gauge, this additional gauge link at infinity is required due to the asymptotic behavior of the gauge potential.  It is possible that Glauber modes are responsible for building up this extra contribution in SCET~\cite{Idilbi:2008vm}.  We note that the total convolution over the hard Wilson coefficient $|C(\omega_1,\omega_2,\mu)|^2$, the purely collinear iBFs,  and the soft function in Eq.~(\ref{kpkmspacefac}) is  just equal to the total perturbative cross-section for gluon-initiated Higgs + multi-parton production and thus gauge independent as required.
%%%%%%%
%%%%%%
%%%%%%
\subsection{Equivalence of zero-bin and soft subtractions}
\label{zero-1}

The purely collinear iBFs $ B_{n,\bn}^{\alpha \beta}(\omega,k^\pm,b_\perp,\mu)$ defined with a zero-bin subtraction can be written as   
\bea
\label{zero-iBF}
B_{n,\bn}^{\alpha \beta}(\omega,k^\pm,b_\perp,\mu) = \tilde{B}_{n,\bn}^{\alpha \beta}(\omega,k^\pm,b_\perp,\mu) -B_{\{n0,\bn0\}}^{\alpha \beta}(\omega,k^\pm,b_\perp,\mu)
\eea
where $\tilde{B}_{n,\bn}^{\alpha \beta}(\omega,k^\pm,b_\perp,\mu) $ is the \textit{naive} iBF or simply the iBF defined without a soft zero-bin subtraction but still with an implicit ultrasoft zero-bin subtraction. The functions $B_{\{n0,\bn0\}}^{\alpha \beta}(\omega,k^\pm,b_\perp,\mu)$ denote the soft zero-bin limit of the iBFs.  It has been demonstrated in several processes~\cite{Lee:2006nr,Idilbi:2007ff, Idilbi:2007yi} that the soft zero-bin subtraction is equivalent to a subtraction of the soft function. The same holds true in this case allowing us to recast the factorization in terms of the iBFs as opposed to the purely collinear iBFs.  In appendix~\ref{equiv-zero} we show that the convolution over the purely collinear iBFs and the soft function can be replaced with a convolution over the naive iBFs with an \textit{Inverse} Soft Function (iSF) so that 
\bea
\label{zero-soft}
 &&\int d\omega_1 d\omega_2|C(\omega_1, \omega_2,\mu)|^2 \int dk_n^+ dk_\bn^- B_n^{\alpha \beta}(\omega_1,k_n^+,b_\perp,\mu)\> B_{\bn \alpha \beta}(\omega_2,k_\bn^-,b_\perp,\mu) \nn \\
&\times& \>{\cal S}(\omega_1-p_h^- - k_\bn^-, \omega_2-p_h^+ - k_n^+,b_\perp,\mu) \nn \\
&=&\int d\omega_1 d\omega_2|C(\omega_1, \omega_2,\mu)|^2 \int dk_n^+ dk_\bn^- \tilde{B}_n^{\alpha \beta}(\omega_1,k_n^+,b_\perp,\mu)\> \tilde{B}_{\bn \alpha \beta}(\omega_2,k_\bn^-,b_\perp,\mu) \nn \\
&\times& \>{\cal S}^{-1}(\omega_1-p_h^- - k_\bn^-, \omega_2-p_h^+ - k_n^+,b_\perp,\mu). \nn \\
\eea
It is useful to recast the factorization theorem in the latter form in terms of iBFs and an iSF and eventually the iBFs will be matched onto the standard QCD PDFs via an operator product expansion. 
The factorization theorem in terms of the iBFs and an inverse soft function takes the form

\bea
\label{kpkmspacefac-1}
\frac{d^2\sigma}{du \>dt} &=& \frac{(2\pi)}{8Q^2(N_c^2-1)^2 } \int dp_h^+ dp_h^- \int d^2k_h^\perp    \int \frac{d^2b_\perp}{(2\pi)^2}  e^{-i \vec{k}_h^\perp \cdot \vec{b}_\perp}  \nn \\
%%%%
&\times & \delta \left [  u - m_h^2 +Q p_h^-\right ] \delta \left [  t - m_h^2 +Q  p_h^+\right ]\delta \left [p_h^+p_h^- - \vec{k}_{h\perp}^2 - m_h^2 \right ] \int d\omega_1 d\omega_2|C(\omega_1, \omega_2,\mu)|^2  \nn \\
&\times&\int dk_n^+ dk_\bn^-\>\tilde{B}_n^{\alpha \beta}(\omega_1,k_n^+,b_\perp,\mu)\> \tilde{B}_{\bn \alpha \beta}(\omega_2,k_\bn^-,b_\perp,\mu) \>{\cal S}^{-1}(\omega_1-p_h^- -k_\bn^-, \omega_2-p_h^+ -k_n^+,b_\perp,\mu). \nn \\
\eea
For later convenience we  switch to the  variables 
\bea
x_1= \frac{\omega_1}{\bn \cdot p_1}=\frac{\omega_1}{Q}, \qquad x_2=\frac{\omega_2}{n\cdot p_2}=\frac{\omega_2}{Q}, \qquad t_n^+ = Qk_n^+, \qquad t_\bn^- = Qk_\bn^-,
\eea
 so that the factorization theorem takes the form
\bea
\label{kpkmspacefac-3}
\frac{d^2\sigma}{du \>dt} &=& \frac{(2\pi)}{8(N_c^2-1)^2 } \int dp_h^+ dp_h^- \int d^2k_h^\perp    \int \frac{d^2b_\perp}{(2\pi)^2}  e^{-i \vec{k}_h^\perp \cdot \vec{b}_\perp}  \nn \\
%%%%
&\times & \delta \left [  u - m_h^2 +Q p_h^-\right ] \delta \left [  t - m_h^2 +Q  p_h^+\right ]\delta \left [p_h^+p_h^- - \vec{k}_{h\perp}^2 - m_h^2 \right ]  \nn \\
&\times&\int_0^1 dx_1 \int_0^1 dx_2 \int  dt_n^+ \int dt_\bn^- H(x_1x_2Q^2,\mu_Q;\mu_T) \>\tilde{B}_n^{\alpha \beta}(x_1,t_n^+,b_\perp,\mu_T)\> \tilde{B}_{\bn \alpha \beta}(x_2,t_\bn^-,b_\perp,\mu_T)\nn \\
&\times&{\cal S}^{-1}(x_1 Q-p_h^- - \frac{t_\bn^-}{Q}, x_2 Q-p_h^+ - \frac{t_n^+}{Q},b_\perp,\mu_T). \nn \\
\eea
We have defined
\bea
\label{def1}
\tilde{B}_n^{\alpha \beta}(x_1,t_n^+,b_\perp,\mu) &\equiv& \tilde{B}_n^{\alpha \beta}(x_1,k_n^+,b_\perp,\mu), \nn \\
\tilde{B}_\bn^{\alpha \beta}(x_2,t_\bn^-,b_\perp,\mu) &\equiv& \tilde{B}_\bn^{\alpha \beta}(x_2,k_\bn^-,b_\perp,\mu), \nn \\
H(x_1x_2Q^2,\mu) &\equiv& |C(x_1x_2Q^2,\mu)|^2, \nn \\ 
\eea
and $H(x_1x_2Q^2,\mu_Q;\mu_T) $ denotes the result of RG evolving the function $H(x_1x_2Q^2,\mu)$  from the scale $\mu_Q \sim m_h$  to the scale $\mu \sim $ p$_T$. In the above equation, the iBFs are written in terms of $t_{n,\bn}^{\pm}$ instead of $k_{n,\bn}^{\pm}$ to make it manifest that the iBFs actually
depend on $t_{n,\bn}^{\pm}$, as demanded by reparameterization invariance.  The choice of the scale $\mu_T \sim $ p$_T$ will become manifest once we perform the Higgs phase space integrals and rewrite the $u$ and $t$ Mandelstam variables in terms of p$_T$ and Y. We will do this in the next section.  The RG evolved $H(x_1x_2Q^2,\mu_Q;\mu_T)$ hard function sums up logarithms of $m_h/$p$_T$.  The iBFs are proton matrix elements and will give rise to logarithms of $\Lambda_{QCD}/p_T$ in the perturbative cross-section that must be resummed. For this reason, as discussed in the next section, the iBFs will be matched onto PDFs and the logarithms of $\Lambda_{QCD}/$p$_T$ will be resummed via the standard DGLAP evolution equations.
%%%%%%%
%%%%%%
%%%%%%

\subsection{iBFs to PDFs}
The matching of the iBF onto the PDF is given by
\bea
\label{iBFPDF}
\tilde{B}_n^{\alpha \beta}(z,t_n^+,b_\perp,\mu)&=& -\frac{1}{z} \sum_{i=g,q,\bar{q}} \int_z^1 \frac{dz'}{z'} \> {\cal I}_{n;g,i}^{\alpha \beta } (\frac{z}{z'}, t_n^+,b_\perp,\mu) f_{i/P}(z',\mu ),
\eea
where ${\cal I}_{g,i}^{\alpha \beta } (\frac{z}{z'}, t_n^+,b_\perp,\mu)$ is the  matching coefficient and  the gluon pdf is defined as
\bea
\label{PDF}
f_{g/P}(z ,\mu)=\frac{-z \bn \cdot p_1}{2} \sum_{\text{spins}} \langle p_1 | \big [ \text{Tr}\{ B_\perp^\mu (0) \delta(\bar{{\cal P}} - z\> \bn \cdot p_1) B_{\perp \mu} (0) \}  \big ] |p_1 \rangle ,
\eea
so that the leading order perturbative expression is normalized as
\bea
f_{g/P}^{(0)} (x) = \delta (1-x).
\eea
A matching equation analogous to Eq.~(\ref{iBFPDF}) holds for the $\bn$-collinear iBF $\tilde{B}_\bn^{\alpha \beta}$. Note that in the iBF to PDF matching in Eq.~(\ref{iBFPDF}), the iBF can match onto quark PDFs beyond tree level. In the initial analysis presented in this paper, we ignore this effect.  It is simple and straightforward to include the effects of the quark PDFs if desired. By noting the the PDF is scaleless and that the infrared structure of the iBF and PDF match, as discussed further in Section \ref{fixedorder}, one obtains the following all orders expression for the Wilson coefficient
\bea
\label{iBFPDF1}
 {\cal I}_{n;g,i}^{\beta \alpha} (\frac{z}{z'}, t_n^+,b_\perp,\mu) = - z \Big [\tilde{B}_n^{\alpha \beta}(\frac{z}{z'},z't_n^+,b_\perp,\mu)\Big ]_{\text{finite part in dim-reg}}.
\eea
We explicitly check this expression at one order beyond tree level in Section \ref{fixedorder}. An analogous expression holds for the $\bn$-collinear Wilson coefficient $ {\cal I}_{\bn;g,i}^{\beta \alpha}$.

Using Eq.~(\ref{iBFPDF}) in Eq.~(\ref{kpkmspacefac-3}) we arrive at the factorization theorem
\bea
\label{below-pT-fac}
\frac{d^2\sigma}{du \>dt} &=& \frac{(2\pi)}{8(N_c^2-1)^2 Q^2 } \int dp_h^+ dp_h^- \int d^2k_h^\perp    \int \frac{d^2b_\perp}{(2\pi)^2}  e^{-i \vec{k}_h^\perp \cdot \vec{b}_\perp} \delta \Big[p_h^+p_h^- - \vec{k}_{h\perp}^2 - m_h^2 \Big ]  \nn \\
%%%%
&\times & \delta \left [u - m_h^2 +Q p_h^-\right ] \delta \left [  t - m_h^2 +Q  p_h^+\right ]\int_0^1 \frac{dx_1}{x_1}\int _0^1 \frac{dx_2}{x_2} \int_{x_1}^1\frac{dx'_1}{x'_1}\int _{x_2}^1 \frac{dx'_2}{x'_2} H(x_1x_2Q^2,\mu_Q;\mu_T)  \nn \\
&\times&\int dt_n^+ dt_\bn^-\>g^\perp_{\alpha \sigma}\> g^\perp_{\beta \omega} \>{\cal I}_{n;g,i}^{ \alpha \beta} (\frac{x_1}{x_1'}, t_n^+,b_\perp,\mu_T)\> {\cal I}_{\bn;g,j}^{\sigma \omega} (\frac{x_2}{x_2'}, t_\bn^-,b_\perp,\mu_T)\>  \>\nn \\
&\times& {\cal S}^{-1}(x_1 Q-p_h^- - \frac{t_\bn^-}{Q}, x_2 Q-p_h^+ - \frac{t_n^+}{Q},b_\perp,\mu_T)  f_{i/P}(x_1',\mu_T ) f_{j/P}(x_2',\mu_T ). \nn \\
\eea
Next we perform the Higgs phase space integrals to get
\bea
\frac{d^2\sigma}{du \>dt} &=& \frac{\pi^2}{4Q^4(N_c^2-1)^2 }     \int_0^1 \frac{dx_1}{x_1}\int _0^1 \frac{dx_2}{x_2}\int_{x_1}^1\frac{dx'_1}{x'_1}\int _{x_2}^1 \frac{dx'_2}{x'_2}\nn \\
&\times&     \nn \\
&\times&  H(x_1x_2Q^2,\mu_Q;\mu_T)\>{\cal G}^{ij}(x_1,x_1',x_2,x_2',u,t,\mu_T)\> f_{i/P}(x_1',\mu_T ) f_{j/P}(x_2' ,\mu_T), \nn \\
\eea
where we have defined the $u$ and $t$ dependent function
\bea
\label{utfunc}
{\cal G}^{ij}(x_1,x_1',x_2,x_2',u,t,\mu_T) &=&  \int dt_n^+ \int  dt_\bn^-\int \frac{d^2b_\perp}{(2\pi)^2} J_0\Bigg [|\vec{b}_\perp|\sqrt{\frac{(m_h^2-u)(m_h^2-t)}{s} - m_h^2}\>\Bigg ]\nn \\
&\times& {\cal I}_{n;g,i}^{\beta \alpha} (\frac{x_1}{x_1'}, t_n^+,b_\perp,\mu_T)\> {\cal I}_{\bn;g,j}^{\beta \alpha} (\frac{x_2}{x_2'}, t_\bn^-,b_\perp,\mu_T)\nn \\
&\times&  {\cal S}^{-1}(x_1 Q-\frac{m_h^2 -u}{Q} - \frac{t_\bn^-}{Q}, x_2 Q-\frac{m_h^2-t}{Q} - \frac{t_n^+}{Q},b_\perp,\mu_T).  \nn \\
\eea

In terms of the $p_T$ and $Y$ variables, related to the Mandelstam $u$ and $t$ variables as in Eq.~(\ref{ut}),  we have
\bea
\label{pT-fac-final}
\frac{d^2\sigma}{d\text{p}_T^2 \>dY} &=& \frac{\pi^2 }{4(N_c^2-1)^2 Q^2 }     \int_0^1\frac{dx_1}{x_1}\int _0^1\frac{dx_2}{x_2} \int_{x_1}^1\frac{dx'_1}{x'_1}\int _{x_2}^1 \frac{dx'_2}{x'_2}  \nn \\
&\times&H(x_1x_2Q^2,\mu_Q;\mu_T) {\cal G}^{ij}(x_1,x_1',x_2,x_2',p_T,Y,\mu_T)f_{i/P}(x_1' ,\mu_T) f_{j/P}(x_2' ,\mu_T), \nn \\
\eea
where we have defined the p$_T$ and $Y$ dependent perturbative function
\bea
\label{pTfunc}
{\cal G}^{ij}(x_1,x_1',x_2,x_2',p_T,Y,\mu_T)&=&  \int dt_n^+ \int dt_\bn^-\> \int \frac{d^2b_\perp  }{(2\pi)^2} J_0 (|\vec{b}_\perp| p_T) \nn \\
&\times& {\cal I}_{n;g,i}^{\beta \alpha} (\frac{x_1}{x_1'}, t_n^+,b_\perp,\mu_T)\> {\cal I}_{\bn;g,j}^{\beta \alpha} (\frac{x_2}{x_2'}, t_\bn^-,b_\perp,\mu_T)\nn \\
&\times&  {\cal S}^{-1}(x_1 Q-e^{Y}\sqrt{\text{p}_T^2+m_h^2}-\frac{t_\bn^-}{Q}, x_2 Q-e^{-Y}\sqrt{\text{p}_T^2+m_h^2}- \frac{t_n^+}{Q},b_\perp,\mu_T) \nn \\
\eea

%%%%%%%%%

\subsection{Momentum space vs impact-parameter space}

As seen in the last section, the factorization formula for the $p_T, Y$ and $u,t$ distributions involved the functions ${\cal G}^{ij}(x_1,x_1',x_2,x_2',p_T,Y,\mu_T)$, defined in Eq.~(\ref{pTfunc}),  and ${\cal G}^{ij}(x_1,x_1',x_2,x_2',u,t,\mu_T)$, defined in Eq.~(\ref{utfunc}), respectively. These functions are defined in terms of impact-parameter space Wilson coefficients ${\cal I}_{n,\bn;g,i}^{\beta \alpha}(z,t,b_\perp) $ and the impact-parameter space iSF $ {\cal S}^{-1}(k^-,k^+,b_\perp)$. We can now introduce momentum space Wilson coefficients and a momentum-space iSF via
\bea
\label{ftI}
{\cal I}_{n,\bn;g,i}^{\beta \alpha}(z,t,k_\perp) &=& \int \frac{d^2b_\perp}{4\pi^2} e^{-i\vec{k}_\perp\cdot \vec{b}_\perp}{\cal I}_{n,\bn;g,i}^{\beta \alpha}(z,t,b_\perp), \nn \\
{\cal S}^{-1}(k^-,k^+,k_\perp)&=&  \int \frac{d^2b_\perp}{4\pi^2} e^{-i\vec{k}_\perp\cdot \vec{b}_\perp} {\cal S}^{-1}(k^-,k^+,b_\perp),
\eea 
so that the function ${\cal G}^{ij}(x_1,x_1',x_2,x_2',p_T,Y,\mu_T) $ can be written as
\bea
\label{momspace-1}
{\cal G}^{ij}(x_1,x_1',x_2,x_2',p_T,Y,\mu_T) &=& \frac{1}{2\pi} \int dt_n^+ \int dt_\bn^-\>\int d^2k_n^\perp \int d^2k_\bn^\perp \int d^2k_{s}^\perp \frac{\delta(p_T - |\vec{k}_n^\perp + \vec{k}_\bn^\perp+\vec{k}_{s}^\perp|)}{p_T}  \nn \\
&\times& {\cal I}_{n;g,i}^{\beta \alpha} (\frac{x_1}{x_1'}, t_n^+,k_n^\perp,\mu_T)\> {\cal I}_{\bn;g,j}^{\beta \alpha} (\frac{x_2}{x_2'}, t_\bn^-,k_\bn^\perp,\mu_T)\nn \\
&\times&  {\cal S}^{-1}(x_1 Q-e^{Y}\sqrt{\text{p}_T^2+m_h^2}-\frac{t_\bn^-}{Q}, x_2 Q-e^{-Y}\sqrt{\text{p}_T^2+m_h^2}- \frac{t_n^+}{Q},k_{s}^\perp,\mu_T) \nn \\
\eea
with no reference to $b_\perp$. The expression for ${\cal G}^{ij}(x_1,x_1',x_2,x_2',u,t,\mu_T) $  can be trivially obtained from Eq.~(\ref{momspace-1}) by rewriting the $p_T$ and $Y$ variables in terms of the $u$ and $t$ variables by using Eq.~(\ref{ut}). Eq.~(\ref{pT-fac-final}) together with Eq.~(\ref{momspace-1}) gives the factorization theorem entirely in momentum space with no reference to the impact parameter. This form of the factorization  makes manifest the $p_T$ dependence and how it is related to the momenta of the final state particles involved.

However, the impact-parameter space formulation is necessary for the matching of the iBF onto the PDF.  This matching must be done in impact-parameter space as in Eq.~(\ref{iBFPDF}). It is only in impact-parameter space that the infrared singular structure of the iBFs and the PDFs becomes manifest. Thus, in order to obtain the momentum space factorization, one must first go to the impact-parameter space formulation and obtain the iBF to PDF matching coefficients ${\cal I}^{\alpha \beta}_{n,\bn; g, i}(z,t,b_\perp)$  and then fourier transform to momentum space.

%%%%%%%%%%%%
%%%%%%%%%
\section{Factorization theorem: summary and discussion}
\label{sec:eftsum}

\subsection{General discussion}
In this section we summarize the main points and give a more detailed discussion about the structure of the factorization theorem.  We also compare our results with both previous SCET studies of low-$p_T$ resummation and with the standard QCD approach.  In the remaining sections we present fixed order results for the various perturbative functions appearing in the factorization theorem, anomalous dimension computations and RG equations,  various internal consistency checks of the factorization theorem, and checks of the factorization by comparing with known fixed order QCD results.

To recap, the factorization theorem for the Higgs transverse momentum p$_T$ and rapidity $Y$ has the form
\bea
\label{fac-summary}
\frac{d^2\sigma}{d\text{p}_T^2 \>dY} &=& \frac{\pi^2 }{4(N_c^2-1)^2 Q^2 }     \int_0^1\frac{dx_1}{x_1}\int _0^1\frac{dx_2}{x_2} \int_{x_1}^1\frac{dx'_1}{x'_1}\int _{x_2}^1 \frac{dx'_2}{x'_2}  \nn \\
&\times&H(x_1x_2Q^2,\mu_Q;\mu_T) {\cal G}^{ij}(x_1,x_1',x_2,x_2',p_T,Y,\mu_T)f_{i/P}(x_1' ,\mu_T) f_{j/P}(x_2' ,\mu_T), \nn \\
\eea
where $H$ and $ {\cal G}^{ij}$ are perturbative coefficients and $f_{i/P}$ are the standard QCD proton PDFs. The indices $i,j$ run over the gluon, quark, and anti-quark species. In our initial analysis we have ignored contributions from quark and anti-quark initiated Higgs production since they are numerically small, but it is straightforward to generalize the analysis and include these effects if desired. 

The factorization theorem achieves a clear separation of the physics associated with the scales  $\hat{Q}\sim m_h\gg p_T\gg \Lambda_{QCD}$ which are now encoded in the hard function $H(x_1x_2Q^2,\mu_Q;\mu_T)$, the perturbative function ${\cal G}^{ij}(x_1,x_1',x_2,x_2',p_T,Y,\mu_T)$, and the non-perturbative PDFs respectively. The hard coefficient $H(x_1x_2Q^2,\mu_Q;\mu_T)$ is obtained by matching $n_f=5$ QCD onto SCET$_{p_T}$ at the scale $\mu_Q \sim \hat{Q} \sim m_h$, as in Eqs.~(\ref{QCDtoSCETmatch}) and (\ref{def1}), followed by RG running down to the scale $\mu_T \sim $ p$_T$. The RG running  sums up the logarithms of $m_h/$p$_T$.  The perturbative function  ${\cal G}^{ij}(x_1,x_1',x_2,x_2',p_T,Y,\mu_T)$ is a convolution of two perturbative collinear functions  and an iSF(inverse Soft Function). The result for ${\cal G}^{ij}$ involves logarithms of $p_T/\mu_T$ and it is thus natural to choose $\mu_T \sim p_T$ to minimize these logarithms. The QCD PDFs $f_{i/P}(x,\mu_T)$ are also evaluated at this scale $\mu_T \sim p_T$ and the logarithms of $\Lambda_{QCD}/p_T$ are summed via the standard  DGLAP evolution of the PDFs. This picture is roughly summarized in Fig.~\ref{cartoon}.  We note that in principle, the iSF could depend on an additional scale $\mu_S$.  In this analysis we set $\mu_S=\mu_T$, 
since we see no evidence of large logarithms associated with the additional scale $Q$ that must be resummed via an additional renormalization-group evolution in $\mu_S$.  We show this when calculating the iSF in Section~\ref{fixedorder}.  If a 
higher-order analysis indicates such a need for a separate $\mu_S$, it is straightforward to include.

The perturbative function ${\cal G}^{ij}$  defined in Eq.~(\ref{pTfunc}) has an intricate structure on which we now elaborate. As seen in Eq.~(\ref{pTfunc}), ${\cal G}^{ij}$ is a convolution, in light-cone momenta and an impact parameter, over a n-collinear function ${\cal I}_{n;g,i}^{\alpha \beta}$, $\bn$-collinear function ${\cal I}_{\bn;g,i}^{\alpha \beta}$ and the iSF ${\cal S}^{-1}$. The ${\cal I}_{n,\bn;g,i}^{\alpha \beta}$ functions are obtained from matching the iBFs (impact-parameter Beam Functions) $\tilde{B}_{n,\bn}^{\alpha \beta}$ onto the standard QCD PDFs as defined in Eq.~(\ref{iBFPDF}) and are related to the iBFs as in Eq.~(\ref{iBFPDF1}). The iBFs are  defined without  a soft zero-bin subtraction but with an ultrasoft zero-bin subtraction.  This is a consequence of the equivalence of soft zero-bin and soft subtractions~\cite{Lee:2006nr,Idilbi:2007ff, Idilbi:2007yi} which we have discussed in some detail in Section \ref{zero-1}.  If one  works with iBFs defined with a soft zero-bin subtraction, the purely collinear iBFs defined earlier in Section ~\ref{zero-1}, then these objects would naturally match onto \textit{purely collinear} PDFs which differ from the standard QCD PDFs by soft zero-bin subtraction terms. For more details on the relationship between standard PDFs and purely collinear PDFs see~\cite{Idilbi:2007ff, Idilbi:2007yi}.

The n-collinear and $\bn$-collinear iBFs describe the physics of collinear emissions in the n and $\bn$ directions respectively  but in addition, as already mentioned, they each encode the \textit{same} soft region due to the absence of a  soft zero-bin subtraction.  As a result the soft region, which encodes the physics of soft emissions and soft virtual effects,  ends up being double-counted between the two iBFs. The iSF plays the role of precisely \text{subtracting} a soft region so that in the convolution over the two iBFs and the iSF there is only one soft region as required. After the iBFs are matched onto PDFs as in Eq.~(\ref{iBFPDF}), we end up with a convolution over the two collinear functions ${\cal I}_{n,\bn;g,i}^{\alpha \beta}$  and the iSF which defines ${\cal G}^{ij}$ as in Eq.~(\ref{pTfunc}).

The iBFs $\tilde{B}_{n,\bn}$ are defined in covariant or non-singular gauges where the gauge potential vanishes at infinity. Under non-singular gauge transformations, the iBFs are invariant. However, the iBFs are not  invariant, as defined, under singular gauge transformations with boundary conditions where the gauge potential is non-vanishing at infinity. In singular gauges like the light-cone gauge, the definition of the iBF must be modified~\cite{Ji:2002aa,Belitsky:2002sm,Collins:2003fm,Idilbi:2008vm} with a transverse gauge link in order to be invariant under singular gauge transformations. However, the restriction of the  iBFs to covariant gauges suffices since the quantity ${\cal G}^{ij}$ which is a convolution over the two iBFs and the iSF is fully gauge invariant. This can be seen in two ways.  First, since the iBFs and the iSF were obtained after applying the soft-collinear decoupling property of the $\text{SCET}_{p_T}$ Lagrangian on the manifestly gauge invariant operator ${\cal O}$ defined in Eq.~(\ref{vertexop}), the total convolution over the iBFs and the iSF must again be  gauge invariant. Second, the product of the hard coefficient $H$ and the perturbative function ${\cal G}^{ij}$, appearing in the factorization theorem in Eq.~(\ref{fac-summary}),  just gives the perturbative cross-section for Higgs + multi-parton production as is manifest before applying the soft-collinear decoupling transformation. Together with the gauge invariance of $H$, the quantity ${\cal G}^{ij}$ must again be gauge invariant. The additional functional dependence of the iBF on the variable $t_n$ also removes the singularity that occurs when this variable is fully integrated over, as it is in the transverse-momentum dependent PDF.  In our factorization theorem this variable is instead constrained by the external kinematics of the process.  

The factorization of ${\cal G}^{ij}$ into the convolution ${\cal I}_{n;g,i}^{\alpha \beta}\otimes  {\cal I}_{\bn;g,j \alpha \beta} \otimes {\cal S}^{-1}$ can be quite useful in the implementation of higher order radiative corrections in $\alpha_s(p_T)$. This is because the computation of the individual functions ${\cal I}_{n,\bn;g,i}^{\alpha \beta}$, defined in covariant gauges, and ${\cal S}^{-1}$ is much simpler than the full computation of ${\cal G}^{ij}$ without factorization.

Another property of the iBFs $\tilde{B}_{n,\bn}^{\alpha \beta}$ and the matching coefficients ${\cal I}_{n,\bn;g,i}^{\alpha \beta}$ is their index structure. The indices $\alpha, \beta$ run over the transverse directions. Note that these indices remain uncontracted in each iBF  due to their non-trivial dependence of the iBFs on the impact parameter $b_\perp^\alpha$. Instead, these indices are contracted between the n-collinear and the $\bn$-collinear functions as seen in Eqs.~(\ref{kpkmspacefac-3}) and (\ref{pTfunc}). This contraction of the indices between the n and $\bn$ collinear functions is necessary to reproduce dot products between the transverse momenta of final state particles in the  n and $\bn$ directions which will appear at higher orders in perturbation theory.

We next discuss the formulation of  factorization in impact-parameter space vs momentum space. So, far we have discussed the factorization theorem in terms of the impact-parameter space iBFs and iSF as seen in Eq.~(\ref{pTfunc}). In particular, the matching of the iBFs onto the standard QCD PDFs, as defined in Eq.~(\ref{iBFPDF}), must be done in impact-parameter space.  It is only in impact-parameter space that the infrared singularities of the iBFs become manifest as poles in 
dimensional regularization, and can therefore be seen to be the same as those of the standard PDFs.  However, once the impact-parameter space matching coefficients are obtained one can Fourier transform to momentum space as in Eq.~(\ref{ftI}) and obtain the perturbative function ${\cal G}^{ij}$ entirely in terms of the momentum space matching coefficients and iSF as seen in Eq.~(\ref{momspace-1}). In this momentum space factorization there is no reference to any impact-parameter. This form also makes manifest that the Landau poles which appear in the impact-parameter space CSS formulation ~\cite{Collins:1984kg} are avoided here.  We discuss this point further below in our comparison with the standard approach.

\subsection{Comparison with CSS approach}

The classic QCD analysis of resummation of low transverse momentum logarithms expresses the cross section in the low $p_T$ region as~\cite{Collins:1984kg}
\begin{eqnarray}
\frac{d^2 \sigma}{dp_T \,dY} &=& \sigma_0 \int \frac{d^2 b_{\perp}}{(2\pi)^2} e^{-i \vec{p}_T \cdot \vec{b}_{\perp}} \sum_{a,b} \left[ C_a \otimes f_{a/P} \right] (x_A, b_0/b_{\perp} )\left[ C_b \otimes f_{b/P} \right] (x_B, b_0/b_{\perp} ) \nn \\
	&\times& \text{exp}\left\{ \int_{b_0^2/b_{\perp}^2}^{\hat{Q}^2}  \frac{d \mu^2}{\mu^2} \left[ \text{ln}\frac{\hat{Q}^2}{\mu^2} A(\alpha_s (\mu^2)) +B(\alpha_s (\mu^2))\right] \right\}.
\end{eqnarray}
The sum is over parton species labeled by $a,b$, while $x_{A,B}$ denote the equivalent parton fractions $x_{A,B} = e^{\pm Y} m_h/Q$ respectively.  The functions $A$, $B$, and $C$ have perturbative expansions in $\alpha_s$, while $b_0$ is an arbitrary constant chosen for computational convenience.  

One significant difference between this result and our approach outlined in the previous section is the appearance of the Landau pole of the strong coupling constant when $\mu^2=0$ in the exponent.  To deal with this singularity, several modifications of this formula are employed, including a deformation of the $b_{\perp}$ integration contour~\cite{Laenen:2000de,Kulesza:2002rh,Bozzi:2005wk} and the introduction of a phenomenological model to cut off the $b_{\perp} \to \infty$ region~\cite{Landry:1999an}.  In our approach the most natural choice for the scale which 
controls the lower limit of the RG evolution is $\mu_L = p_T$.  This can also be understood by noting that the perturbative function ${\cal G}^{ij}$ is independent of the impact parameter, in both the impact-parameter and momentum-space formulations of the factorization theorem, and depends on $p_T$ and $\mu$ and no other dimensionful scales. Furthermore from the structure of the factorization theorem, we see that the logarithms of $m_h/p_T$ are summed by the RG evolution of the hard coefficient $H(x_1,x_2Q^2,\mu)$ which multiplies the function ${\cal G}^{ij}$ and also has no reference to an impact parameter.  In the effective  theory, non-perturbative effects such as those indicated by the appearance of the Landau pole are encoded in operators suppressed by $\Lambda_{QCD}/p_T$.  When $p_T \sim \Lambda_{QCD}$, the expansion in this parameter breaks down, and  a model of ${\cal G}^{ij}$ fit to data can be used analogous to the standard approach.  However, no reference to a non-pertubative function is needed above $\Lambda_{QCD}$.  Previous comparisons of $b$-space and momentums-space resummation formalisms have indicated numerical agreement between the obtained results down to $p_T \sim \text{few GeV}$~\cite{Ellis:1997ii}.  At this stage, power-suppressed operators presumably give important contributions.  The use of SCET allows such effects to be studied in a systematic way.  The avoidance of the Landau singularity also simplifies the matching of the resummed result to the fixed-order expression.  In the usual approach, a large cancellation between the resummed component and the fixed-order QCD contribution occurs, leading to potential instabilities in the matched distribution.  This cancellation typically occurs numerically because of the introduction of a non-perturbative model for the large $b_{\perp}$ region.  Since it can be arranged analytically if the $b_{\perp}$ integrals can be done, avoidance of the Landau pole is useful for this purpose also (we note that the matching to fixed-order QCD results can be made smoother in the CSS approach by an appropriate change in the logarithm resummed~\cite{Bozzi:2003jy}).  We note that because of the effective theory expansion in $\Lambda_{QCD}/p_T$, it is not possible to calculate exactly the $p_T=0$ result for very large $\hat{Q}$ as in the $b$-space approach~\cite{Parisi:1979se}.  However, the model of the non-perturbative region can be arranged to reproduce this result~\cite{Ellis:1997ii}.

\subsection{Comparison with previous SCET analyses}

Resummation of the of  the Higgs production transverse momentum distribution has been studied previously in~\cite{Idilbi:2005er} by  a perturbative matching calculation onto SCET followed RG running. An explicit all orders factorization formula at the $p_T$ scale, corresponding  to our Eq.(\ref{fac-intro-pT}), was written in~\cite{Gao:2005iu}. The structure of our factorization theorem differs from that in~\cite{Gao:2005iu} in several key ways.  Our factorization theorem involves a convolution over the residual light-cone momenta that connects the collinear and soft sectors necessary by momentum conservation which is absent in~\cite{Gao:2005iu}.  This additional dependence is required in order to extend the factorization and resummation to more exclusive quantities where final-state hadronic quantities are also restricted, as is clear from the dependence on the light-cone momenta in the beam functions of 
Ref.~\cite{Stewart:2009yx}.  The iBF in our factorization theorem has transverse indices which are contracted between the n-collinear and $\bn$-collinear sectors necessary to generate dot products between the transverse momenta of partons in the $n$ and $\bn$ directions in the final state. This index structure is not present in~\cite{Gao:2005iu}.  We disagree with this result, as it is not the structure that arises from the SCET analysis.  In our formulation, the matching onto the standard QCD PDFs was done separately in the n and $\bn$ sectors making full use of the SCET factorization. The work of~\cite{Gao:2005iu} does the matching to PDFs by matching the full product of soft and collinear functions onto the QCD result, as required if the index structure of the collinear sectors is dropped. Finally, our work incorporates the formalism of zero-bin subtractions necessary to avoid double counting the soft region which was not addressed in Ref.~\cite{Gao:2005iu}.

%%%%%%%%%%%%%%%%%%%%%%%%%

\section{Fixed order analysis }
\label{fixedorder}

In this section we give details of the perturbative computation of the hard Wilson coefficient, the n-collinear and $\bn$-collinear iBFs, and the iSF which appear in the factorization theorem. We emphasize that the iBF and iSF derived here are universal, and can be applied to other gluon-initiated processes at hadron colliders.

\subsection{Calculation of the $\text{QCD} \to \text{SCET}_{p_T}$ Wilson coefficient}

\begin{figure}
\includegraphics[scale=1.25]{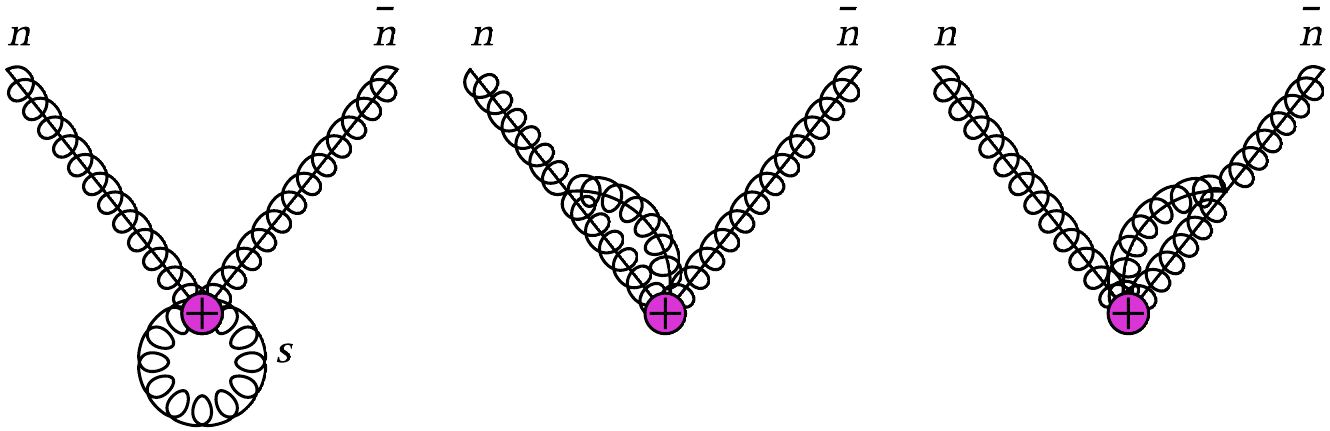}
\caption{The SCET diagrams contributing to the calculation of the Wilson coefficient $C(\omega_1, \omega_2, \mu)$.  The purple cross denotes the $n$ and $\bn$ collinear Wilson lines and the soft Wilson lines, while gluons with lines drawn through them are collinear gluons as the $n$ and $\bn$ labels indicate.  The $S$ label denotes a soft gluon in the first diagram.}
\label{QCDmatchSCETdiags}
\end{figure}

We begin by discussing the matching of QCD onto $\text{SCET}_{p_T}$ in order to extract the Wilson coefficient $C(\omega_1, \omega_2, \mu)$.  The Wilson coefficient can be extracted from the relation presented in Eq.~(\ref{QCDtoSCETmatch}) by computing radiative corrections to the matrix elements of both the QCD and SCET operators and encode their difference in $C(\omega_1,\omega_2)$. For the tree level and one loop matching one can compute the matrix elements $\langle h| O_{QCD}| \hat{p}_1, \hat{p_2}\rangle$ and $\langle h| {\cal O}| \hat{p}_1, \hat{p_2}\rangle$ in QCD and SCET$_{p_T}$ respectively where $\hat{p_1}^\mu = \bn \cdot \hat{p}_1 \frac{n^\mu}{2}$ and  $\hat{p}_2^\mu = n\cdot \hat{p}_2\frac{\bn^\mu}{2}$ denote the momenta of the initial state n-collinear and $\bn$-collinear gluons. The diagrams contributing at next-to-leading order in $\alpha_s$ in SCET$_{p_T}$ are shown in Fig.~\ref{QCDmatchSCETdiags}.  Labeling these graphs from left to right they take the form 
\bea
\text{Fig.~\ref{QCDmatchSCETdiags}a} &=& V_a(\hat{p}_1,\hat{p}_2)\> {\cal O} (\bn \cdot \hat{p}_1, n \cdot \hat{p}_2), \nn \\
\text{Fig.~\ref{QCDmatchSCETdiags}b} &=& [V_b(\hat{p}_1)- V_{b0}(\hat{p}_1)]\> {\cal O} (\bn \cdot \hat{p}_1, n \cdot \hat{p}_2),  \nn \\
\text{Fig.~\ref{QCDmatchSCETdiags}c} &=& [V_b (\hat{p}_2)-V_{b0}(\hat{p}_2)]\> {\cal O} (\bn \cdot \hat{p}_1, n \cdot \hat{p}_2),
\eea
so that the SCET$_{p_T}$ operator is multiplicatively renormalized. With on-shell external gluons and using Feynman gauge, the quantities $V_{a,b,b0}$ take the form
\bea
\label{Vscaleless}
 V_a(\hat{p}_1,\hat{p}_2) &=&  (- i g^2 C_A)   \int \frac{d^d \ell}{(2\pi)^d} \frac{2 }{\ell^2   \> n\cdot \ell\>  \>\bn \cdot \ell } , \nn \\
 V_b(\hat{p}_1) &=&  (-i g^2 C_A) \int \frac{d^d\ell}{(2\pi)^d}\frac{(\bn \cdot \ell)^2 + (\bn \cdot \hat{p}_1)^2 + \bn \cdot \ell \> \bn \cdot \hat{p}_1}{\ell^2 (\ell +\hat{p}_1)^2 \bn\cdot (\ell +\hat{p}_1)\bn \cdot \ell}, \nn \\
  V_{b0}(\hat{p}_1) &=& (- i g^2 C_A)   \int \frac{d^d \ell}{(2\pi)^d} \frac{2 }{\ell^2   \> n\cdot \ell\>  \>\bn \cdot \ell }.
\eea
We note that the collinear graphs in Figs.~\ref{QCDmatchSCETdiags}b and \ref{QCDmatchSCETdiags}c require a zero-bin~\cite{Manohar:2006nz} subtraction given by the $V_{b0}$ term in order to avoid over-counting the soft region.  These integrals are all scaleless and vanish in dimensional regularization.  The Wilson coefficient can therefore be extracted directly from the finite part of the full QCD result in dimensional regularization and is well known in the literature~\cite{Harlander:2000mg,Ahrens:2008qu}. Through next-to-leading order, it is given by
\begin{equation}
C(\bn \cdot \hat{p}_1 n\cdot \hat{p}_2, \mu) = \frac{c \, \bn \cdot \hat{p}_1n\cdot \hat{p}_2}{v} \left\{ 1+\frac{\alpha_s}{4\pi} C_A \left[ 11 +\frac{\pi^2}{6} - \text{ln}^2 \left(-\frac{\bn \cdot \hat{p}_1n\cdot \hat{p}_2}{\mu^2}\right) \right] \right\},
\label{NLOwilson}
\end{equation}
where the first term is just the result of tree level matching quoted earlier in Eq.~(\ref{tree-wil}) and we have used the property   
 $C(\omega_1,\omega_2,\mu)=C(\omega_1\omega_2,\mu)$ to write the LHS above. We note that the $11/2$ in the next-to-leading order expression arises from integrating out the top quark loop to produce an effective Higgs-glue vertex.  The hard Wilson coefficient $H(x_1x_2Q^2,\mu)=|C(x_1x_2Q^2,\mu)|^2$ appearing in the factorization theorem can be obtained from the above equation at next-to-leading order.
 
\subsection{Calculation of the iBF}

\begin{figure}
\includegraphics[]{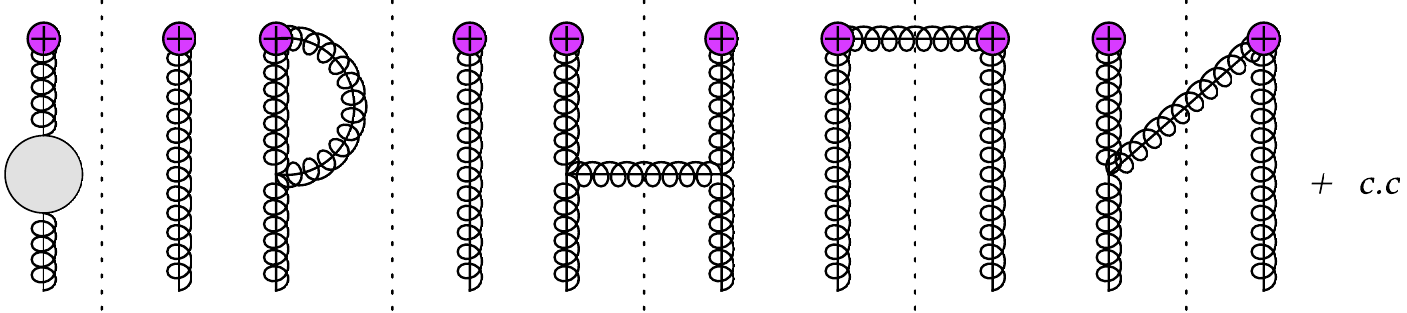}
\caption{The diagrams contributing to the next-to-leading order jet function.  The purple cross denotes the collinear Wilson lines associated with the $B_{\perp}$ field.  The blob in the left-most diagram denotes the wave-function graphs.  We note that the momentum $p_1$ 
is incoming on the left-hand side of the cut and outgoing on the right.}
\label{jetdiags}
\end{figure}

In this section we present results for the calculation of the iBFs as defined in Eqs.~(\ref{beamsoftdef}),  (\ref{jetsoftdef}), and~(\ref{def1}). We compute the n-collinear iBF by inserting a complete set of states  as
\bea
\tilde{B}_{n}^{\alpha\beta}(x_1,t_n^+ ,b_{\perp},\mu) &=& \int \frac{db^-}{4\pi} e^{\frac{i}{2}\frac{t_n^+b^-}{Q}}\sum_{\text{initial pols.}}\sum_{X_n}\langle p_1 |  \big [ g B_{1n \perp \beta}^A  (b^-,b_\perp)|X_n \rangle \nn \\
&\times& \langle X_n |\delta(\bar{{\cal P}} -x_1\bn \cdot p_1 )g B_{1n \perp \alpha}^A  (0)  \big ] | p_1 \rangle, \nn \\
\eea
and then computing the product of matrix elements. Recall that $\tilde{B}_{n}^{\alpha\beta}$ denotes the iBF without a soft zero-bin subtraction as opposed to $B_{n}^{\alpha\beta}$ which is defined with a soft zero-bin subtraction.  An analogous expression exists for the $\bn$-collinear iBF. In this section we focus on the n-collinear iBF, since the $\bn$-collinear iBF can be calculated in an analogous fashion. The lowest order result for the iBF is obtained by choosing $|X_n\rangle = |0\rangle$ and computing the tree level matrix elements to get
\bea
\label{iBFt}
\tilde{B}_n^{(0)\alpha \beta} (x_1,t_n^+,b_\perp) &=& B_n^{(0)\alpha \beta} (x_1,t_n^+,b_\perp)=  - g^2 g_\perp^{\alpha \beta} \delta(t_n^+) \delta(1-x_1). 
\eea
Higher order radiative corrections to the term in the iBF with $|X_n\rangle = |0\rangle $ correspond to pure virtual corrections with no real emissions in the final state. The one loop virtual correction corresponds to the first diagram in Fig.~\ref{jetdiags} and its conjugate and is given by
\bea
\label{iBFV}
  \tilde{B}_n^{V(1)\alpha \beta} (x_1,t_n^+,b_\perp) &=&  \tilde{B}_n^{(0)\alpha \beta} (x_1,t_n^+,b_\perp)   \Big [  V_b(p_1) + V_b(-p_1) \Big ]+w.f. , 
\eea
where $V_b$ is given in Eq.~(\ref{Vscaleless}) and $w.f.$ denotes the external wave-function graphs. Thus, this virtual correction is also scaleless and vanishes in pure dimensional regularization. 

Next we compute the single real gluon emission contribution to the iBF corresponding to a single gluon with momentum $p_g$ in the final state: $|X_n\rangle = |p_g\rangle$. The single gluon emission contribution to the iBF corresponds to the last three diagrams in Fig.~\ref{jetdiags}.  Only the second diagram in Fig.~\ref{jetdiags} contributes if we use the physical gluon polarization sum
\begin{equation}
\sum_{\text{pols.}} \epsilon^{\mu}(p_g) \epsilon^{\nu}(p_g) = -g^{\mu\nu} + \frac{p_g^{\mu} \bn^{\nu} + p_g^{\nu} \bn^{\mu}}{\bn \cdot p_g}.
\end{equation}
Direct calculation of this graph gives
\bea
\tilde{B}_{n}^{R(1)\alpha \beta}(x_1,t_n^+, b_\perp ) &=& -\frac{4 g^4 C_A}{(2\pi)^{d-1}} \int d^dp_g\, \delta(p_g^2)\delta(p_1^{-}-x_1Q -p_g^{-})\, \delta(\frac{t_n^+}{Q}-p_g^+) \>e^{i \vec{b}_{\perp} \cdot \vec{p}_{g\perp}} \nonumber \\
&\times& \frac{1}{(\vec{p}_{g\perp})^2} \left\{ g_{\perp}^{\alpha\beta}\left[1+(1-x_1)^2\right] +(d-2)\frac{(1-x_1)^2}{x_1^2} \frac{p_{g\perp}^{\alpha} p_{g\perp}^{\beta}}{p_{g\perp}^2} \right\}, \nonumber \\
\eea
where we work with the convention $d=4-2\epsilon$.
After performing the integrations over the light cone momenta $p_g^+$ and $p_g^-$ the single gluon emission contribution to the iBF can be brought into the form
\begin{eqnarray}
\tilde{B}_{n}^{R(1)\alpha\beta }(x_1,t_n^+ ,b_{\perp},\mu) &=& - \frac{2g^4 \mu^{2\epsilon} C_A}{(2\pi)^{d-1}} \int d^{2-2\epsilon} p_{g\perp} \frac{\delta[t_n^+ (1-x_1 )-\vec{p}^{\,2}_{g\perp}]}{t_n^+} e^{i \vec{b}_{\perp} \cdot \vec{p}_{g\perp}} \nonumber \\
&\times& \left\{  g_{\perp}^{\alpha\beta} \left[ 1-x_1 +\frac{1}{1-x_1}\right] +(d-2) \frac{1-x_1}{x_1^2} \frac{p_{g\perp}^{\alpha} p_{g\perp}^{\beta}}{p_{g\perp}^2}\right\}.
\label{jetint}
\end{eqnarray}
Three-dimensional quantities have been denoted by a vector symbol.  The result is the same in any $R_{\xi}$ gauge due to the structure of the Wilson-line coupling.  We note that if $t_n^+$ was integrated over, as it is when considering transverse-momentum dependent PDFs, singularities would occur at the integration boundaries.  This is avoided in the iBF.  The variable $t_n^+$ is instead set by the external kinematics through the convolution structure of the factorization theorem, as we will see in a later section when comparing to fixed-order QCD.

In order to perform the integral appearing in Eq.~(\ref{jetint}), we require the following two integrals:
\begin{eqnarray}
\mathcal{I}_1 &=&  \int d^{2-2\epsilon} p_{g\perp} \frac{\delta[t_n^+ (1-x_1 )-\vec{p}^{\,2}_{g\perp}]}{t_n^+} e^{i \vec{b}_{\perp} \cdot \vec{p}_{g\perp}} \nonumber \\
	&=& \frac{1}{2}(2\pi)^{1-\epsilon} b_{\perp}^{\epsilon} (t_n^+)^{-1-\epsilon/2} (1-x_1 )^{-\epsilon/2} J_{-\epsilon} [b_{\perp} \sqrt{t_n^+ (1-x_1 )}], \nonumber \\
\mathcal{I}_2 &=&  \int d^{2-2\epsilon} p_{g\perp} \frac{\delta[t_n^+ (1-x_1 )-\vec{p}^{\,2}_{g\perp}]}{t_n^+} e^{i \vec{b}_{\perp} \cdot \vec{p}_{g\perp}} \frac{(\vec{b}_{\perp} \cdot \vec{p}_{g\perp})^2}{\vec{p}_{g\perp}^{\,2}} \nonumber \\
	&=& \frac{b_{\perp}^2}{2} 	(2\pi)^{1-\epsilon} b_{\perp}^{\epsilon} (t_n^+)^{-1-\epsilon/2} (1-x_1 )^{-\epsilon/2} \left\{  J_{-\epsilon} [b_{\perp} \sqrt{t_n^+ (1-x_1 )}]
	-(d-3)\frac{J_{1-\epsilon} [b_{\perp} \sqrt{t_n^+ (1-x_1 )}]}{b_{\perp} \sqrt{t_n^+ (1-x_1 )}}\right\}, \nonumber \\
\label{ibfints}	
\end{eqnarray}
and we have used the notation $|\vec{b}_{\perp}| = b_{\perp}$.  To utilize these results, it is most useful to expand the integral over the second piece of Eq.~(\ref{jetint}) in terms of form factors.  We can schematically write this as
\begin{equation}
\label{ibfints-1}
\int d^{2-2\epsilon} p_{g\perp}\frac{p_{g\perp}^{\alpha} p_{g\perp}^{\beta}}{p_{g\perp}^2} = \mathcal{G}_1 \,b_{\perp}^2g_\perp^{\alpha \beta} +  \mathcal{G}_2 \,\vec{b}_{\perp}^{\alpha} \vec{b}_{\perp}^{\beta}. 
\end{equation}
We can express these in terms of the two integrals calculated in Eq.~(\ref{ibfints}):
\begin{equation}
\label{ibfints-2}
{\cal G}_1 = \frac{b_{\perp}^2 \mathcal{I}_1 -\mathcal{I}_2}{b_{\perp}^4 (d-3)}, \;\;\; {\cal G}_2 = \frac{b_{\perp}^2 \mathcal{I}_1 -(d-2)\mathcal{I}_2}{b_{\perp}^4 (d-3)}.
\end{equation}
Furthermore it is convenient to parameterize the iBF in terms of two form factors ${\cal F}_1$ and ${\cal F}_2$ as %
\begin{equation}
\tilde{B}_{n}^{R(1)\alpha\beta}(x_1,t_n^+, b_{\perp},\mu) = \mathcal{F}_1 (x_1,t_n^+,b_{\perp},\mu) g_{\perp}^{\alpha\beta} +\mathcal{F}_2(x_1,t_n^+,b_{\perp},\mu) \left[ g_{\perp}^{\alpha\beta} +(d-2)\frac{\vec{b}_{\perp}^{\alpha} \vec{b}_{\perp}^{\beta}}{b_{\perp}^2}\right].
\label{ffdecomp}
\end{equation}
Note that if the indices $\alpha$, $\beta$ are contracted the term proportional to $\mathcal{F}_2$ vanishes.  Using Eqs.~(\ref{jetint}), (\ref{ibfints}), (\ref{ibfints-1}),  and (\ref{ibfints-2}) we find the following results for the form factors ${\cal F}_1$ and ${\cal F}_2$
\begin{eqnarray}
\label{f1f2}
\mathcal{F}_1 &=& - \left[\frac{2g^4 \mu^{2\epsilon} C_A}{(2\pi)^{d-1}}\right] \frac{1}{2}(2\pi)^{1-\epsilon} b_{\perp}^{\epsilon} (t_n^+)^{-1-\epsilon/2} (1-x_1 )^{-\epsilon/2} J_{-\epsilon} [b_{\perp} \sqrt{t_n^+ (1-x_1 )}] \nonumber \\ 
&\times& \left\{1-x_1+\frac{1-x_1}{x_1^2}+\frac{1}{1-x_1}\right\}, \nonumber \\
\mathcal{F}_2 &=& - \left[\frac{2g^4 \mu^{2\epsilon} C_A}{(2\pi)^{d-1}}\right] \frac{1}{2}(2\pi)^{1-\epsilon} b_{\perp}^{\epsilon} (t_n^+)^{-1-\epsilon/2} (1-x_1)^{-\epsilon/2} J_{2-\epsilon} [b_{\perp} \sqrt{t_n^+ (1-x_1 )}] \frac{1-x_1}{x_1^2}. \nn \\
\end{eqnarray}
The form factor $\mathcal{F}_1$ has ultraviolet and infrared divergences that are regulated by $\epsilon$ in our pure dimensional regularization calculation. On the other hand, the form factor $\mathcal{F}_2$ is finite.  It is convenient to remove the fractional powers of the Bessel-function arguments by writing them in terms of hypergeometric functions in the following way:
\begin{equation}
J_{\nu}(z) \equiv \left(\frac{z}{2}\right)^{\nu} \frac{1}{\Gamma(1+\nu)} {_{0}F_{1}} \left(1+\nu ; -\frac{z^2}{4}\right).
\label{hypergeo}
\end{equation}
Since ${_{0}F_{1}} (\nu ,0)=1$, and therefore $J_2(z) \sim z^2$, it is clear that $\mathcal{F}_2$ is finite and that we can safely set $\epsilon=0$ to get
\begin{equation}
\mathcal{F}_2 = -g^2 C_A \frac{\alpha_s}{8\pi} \frac{(1-x_1 )^2}{x_1^2} b_\perp^2 \>{_{0}F_{1}} \left(3 ; -\frac{b_{\perp}^2 t_n^+ (1-x_1 )}{4}\right).
\label{f2final}
\end{equation}
For the form factor ${\cal F}_1$, we first rewrite its expression in Eq.~(\ref{f1f2}) as
\begin{eqnarray}
\mathcal{F}_1 &=& -g^2 C_A \frac{\alpha_s}{\pi} \frac{e^{\epsilon\gamma}}{\Gamma(1-\epsilon)} (1-x_1 )^{-\epsilon} \left( \frac{\hat{Q}^2}{\mu^2}\right)^{-\epsilon}\frac{1}{\hat{Q}^2}\left( \frac{t_n^+}{\hat{Q}^2}\right)^{-1-\epsilon}  \left\{1-x_1+\frac{1-x_1}{x_1^2}+\frac{1}{1-x_1}\right\} \nonumber \\ &\times& {_{0}F_{1}} \left(1-\epsilon ; -\frac{b_{\perp}^2 t_n^+ (1-x_1 )}{4}\right),
\end{eqnarray}
where we have  have switched to an $\overline{MS}$ definition of $\mu$ by replacing $\mu^{2\epsilon} \to \mu^{2\epsilon} (4\pi)^{-\epsilon} e^{\epsilon \gamma}$.  We have also used the partonic center-of-mass energy squared $\hat{Q}^2 = x_1 x_2 Q^2$ to form dimensionless ratios where required; this is convenient for the comparison with fixed-order QCD calculations, which is discussed in a later section.  We note that the ratio of scales which appears in this expression is $t_n^{+}/\mu^2$, indicating that we should choose $\mu \sim p_T$ to minimize logarithms since in the partonic calculation $t_n^+ \sim p_T^2$.  We write the expansion of the form factor $\mathcal{F}_1$ in $\epsilon$ as
\begin{equation}
\label{f1f2-1}
\mathcal{F}_1 = \frac{\mathcal{F}_{1;2}}{\epsilon^2}+\frac{\mathcal{F}_{1;1}}{\epsilon}+\mathcal{F}_{1;0},
\end{equation}
and derive the following expressions
\begin{eqnarray}
\mathcal{F}_{1;2} &=& -g^2 C_A \frac{\alpha_s}{\pi} \delta(t_n^+ ) \, \delta(1-x_1 ), \nonumber \\
\mathcal{F}_{1;1} &=& g^2 C_A \frac{\alpha_s}{\pi} \left\{ \delta(1-x_1 ) \frac{1}{\hat{Q}^2}\left[ \frac{\hat{Q}^2}{t_n^+}\right]_+ + \delta(t_n^+ ) \left[ 1-x_1+\frac{1-x_1}{x_1^2} 
	+\frac{1}{[1-x_1]_+}\right] \right. \nonumber \\ && \left. +\delta(t_n^+ )\delta(1-x_1 ) \text{ln}\frac{\hat{Q}^2}{\mu^2}\right\}, \nonumber \\
\mathcal{F}_{1;0} &=& -g^2 C_A \frac{\alpha_s}{\pi} \left\{ -\frac{\pi^2}{12} \delta(t_n^+ ) \, \delta(1-x_1 ) + \delta(1-x_1 ) \frac{1}{\hat{Q}^2}\left[\frac{\hat{Q}^2}{t_n^+}\text{ln}	\left(\frac{t_n^+}{\hat{Q}^2}\right)\right]_+  \right. \nonumber \\
	&+&\delta(t_n^+ )\left[ \left( 1-x_1 +\frac{1-x_1}{x_1^2}\right) \text{ln}(1-x_1 ) + \left[ \frac{\text{ln}(1-x_1 )}{1-x_1 }\right]_+ \right] +\frac{1}{2}\delta(t_n )\delta(1-x_1 ) 	\text{ln}^2 \frac{\hat{Q}^2}{\mu^2}\nonumber \\
	&+&  \frac{1}{\hat{Q}^2}\left[ \frac{\hat{Q}^2}{t_n^+}\right]_+ \left[1-x_1 +\frac{1-x_1}{x_1^2}+\frac{1}{[1-x_1 ]_+}\right] {_{0}F_{1}} \left(1; -\frac{b_{\perp}^2 t_n^+ (1-x_1 )}{4}\right) \nonumber \\
	 &+& \left. \delta(t_n^+ ) \left[ 1-x_1 +\frac{1-x_1}{x_1^2} +\frac{1}{[1-x_1 ]_+}\right] \text{ln}\frac{\hat{Q}^2}{\mu^2}+\delta(1-x_1 )  \frac{1}{\hat{Q}^2}\left[\frac{\hat{Q}^2}{t_n^+}\right]_+ \text{ln} \frac{\hat{Q}^2}{\mu^2} \right\}.
\label{f1final}
\end{eqnarray}
The result for the single gluon emission contribution to the n-collinear iBF in pure dimensional regularization is given by Eqs.~(\ref{ffdecomp}), (\ref{f2final}), (\ref{f1f2-1}), and (\ref{f1final}). The expression for the $\bn$-collinear iBF is obtained by repacing $t_n^+ \to t_\bn^-$ and $x_1\to x_2$ in the expression for the n-collinear iBF.

\subsection{Soft function: real emission}

In this section we give results for the computation of the iSF which was defined earlier as
\begin{equation}
\label{iSF-4}
\mathcal{S}^{-1}(\tilde{\omega}_1,\tilde{\omega}_2,b_{\perp},\mu) = \int \frac{db^+ db^-}{16\pi^2} e^{ib^+ \tilde{\omega}_1 /2} e^{i b^- \tilde{\omega}_2 /2} 
	S^{-1}(b^+ ,b^-, b_{\perp}),
\end{equation}
where the position space soft function that appears on the RHS above is defined in Eq.~(\ref{jetsoftdef}). The iSF  in the factorization theorem, before doing the Higgs phase space integrals, has the arguments
\bea
\tilde{\omega}_1 = \omega_1 -p_H^- -k_{\bar{n}}^- , \qquad \tilde{\omega}_2 &=& \omega_2 -p_H^+ -k_{n}^+ ,
\eea
as seen in Eq.~(\ref{below-pT-fac}). For convenience we introduce the notation
\bea
 t_{n}^+ = Q k_n^+, \qquad t_{\bar{n}}^- = Q k_{\bar{n}}^-,\qquad t_{n}^{max} &=& Q(\omega_2 -p_H^+ ) , \qquad t_{\bar{n}}^{max} = Q (\omega_1 -p_H^- ),
\eea
which we will often use in this section. We compute the iSF by inserting a complete  set of soft states in the position space soft function as
\begin{equation}
\label{soft-4}
S(b,\mu) = \sum_{X_{s}} \langle 0 | \bar{T} \left[\text{Tr} \left ( S_\bn T^D Y_\bn^\dagger S_n T^C S_n^\dagger \right ) (b)\right] |X_{s}\rangle \langle X_{s} | T \left[\text{Tr} \left ( S_n T^C S_n^\dagger S_\bn T^D S_\bn^\dagger \right ) (0)\right] | 0 \rangle ,
\end{equation}
and compute the product of matrix elements and use these results in Eq.~(\ref{iSF-4}).  Through next-to-leading order in the QCD coupling, the position space inverse soft function $S^{-1}(b)$ that appears in Eq.~(\ref{iSF-4}) is obtained  by inserting an overall minus sign in the $\mathcal{O}(\alpha_s)$ correction to the soft function $S(b)$ of Eq.~(\ref{soft-4}).  %

\begin{figure}
\includegraphics[]{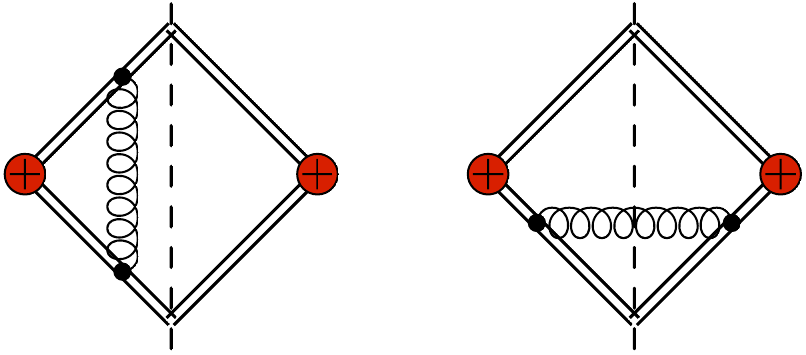}
\caption{Example diagrams contributing to the next-to-leading order iSF.  The four lines at each vertex schematically denote the soft Wilson lines appearing in the definition of the iSF $\mathcal{S}^{-1}$.  The diagram on the left corresponds to a virtual correction to the iSF and the diagram on the right corresponds to a real emission as seen by the cut through the gluon.   }
\label{SCETsoft}
\end{figure}

The lowest order result for the iSF comes from choosing $|X_{s}\rangle =|0\rangle $ and computing the tree level result which gives
\begin{equation}
\mathcal{S}^{-1 (0)}(t_n^{max}-t_n^+, t_{\bn}^{max}-t_\bn^-,b_{\perp},\mu) = \frac{N_c^2-1}{4} Q^2 \,\delta(t_{\bar{n}}^{max}-t_{\bar{n}}^-)  \delta(t_n^{max}-t_n^+) .
\end{equation}
Higher order corrections to the term with $|X_{s}\rangle =|0\rangle$ corresponds to virtual graphs with no real emissions in the final state.
At one loop, the virtual corrections correspond to the first diagram and its permutations in Fig.~\ref{SCETsoft} which gives the result
\bea
{\cal S}^{-1V(1)}(\omega_1-p_h^- , \omega_2-p_h^+ ,b_\perp) &=& -{\cal S}^{V(1)}(\omega_1-p_h^- , \omega_2-p_h^+ ,b_\perp) \nn \\
&=& {\cal S}^{(0)}(\omega_1-p_h^- , \omega_2-p_h^+ ,b_\perp) (-2i g^2 C_A) I_s\nn \\
 \eea
 where $I_s$ is the scaleless integral
 \bea
 I_s &=& 2 \int \frac{d^d \ell}{(2\pi)^d} \frac{1}{(\ell^2 +i0) \>(\bn \cdot \ell - i0) \> (n\cdot \ell +i0)} ,
 \eea
and vanishes in pure dimensional regularization.

Next we compute the contribution to the iSF from the real emission of an soft gluon corresponding to choosing $|X_{s}\rangle = | k\rangle$ for a gluon of momentum $k$, as shown in the second diagram of Fig.~\ref{SCETsoft}. Explicit computation gives
\bea
\mathcal{S}^{-1 R(1)}(\tilde{\omega}_1,\tilde{\omega}_2,b_{\perp},\mu ) &=& -\mathcal{S}^{R(1)}(\tilde{\omega}_1,\tilde{\omega}_2,b_{\perp},\mu ) \nn \\
 &=&-  \frac{N_c^2-1}{4} \frac{g^2 \mu^{2\epsilon} C_A}{(2\pi)^{d-1}}
	 \int \frac{db^+ db^-}{16\pi^2} e^{ib^+ \tilde{\omega}_1 /2} e^{i b^- \tilde{\omega}_2 /2} \int d^d k \,\delta(k^2) \frac{4}{k^+ k^-} e^{-ib \cdot k}. \nn \\
\eea
Switching to an $\overline{MS}$ definition of $\mu$ and performing integrals as before, we can derive the following expression:
\begin{equation}
\mathcal{S}^{-1 R(1)}(\tilde{\omega}_1,\tilde{\omega}_2,b_{\perp},\mu ) = -  \frac{N_c^2-1}{4} \frac{\alpha_s C_A}{\pi} \frac{e^{\epsilon\gamma}}{\Gamma(1-\epsilon)} \mu^{2\epsilon} \tilde{\omega}_1^{-1-\epsilon} \tilde{\omega}_2^{-1-\epsilon}  {_{0}F_{1}}\left( 1-\epsilon; -\frac{b_{\perp}^2 \tilde{\omega}_1 \tilde{\omega}_2 }{4}\right).
\end{equation}
The expansion in $\epsilon$ proceeds identically to that for the iBF.  Defining the expansion
\begin{equation}
\label{iSFreal-1}
\mathcal{S}^{-1 R(1)}(\frac{t_n^{max}-t_n^+}{Q}, \frac{t_{\bn}^{max}-t_\bn^-}{Q},b_{\perp},\mu ) = \frac{S_2}{\epsilon^2}+\frac{S_1}{\epsilon}+S_0,
\end{equation}
we arrive at the result
\begin{eqnarray}
\label{iSFreal-2}
S_2 &=& - \frac{N_c^2-1}{4} \frac{\alpha_s C_A}{\pi} Q^2\,\delta(t_{\bar{n}}^{max}-t_{\bar{n}}^-)  \delta(t_n^{max}-t_n^+), \nonumber \\
S_1 &=&  \frac{N_c^2-1}{4} \frac{\alpha_s C_A}{\pi} Q^2 \left\{ \delta(t_{\bar{n}}^{max}-t_{\bar{n}}^-) \frac{1}{Q^2}\left[ \frac{Q^2}{t_n^{max}-t_n^+}\right]_+ \
	+\delta(t_n^{max}-t_n^+) \frac{1}{Q^2}\left[ \frac{Q^2}{t_{\bar{n}}^{max}-t_{\bar{n}}^-} \right]_+ \right. \nonumber \\
	 &+& \left. \delta(t_{\bar{n}}^{max}-t_{\bar{n}}^-)  \delta(t_n^{max}-t_n^+) \,\text{ln} \frac{Q^2}{\mu^2} \right\}, \nonumber \\
S_0 &=& -\frac{N_c^2-1}{4} \frac{\alpha_s C_A}{\pi} Q^2 \left\{ -\frac{\pi^2}{12} \delta(t_{\bar{n}}^{max}-t_{\bar{n}}^-)  \delta(t_n^{max}-t_n^+) +\frac{1}{2} 	\delta(t_{\bar{n}}^{max}-t_{\bar{n}}^-)\delta(t_n^{max}-t_n^+) \text{ln}^2 \frac{Q^2}{\mu^2} \right. \nonumber \\
	&+& \delta(t_{\bar{n}}^{max}-t_{\bar{n}}^-) \frac{1}{Q^2}\left[ \frac{Q^2}{t_n^{max}-t_n^+}\right]_+ \text{ln} \frac{Q^2}{\mu^2} + \delta(t_n^{max}-t_n^+)  \frac{1}{Q^2}\left[ \frac{Q^2}{t_{\bar{n}}^{max}-	t_{\bar{n}}^-} \right]_+ \text{ln} \frac{Q^2}{\mu^2} \nonumber \\
	&+& \delta(t_{\bar{n}}^{max}-t_{\bar{n}}^-) \frac{1}{Q^2}\left[ \frac{Q^2}{t_n^{max}-t_n^+}\text{ln}\frac{t_n^{max}-t_n^+}{Q^2}\right]_+ + \delta(t_n^{max}-t_n^+)  \frac{1}{Q^2}\left[ \frac{Q^2}{t_{\bar{n}}^{max}-	t_{\bar{n}}^-}\text{ln}\frac{t_{\bar{n}}^{max}-t_{\bar{n}}^-}{Q^2}\right]_+ \nonumber \\
	&+&  \left. \frac{1}{Q^4}\left[ \frac{Q^2}{t_n^{max}-t_n^+}\right]_+  \left[ \frac{Q^2}{t_{\bar{n}}^{max}-t_{\bar{n}}^-} \right]_+  {_{0}F_{1}}\left( 1; -\frac{b_{\perp}^2 (t_n^{max}-t_n^+)	(t_{\bar{n}}^{max}-t_{\bar{n}}^-)}{4Q^2}\right) \right\}.
\label{softfunc}
\end{eqnarray}
We have used the scale $Q$ to define dimensionless ratios in logarithms and plus distributions, as in the calculation of the iBF.  We note that when this result is plugged into the full expression for ${\cal G}^{ij}$ in Eq.~(\ref{utfunc}), the logarithm that appears is $\text{ln} \left [\hat{t} \hat{u} / (\mu^2 \hat{Q}^2) \right]$, where $\hat{t},\hat{u}$ are the partonic Mandelstam variables.  This indicates that we should choose $\mu \sim p_T$ in the soft function, and is the motivation for our decision to set $\mu_S = \mu_T$ in Section~\ref{sec:eftsum}.  

\subsection{iBFs to PDFs}
\label{iBFPDFfix}

In order to be able to sensibly match the iBFs onto the PDFs with a finite Wilson coefficient, as indicated in Eq.~(\ref{iBFPDF}), the infrared divergences in the iBF and PDF must match.  In this section, we show that this can be manifestly seen from the integrand level expressions for the iBF and PDF at the order we are working. We first note that the tree level expression for the iBF to PDF matching is given by
\bea
\label{Itree}
 {\cal I}_{n;g,i}^{(0)\beta \alpha} (\frac{z}{z'}, t_n^+,b_\perp,\mu) &=& g^2 g_\perp^{\alpha \beta}\delta(t_n^+) \delta(1 - \frac{z}{z'}),
\eea
which can be straightforwardly verified from Eq.~(\ref{iBFPDF}).
Next we consider the integrand-level expression for the real gluon emission contribution to the iBF in Eq.~(\ref{jetint}) where the integrals over the light cone momentum components of the final state gluon have already been performed. The infrared divergences now occur in the limit where the perpendicular momentum of the final state gluon vanishes $p_{g\perp} \to 0$.  In this limit, the iBF real-emission contribution in Eq.~(\ref{jetint}) becomes
\begin{equation}
\label{IRlim}
\left[B_{n}^{R(1)\alpha\beta}(x_1,t_n^+ ,b_{\perp},\mu)\right]_{\text{IR}} = -g_{\perp}^{\alpha\beta} \delta(t_n^+) \frac{2g^4 \mu^{2\epsilon} C_A}{(2\pi)^{d-1}} \int \frac{d^{2-2\epsilon} p_{g\perp}}{\vec{p}_{g\perp}^{\,2}} \left\{ 1-x_1 +\frac{1}{1-x_1} + \frac{1-x_1}{x_1^2} \right\}.
\end{equation}
A calculation of the real-emission contribution to the PDF defined in Eq.~(\ref{PDF}) can be done inserting a single-gluon state between the field operators in the same way as was done for the iBF.  The result at the integrand level is 
\begin{equation}
f^{(R)}_{g/P}(x_1) = \frac{2g^2 \mu^{2\epsilon} C_A}{(2\pi)^{d-1}} \int \frac{d^{2-2\epsilon} p_{g\perp}}{\vec{p}_{g\perp}^{\,2}} x_1  \left\{ 1-x_1 +\frac{1}{1-x_1} + \frac{1-x_1}{x_1^2} \right\},
\end{equation}
from which we see that
\begin{equation}
\label{ircomp}
\left[B_{n}^{R(1)\alpha\beta}(x_1,t_n^+ ,b_{\perp},\mu)\right]_{\text{IR}} = -g^2 g_{\perp}^{\alpha\beta} \delta(t_n^+ ) \frac{f^{R(1)}_{g/P}(x_1)}{x_1}.
\end{equation}
Thus, the infrared divergences which appear upon integration over $p_{g\perp}$ are identical for both the iBF and the PDF, up to an overall multiplicative factor. Similarly, we note from Eqs.~(\ref{iBFV}) and (\ref{iBFt}) that the contribution of one loop virtual corrections to the iBF take the form 
\bea
\label{IRV}
B_{n}^{V(1)\alpha\beta}(x_1,t_n^+ ,b_{\perp},\mu) &=& -g^2 g_{\perp}^{\alpha\beta} \delta(t_n^+ ) \frac{f^{V(1)}_{g/P}(x_1)}{x_1},
\eea
so that once again the infrared structure is the same for the iBF and PDF virtual contributions up to the same multiplicative factor. Thus, the infrared divergences in the sum of the real and virtual graphs for the iBF can be matched to those in the PDF 
as in Eq.~(\ref{iBFPDF}) with the finite Wilson coefficient given in Eq.~(\ref{Itree}). Additional contributions to the Wilson coefficient will come from the finite terms in the iBF.

In fact, since the PDFs are scaleless and we have shown that the infrared divergences of the iBF and the PDF match up, the Wilson coefficient for the iBF to PDF matching is just given by the finite part of the iBF in pure dimensional regularization as indicated in Eq.~(\ref{iBFPDF1}). For a more detailed explanation of this point see~\cite{Manohar:2003vb}.   Using Eqs.~(\ref{iBFPDF1}) and (\ref{ffdecomp}) we can now write the iBF to PDF matching coefficient at the first order beyond tree level as
\begin{equation}
\mathcal{I}_{n}^{(1)\alpha\beta}(\frac{x_1}{x_1^{'}},t_n^+, b_{\perp},\mu) = -x_1\left\{ \mathcal{F}_{1;0} (\frac{x_1}{x_1^{'}},x_1^{'}t_n^+,b_{\perp},\mu) g_{\perp}^{\alpha\beta} +\mathcal{F}_2(\frac{x_1}{x_1^{'}},x_1^{'}t_n^+,b_{\perp},\mu) \left[ g_{\perp}^{\alpha\beta} +(d-2)\frac{\vec{b}_{\perp}^{\alpha} \vec{b}_{\perp}^{\beta}} {b_{\perp}^2}\right]\right\}, \\ 
\label{beamwilson}
\end{equation}
where the form factors $\mathcal{F}_{1;0}$ and $\mathcal{F}_{2}$ are given in Eq.~(\ref{f1final}) and (\ref{f2final}) respectively.
We perform an explicit check of this expression in section~\ref{sec:consistency}  by using this expression in the factorization theorem and comparing the fixed order result in full QCD.

\section{Running}
\label{running}

In the factorization formula of Eq.~(\ref{fac-summary}), the logarithms of $m_h/p_T$ are summed via the RG evolution of the hard coefficient $H(x_1 x_2 Q^2, \mu)$ between the scales $\mu_Q\sim m_h$ and $\mu_T\sim p_T$. The factor $H(x_1 x_2 Q^2, \mu_Q;\mu_T)$ in the factorization formula denotes this RG evolved hard coefficient. The logarithms of $\Lambda_{QCD}/p_T$ are summed via the DGLAP evolution of the PDFs up to the $\mu_T\sim p_T$ scale. The DGLAP evolved PDFs are denoted by $f_{i/P}(x,\mu_T)$.  In this section we give details of this RG running above and below the $p_T$ scale. 

\subsection{Running above p$_T$}
\label{topdownrunning}

The anomalous dimension of the hard coefficient $H$ can be obtained from the anomalous dimension of the operator
\bea
\label{SCETop}
{\cal O} (\omega_1, \omega_2) =  g_{\mu \nu} h \Tr \big [ ( g B_{n\perp}^\mu)_{\omega_1} (  g B_{\bn \perp}^\nu)_{\omega_2} \big ],
\eea
which was first defined in Eq.~(\ref{vertexop}). The RG equation for ${\cal O}$ has the form
\bea
\label{local-run}
\mu \frac{d}{d\mu} {\cal O} = - \gamma_O \>{\cal O},
\eea
where the anomalous dimension $\gamma_O$ is given by
\bea
\gamma_O = \frac{1}{Z_O} \mu \frac{d}{d\mu} Z_{O},
\eea
and $Z_O$ is the renormalization constant that relates the bare operator ${\cal O}^b$ to the renormalized operator ${\cal O}$ by
\bea
{\cal O}^b = Z_O {\cal O}.
\eea
We use the notation $\omega_{1} = x_{1} \>\bn \cdot p_{1} = \bn \cdot\hat{p}_{1} $, $\omega_{2} = x_{2} \>n\cdot p_{2} = n\cdot\hat{p}_{2} $ and $\bn \cdot \hat{p}_{1}n\cdot \hat{p}_{2}= x_1 x_2 Q^2 = \hat{Q}^2$. The anomalous dimension of ${\cal O}(\bn \cdot\hat{p}_1, n\cdot \hat{p}_2)$ is typically written in the general form
\bea
\label{anom-gen-1}
\gamma_{O} = \Gamma_{O} [\alpha_s] \ln \frac{-\hat{Q}^2}{\mu^2} + \gamma_O [\alpha_s],
\eea
where the cusp ($\Gamma_O$) and non-cusp ($\gamma_O$) pieces have expansions in $\alpha_s$ as
\bea
\label{anom-gen-2}
\Gamma_{O} [\alpha_s] = \frac{\alpha_s}{4\pi} \Gamma_0^O + \Big [  \frac{\alpha_s}{4\pi} \Big ]^2 \Gamma_1^O +\cdots, \qquad \gamma_{O} [\alpha_s] = \frac{\alpha_s}{4\pi} \gamma_0^O + \Big [  \frac{\alpha_s}{4\pi} \Big ]^2 \gamma_1^O + \cdots
\eea
Equivalently, one can think of the running of the Wilson coefficient $C$ of the operator ${\cal O}$ which has the RG equation
\bea
\label{anom-gen-3}
\mu \frac{d}{d\mu} C= \gamma_c \> C, \qquad \qquad \gamma_c =-\frac{1}{Z_c} \mu \frac{d}{d\mu} Z_{c}, \qquad Z_c = Z_O^{-1},\qquad \gamma_c = \gamma_O.
\eea
The running of absolute value squared of the Wilson coefficient $H(\hat{Q}^2,\mu) = |C(\hat{Q}^2,\mu)|^2$, which is the quantity that appears in the factorization theorem, is given by
\bea
\label{anom-gen-4}
\mu \frac{d}{d\mu} H = \gamma_H \>H, \qquad \gamma_H =-\frac{1}{Z_H} \mu \frac{d}{d\mu} Z_{H}, \qquad Z_H = (Z_O +Z_O^*)^{-1}\qquad \gamma_H = \gamma_c + \gamma_c^*, 
\eea
so that
\bea
\label{anom-gen-5}
\gamma_H  = \Gamma_H[\alpha_s] \ln \frac{\hat{Q}^2}{\mu^2} + \gamma_H[\alpha_s],
\eea
where $\Gamma_H$ and $\gamma_H$ have the $\alpha_s$ expansions
\bea
\label{anom-gen-6}
\Gamma_{H} [\alpha_s] = \frac{\alpha_s}{4\pi} \Gamma_0^H + \Big [  \frac{\alpha_s}{4\pi} \Big ]^2 \Gamma_1^H +\cdots, \qquad \gamma_{H} [\alpha_s] = \frac{\alpha_s}{4\pi} \gamma_0^H + \Big [  \frac{\alpha_s}{4\pi} \Big ]^2 \gamma_1^H +\cdots
\eea
The result for the anomalous dimension $\gamma_{O}$ is well-known~\cite{ Harlander:2000mg, Ravindran:2004mb, Idilbi:2005er, Becher:2007ty, Ahrens:2008nc}. Since the SCET$_{p_T}$ graphs that renormalize the operator ${\cal O}$ are scaleless, as discussed in Section~\ref{fixedorder}, the anomalous dimension can be read off~\cite{Manohar:2003vb} from the infrared pole structure of loop corrections~\cite{Harlander:2000mg, Ravindran:2004mb} to the QCD operator $O_{QCD}$ defined in Eq.~(\ref{QCDop}). In this section we reproduce this known result at one loop by explicit calculation of the SCET$_{p_T}$ graphs using off-shell external gluons to regulate the infrared divergences and dimensional regularization to regulate the ultraviolet divergences. We do this calculation in both Feynman gauge and using the background field method as an additional check on our calculation.

%%%%%%%%%%%
%%%%%%%%%%%
%%%%%%%%
\subsubsection{Anomalous dimension computation in Feynman gauge}

At one loop, there are three graphs in SCET$_{p_T}$, in addition to the wave function graphs, which determine the anomalous dimension $\gamma_{O}=\gamma_c$, as shown in Fig.~\ref{QCDmatchSCETdiags}. The first graph involves a soft gluon loop emanating from the soft Wilson lines and the remaining two graphs
involve a (n,$\bn$)-collinear gluon originating from the (n,$\bn$)-collinear Wilson line and attaching to the (n,$\bn$)-collinear
 gluon. These diagrams are given by
\bea
\text{Fig. \ref{QCDmatchSCETdiags}a} &=& V_a(\hat{p}_1,\hat{p}_2)\> {\cal O} (\bn \cdot \hat{p}_1, n \cdot \hat{p}_2), \nn \\
\text{Fig. \ref{QCDmatchSCETdiags}b} &=& \big [ V_b(\hat{p}_1) - V_{b0}(\hat{p}_1) \big ] \> {\cal O}(\bn \cdot \hat{p}_1, n \cdot \hat{p}_2),  \nn \\
\text{Fig. \ref{QCDmatchSCETdiags}c} &=&\big [ V_b (\hat{p}_2)-  V_{b0}(\hat{p}_2) \big ] \> {\cal O} (\bn \cdot \hat{p}_1, n \cdot \hat{p}_2), \nn \\
\eea
where the quantities $V_a(\hat{p}_1,\hat{p}_2), V_b(\hat{p}_1), $ and $ V_{b0}(\hat{p}_1)$ with off-shell momenta $\hat{p}_{1,2}$ are given by
\bea
\label{VaVb-1}
 V_a(\hat{p}_1,\hat{p}_2) &=& (ig^2 C_A)  \> \int \frac{d^d \ell}{(2\pi)^d} \frac{2 \bn \cdot \hat{p}_1 n\cdot \hat{p}_2}{(\ell^2 + i0) [ n\cdot \ell \>\bn \cdot \hat{p}_1+ \hat{p}_1^2+i0] [-\bn \cdot \ell \>n\cdot \hat{p}_2+ p_2^2+ i0]} \nn \\
 V_b(\hat{p}_1) &=&  (-i g^2 C_A) \int \frac{d^d\ell}{(2\pi)^d}\frac{(\bn \cdot \ell)^2 + (\bn \cdot \hat{p}_1)^2 + \bn \cdot \ell \> \bn \cdot \hat{p}_1}{(\ell^2 + i0) [(\ell +\hat{p}_1)^2 +i0] [\bn\cdot (\ell +\hat{p}_1) +i0][\bn \cdot \ell+i0]}\nn \\
 V_{b0}(\hat{p}_1) &=& (- i g^2 C_A)   \int \frac{d^d \ell}{(2\pi)^d} \frac{2 \bn \cdot \hat{p}_1}{(\ell^2 + i0)  \> (n\cdot \ell\> \bn \cdot \hat{p}_1 + \hat{p}_1^2 + i0)\>[\bn \cdot \ell +i0]},
 \eea
and note that in the limit $p_1^2,p_2^2\to 0$ these integrals reduce to the expected results of Eq.(\ref{Vscaleless}). 
Note that the (n,$\bn$)-collinear graphs require the zero-bin~\cite{Manohar:2006nz} subtraction terms $V_{b0}(\hat{p}_1)$ and $V_{b0}(\hat{p}_2)$ respectively.  These zero-bin subtractions are necessary to remove the soft region in the collinear graphs and thus avoid double counting the soft region. The results for these graphs and the zero-bin subtraction term are
 \bea
 \label{VaVb-2}
 V_{a}(\hat{p}_1,\hat{p}_2) &=&  \frac{\alpha_s C_A}{4 \pi} \Big [- \frac{2}{\epsilon_{\text{UV}}^2} - \frac{2}{\epsilon_{\text{UV}}} \ln\Big ( \frac{-\mu^2 \hat{Q}^2}{\Delta^4}\Big )-  \ln^2 \Big (\frac{- \mu^2 \hat{Q}^2}{\Delta^4} \Big ) - \frac{\pi^2}{2} \Big ], \nn \\
%%%%%%%%%%%
 V_{b}(\hat{p}_1) &=& \frac{\alpha_s C_A}{4 \pi} \Big [\frac{2}{\epsilon_{\text{UV}}\epsilon_{\text{IR}}} + \frac{1}{\epsilon_{\text{UV}}} + \frac{2}{\epsilon_{\text{IR}}} \ln\Big ( \frac{-\mu^2}{\Delta^2}\Big ) + ( \frac{2}{\epsilon_{\text{UV}}} - \frac{2}{\epsilon_{\text{IR}}}\Big ) \ln \Big ( \frac{\mu}{\bn \cdot \hat{p}_1}\Big ) \nn \\
 &+& \frac{1}{2} \ln^2 \Big (\frac{-\mu^2}{\Delta^2}\Big ) +  \ln \Big (\frac{-\mu^2}{\Delta^2} \Big ) - \frac{\pi^2}{6} + \frac{1}{2} \Big ], \nn \\
 %%%%
 V_{b0}(\hat{p}_1) &=&-\frac{\alpha_sC_A}{4\pi} \Big [\Big ( \frac{2}{\epsilon_{\text{UV}}} - \frac{2}{\epsilon_{\text{IR}}}\Big )\Big \{ \frac{1}{\epsilon_{\text{UV}}} + \ln   \Big ( \frac{\mu^2}{-\Delta^2}\Big)  - \ln \Big ( \frac{\mu}{\bn \cdot \hat{p}_1}\Big) \Big \} \Big ] ,
  \eea
where we have used an off-shell infrared regulator $\Delta^2 = \hat{p}_1^2 =\hat{p}_2^2$.  We note that the off-shellness of the external gluons does not completely regulate the infrared divergences individually  in $V_{b}$ and $V_{b0}$. Furthermore, $V_b$ contains a UV-IR mixed double pole. However,  in the  zero-bin subtracted combination $V_b - V_{b0}$ the infrared divergences are completely regulated by off-shell external gluons and all pole terms are pure UV. This is critical for being able to properly extract the anomalous dimension from the pole structure.  The sum of these graphs is
\bea
\label{sum-ver}
\text{Fig. \ref{QCDmatchSCETdiags}a} + \text{Fig. \ref{QCDmatchSCETdiags}b}+\text{Fig. \ref{QCDmatchSCETdiags}c}&=&  \frac{\alpha_s C_A}{4 \pi} \Big [ \frac{2}{\epsilon_{\text{UV}}^2} + \frac{2}{\epsilon_{\text{UV}}} + \frac{2}{\epsilon_{\text{UV}}} \ln\Big ( \frac{- \mu^2}{\hat{Q}^2}\Big ) +   \ln^2 \Big (\frac{-\mu^2}{ \Delta^2} \Big ) -  \ln^2\Big ( \frac{-\mu^2 \hat{Q}^2}{\Delta^4}\Big ) \nn \\
%%%%
& +& 2 \ln \Big ( \frac{-\mu^2} {\Delta^2}\Big )  -\frac{2\pi^2}{3}  +1 \Big ].
\eea
Note that the mixed UV-IR terms proportional to $\frac{1}{\epsilon_{\text{UV}}}\ln \Delta^2$  cancel between the soft and collinear graphs so that the UV and IR divergences are clearly separated in the sum of all graphs. Along with the gluon wave-function  and strong  coupling renormalization constants, we can now extract the anomalous dimension of ${\cal O}$. The bare operator $\hat{O}^b$ obtained  by expanding the bare $B_\perp^b$ fields in the strong coupling and keeping the leading term, can be related to the  renormalized operator $\hat{O}$ as
\bea
\label{bareO}
\hat{O}^b =  g_{\mu \nu} h \Tr \big [ (g A_{n\perp}^{b\mu})_{\omega_1} (g  A_{\bn \perp}^{b\nu})_{\omega_2} \big ] =  Z_O \> \hat{O}, 
\eea
so that the renormalized operator can be written as
\bea
\label{bareO-1}
\hat{O} &=&  g_{\mu \nu} h \Tr \big [ (g  A_{n\perp}^{\mu})_{\omega_1} (g A_{\bn \perp}^{\nu})_{\omega_2} \big ] + \Big (\frac{Z_g^2 Z_3}{Z_O} - 1 \Big ) g_{\mu \nu} h \Tr \big [ ( gA_{n\perp}^{\mu})_{\omega_1} ( g A_{\bn \perp}^{\nu})_{\omega_2} \big ] , \nn \\
&=&  g_{\mu \nu} h \Tr \big [ (g A_{n\perp}^{\mu})_{\omega_1} (g A_{\bn \perp}^{\nu})_{\omega_2} \big ] + \Big (2\delta Z_g + \delta Z_3 - \delta Z_O \Big ) g_{\mu \nu} h \Tr \big [ ( gA_{n\perp}^{\mu})_{\omega_1} (g A_{\bn \perp}^{\nu})_{\omega_2} \big ] +\cdots, \nn \\
\eea
where $Z_3$ and $Z_g$ are the gluon wave function renormalization and coupling renormalization constants respectively
\bea
A^b = Z_3 A, \qquad g_b = g Z_g \mu^\epsilon ,
\eea
and we have defined
\bea
Z_3 = 1 + \delta Z_3, \qquad Z_g = 1+ \delta Z_g \qquad Z_O = 1 + \delta Z_{O}.
\eea
At one loop, in the $\overline{\text{MS}}$ scheme,  the gluon wave function and the strong coupling  renormalization constants are known~\cite{Buras:1998raa}:
\bea
\label{wf}
\delta Z_3^{(1)} =  -\frac{\alpha_s}{4\pi} \Big [ \frac{2}{3} n_f - \frac{5}{3}N_c \Big ]\frac{1}{\epsilon}, \qquad \delta Z_g^{(1)} =  -\frac{\alpha_s}{4\pi}\Big [ \frac{11}{6} N_c - \frac{2}{6}n_f \Big ]\frac{1}{\epsilon}.
\eea
We see from the above results that in the $\overline{\text{MS}}$ scheme  $\delta Z_{O}$ at one loop is given by
\bea
 \delta Z_O^{(1)} &=& 2\>\delta Z_g^{(1)} + \delta Z_3^{(1)} + \big [ V_a(\hat{p}_1,\hat{p}_2) + V_b(\hat{p}_1)- V_{b0}(\hat{p}_1) + V_b(\hat{p}_2)-V_{b0}(\hat{p}_2)\big ]_{\text{div}},
\eea
where the subscript ``div" above indicates the UV divergent part of the sum of graphs. Putting these results together we get the one loop result\bea 
 \delta Z_O^{(1)}&=& -\frac{\alpha_s N_c}{4\pi} \frac{2}{\epsilon} + \frac{\alpha_s C_A}{4\pi} \Big [ \frac{2}{\epsilon^2} + \frac{2}{\epsilon} + \frac{2}{\epsilon} \ln \Big ( \frac{-\mu^2}{\hat{Q}^2}\Big ) \Big ].
\eea
Using $N_c=C_A = 3$ this result simplifies to
\bea
\label{zO-1}
\delta Z_O^{(1)} =   \frac{\alpha_s C_A}{4\pi} \Big [ \frac{2}{\epsilon^2} + \frac{2}{\epsilon} \ln \Big ( \frac{-\mu^2}{\hat{Q}^2}\Big ) \Big ],
\eea
from which we get the well known result for the one loop anomalous dimension $\gamma_O^{(1)}$ 
\bea
\label{cusp-anom}
\gamma_O^{(1)} = \frac{\alpha_s C_A}{\pi} \ln \Big (\frac{-\hat{Q}^2}{\mu^2} \Big).
\eea
In the notation of Eqs.~(\ref{anom-gen-1}) and (\ref{anom-gen-2}) this result for the one loop anomalous dimension corresponds to
\bea
\Gamma_0^O = 4, \qquad \gamma_0^O =0.
\eea
Thus, the non-cusp piece $\gamma_0^O$ vanishes. Putting this together with the notation in Eqs.~(\ref{anom-gen-4}),(\ref{anom-gen-5}), and (\ref{anom-gen-5}) the anomalous dimension of the hard coefficient $H$ at one loop is by
\bea
 \Gamma_0^H = -2 \Gamma_0^O = -8, \qquad \gamma_0^H = -2 \gamma_0^O =0,
\eea
and also contains only a cusp contribution.

\subsubsection{Anomalous dimension computation in background field method}

One can also compute the anomalous dimension using the background field method with Feynman gauge for internal lines in the loop graphs, which we show as an additional check on our SCET computational techniques. In this method, the combination $g A_{n\perp}^{\mu} $ is not renormalized. In this case, the result for the soft graph $V_a^{\text{bfg}}$ is unaffected
\bea
V_{a}^{\text{bfg}}(\hat{p}_1,\hat{p}_2)= V_a(\hat{p}_1,\hat{p}_2) &=&  \frac{\alpha_s C_A}{4 \pi} \Big [- \frac{2}{\epsilon_{\text{UV}}^2} - \frac{2}{\epsilon_{\text{UV}}} \ln\Big ( \frac{-\mu^2 \hat{Q}^2}{\Delta^4}\Big )-  \ln^2 \Big (\frac{- \mu^2 \hat{Q}^2}{\Delta^4} \Big ) - \frac{\pi^2}{6} \Big ].
\eea
 The collinear graph $V_b$ is modified to
 \bea
  V_b^{\text{bfg}}(\hat{p}_1) &=&  (-i g^2 C_A) \int \frac{d^d\ell}{(2\pi)^d}\frac{ (\bn \cdot \hat{p}_1)^2 }{(\ell^2 + i0) [(\ell +\hat{p}_1)^2 +i0] [\bn\cdot (\ell +\hat{p}_1) +i0][\bn \cdot \ell+i0]}\nn \\
  &=& \frac{\alpha_s C_A}{4 \pi} \Big [\frac{2}{\epsilon_{\text{UV}}\epsilon_{\text{IR}}} + \frac{2}{\epsilon_{\text{IR}}} \ln\Big ( \frac{-\mu^2}{\Delta^2}\Big ) +  \ln^2 \Big (\frac{-\mu^2}{\Delta^2}\Big )  - \frac{\pi^2}{6}  \Big ], \nn \\
 \eea
 and the zero-bin subtraction term is unaffected
\bea
V_{b0}^{\text{bfg}} &=&V_{b0}.
\eea
One can now extract the anomalous dimension following the same procedure as in the Feynman gauge computation of the last section. The bare operator obtained at leading order by expanding the bare $B_\perp^b$ field can again be written in terms of the renormalized operator as in Eq.~(\ref{bareO}),
so that the renormalized operator can be written as
\bea
\hat{O} &=&  g_{\mu \nu} h \Tr \big [ (g  A_{n\perp}^{\mu})_{\omega_1} (g A_{\bn \perp}^{\nu})_{\omega_2} \big ] + \Big (\frac{1}{Z_O} - 1 \Big ) g_{\mu \nu} h \Tr \big [ (g A_{n\perp}^{\mu})_{\omega_1} (g A_{\bn \perp}^{\nu})_{\omega_2} \big ] , \nn \\
&=&  g_{\mu \nu} h \Tr \big [ (g A_{n\perp}^{\mu})_{\omega_1} (g A_{\bn \perp}^{\nu})_{\omega_2} \big ]  - \delta Z_O \> g_{\mu \nu} h \Tr \big [ (g A_{n\perp}^{\mu})_{\omega_1} (g A_{\bn \perp}^{\nu})_{\omega_2} \big ] +\cdots.
\eea
Note that in the background field method, compared to Eq.~(\ref{bareO-1}), the wave function ($Z_3$) and coupling constant ($Z_g$) renormalization constants cancel out and do not appear. In other words, the combination $g A_{n,\bn \perp}^{\mu}$ does not undergo renormalization in the background field gauge. Thus, we get
\bea
\delta Z_O^{(1)} =  \big [ V_a^{\text{bfg}}(\hat{p}_1,\hat{p}_2) + V_b^{\text{bfg}}(\hat{p}_1)- V_{b0}^{\text{bfg}}(\hat{p}_1) + V_b^{\text{bfg}}(\hat{p}_2)-V_{b0}^{\text{bfg}}(\hat{p}_2)\big ]_{\text{div}},
\eea
which gives the same result as in Eq.~(\ref{zO-1}) and thus leads to the same anomalous dimension as derived from the Feynman gauge calculation.

\subsection{Running below p$_T$}

Logarithms of $\Lambda_{QCD}/p_T$ are summed via the standard DGLAP equations for the PDFs. For completeness we write down the RG equation for the gluon PDF at leading order
\bea
\mu \frac{d}{d\mu} f_{g/P}(x,\mu) &=& \frac{\alpha_s(\mu)}{\pi} \int_x^1 \frac{dz}{z} P_{g \leftarrow g} (z)  f_{g/P}(\frac{x}{z},\mu),
\eea
where $P_{g \leftarrow g} (z)$ is the standard gluon splitting function and  we have ignored contributions from the quark and antiquark PDFs.  This DGLAP equation is used to evolve the gluon PDF up to the $\mu_T \sim p_T$ scale. These evolved PDFs, denoted by $f_{g/P}(x,\mu_T)$, are the ones that appear in the factorization theorem as in Eq.~(\ref{fac-summary}).

%%%%%%%%%%%%

%%%%%%%%
\section{Consistency checks}
\label{sec:consistency}

\subsection{Cross-section comparisons with full QCD}

In this section we show that the iBFs and iSF computed in fixed-order perturbation theory in Section~\ref{fixedorder} reproduce the QCD cross section up to terms 
suppressed by the ratio $p_T / m_h$.

\subsubsection{Leading-order cross section}

We begin by reproducing the leading-order cross section of perturbative QCD for $gg \to h$, with no final state gluon emissions, using our factorization formula in Eq.~(\ref{pT-fac-final}).  We utilize the leading-order expressions for the Wilson coefficient $H$, the iBFs and the iSF:
\bea
H^{(0)}(x_1 x_2 Q^2) &=& \Big ( \frac{c \>x_1 x_2 Q^2 }{v}\Big )^2, \nn \\
{\cal I}_{n,gg}^{(0)\alpha \beta}(\frac{x_1}{x_1^{'}},t_n^+,b_\perp) &=& -g^2g_\perp^{\alpha \beta}\delta(t_n^+)\delta (1-\frac{x_1}{x_1^{'}}) , \nn \\
{\cal I}_{\bn,gg}^{(0)\alpha \beta}(\frac{x_2}{x_2^{'}},t_{\bn}^-,b_\perp) &=& -g^2g_\perp^{\alpha \beta}\delta(t_{\bn}^-)\delta (1-\frac{x_2}{x_2^{'}}) , \nn \\
\mathcal{S}^{-1 (0)}(\frac{t_n^{max}-t_n^+}{Q}, \frac{t_{\bn}^{max}-t_\bn^-}{Q},b_{\perp}) &=& \frac{N_c^2-1}{4} Q^2 \,\delta(t_{\bar{n}}^{max}-t_{\bar{n}}^-)  \delta(t_n^{max}-t_n^+).
\eea
Inserting these results into Eq.~(\ref{pT-fac-final}), the integrals over $x_{1,2}$, $t_n$, $t_{\bn}$, and $b_{\perp}$ can be easily performed to give
\begin{equation}
\frac{d^2 \sigma^{(0)}_{PP \to h}}{du \,dt} = \frac{\pi}{576 v^2} \left( \frac{\alpha_s}{\pi}\right)^2 \int dx_1 \, dx_2 \, f_{g/P} (x_1 ) f_{g/P}(x_2) \delta(1-z) \delta (u-m_h^2+ x_1Q^2) \delta (t-m_h^2 +x_2 Q^2 ). \\
\end{equation}
We have introduced the variable $z = m_h^2 / (x_1 x_2 Q^2) = m_h^2 /\hat{Q}^2$, which measures the amount of energy released into final-state particles besides the Higgs (for $z=1$, all energy in the partonic scattering process goes to the Higgs).  It is convenient for this expression and for later results to switch to partonic Mandelstam variables $\hat{u}$, $\hat{t}$ defined by $\hat{u} -m_h^2 = x_2 (u-m_h^2 )$ and $\hat{t} -m_h^2 = x_1 (t-m_h^2 )$, and introduce a partonic differential cross section via
\begin{equation}
\sigma_{PP \to h} = \int dx_1\, dx_2 \,f_{g/P} (x_1 ) f_{g/P}(x_2) \int d\hat{u}\, d\hat{t} \;\frac{ d^2 \hat{\sigma}}{d\hat{u} \,d\hat{t}}.
\end{equation}
The partonic cross section at leading order takes the form
\begin{equation}
\frac{d^2 \hat{\sigma}^{(0)}}{d\hat{u} \,d\hat{t}} = \frac{\pi}{576 v^2}  \left( \frac{\alpha_s}{\pi}\right)^2 \delta(1-z) \delta (\hat{u})\delta (\hat{t}).
\end{equation}
Upon integration over the variables $\hat{u}$ and $\hat{t}$, this reproduces the leading order result of Eq.~(\ref{qcdcross}). 

\subsubsection{Next-to-leading order cross section}

The next-to-leading order prediction of QCD has contributions from virtual corrections to $gg\to h$ and from the emission of one final state gluon $gg\to gh$. To compare with this next-to-leading order prediction of QCD, we expand out all factors appearing in Eq.~(\ref{pT-fac-final}) through next-to-leading order in perturbation theory.  After expansion, we can write the cross section in the schematic form
\begin{equation}
\frac{d^2 \sigma^{(1)}}{du \, dt} \sim H^{(1)} \mathcal{I}_n^{(0)} \mathcal{I}_{\bn}^{(0)} \mathcal{S}^{-1 (0)} +  H^{(0)} \mathcal{I}_n^{(1)} \mathcal{I}_{\bn}^{(0)} \mathcal{S}^{-1 (0)}+ H^{(0)} \mathcal{I}_n^{(0)} \mathcal{I}_{\bn}^{(1)} \mathcal{S}^{-1 (0)}+ H^{(0)} \mathcal{I}_n^{(0)} \mathcal{I}_{\bn}^{(0)} \mathcal{S}^{-1 (1)}.
\end{equation}
The next-to-leading order expressions for the terms appearing in this cross section can be found in Eqs.~(\ref{NLOwilson}), (\ref{softfunc}), and~(\ref{beamwilson}).  Before explicitly computing the various contributions, we note several simplifying features that appear when the iSF and iBF terms are combined.
\begin{itemize}

\item When the combinations $\mathcal{I}_n^{(1)} \mathcal{I}_{\bn}^{(0)}$ and $\mathcal{I}_n^{(0)} \mathcal{I}_{\bn}^{(1)}$ are computed, a contraction of the form 
$g_{\perp}^{\alpha\beta} \mathcal{I}^{(1)}_{\alpha\beta}$ occurs, since one of the two Wilson coefficients contributes only at leading order.  This contraction removes the 
form factor $\mathcal{F}_2$ in Eq.~(\ref{beamwilson}), indicating that this structure contributes only at higher orders in perturbation theory.

\item When the sum  $H^{(0)} \mathcal{I}_n^{(1)} \mathcal{I}_{\bn}^{(0)} \mathcal{S}^{-1 (0)}+ H^{(0)} \mathcal{I}_n^{(0)} \mathcal{I}_{\bn}^{(1)} \mathcal{S}^{-1 (0)}+ H^{(0)} \mathcal{I}_n^{(0)} \mathcal{I}_{\bn}^{(0)} \mathcal{S}^{-1 (1)}$ is performed, several terms that appear in the individual expressions cancel.  In particular, it is simple to check that the following combination that appears in the the cross section vanishes upon integration over $t_n$ and $t_{\bn}$:
\begin{eqnarray}
&& \delta (t_{\bn}^-) \delta (t_n^{max} - t_n^+ ) \delta (t_{\bn}^{max} - t_{\bn}^- ) \left[\frac{\hat{Q}^2}{t_n^+} \text{ln} \left(\frac{t_n^+}{\hat{Q}^2}\right) \right]_+
+\,\delta (t_n^+) \delta (t_n^{max} - t_n^+ ) \delta (t_{\bn}^{max} - t_{\bn}^- ) \nonumber \\ &\times& \left[\frac{\hat{Q}^2}{t_{\bn}^-} \text{ln} \left(\frac{t_{\bn}^-}{\hat{Q}^2}\right) \right]_+
 - \delta (t_{\bn}^-) \delta (t_n^+) \left\{  \delta (t_{\bn}^{max} - t_{\bn}^- )\left[ \frac{\hat{Q}^2}{t_n^{max}-t_n^+}\text{ln}\frac{t_n^{max}-t_n^+}{\hat{Q}^2}\right]_+ \right. \nonumber \\
 &+& \left. \delta (t_n^{max} - t_n^+ )\left[ \frac{\hat{Q}^2}{t_{\bn}^{max}-t_{\bn}^-}\text{ln}\frac{t_{\bn}^{max}-t_{\bn}^-}{\hat{Q}^2}\right]_+\right\} \nonumber
\end{eqnarray}

\end{itemize}
After making these simplifications, we have four contributions to the cross section at next-to-leading order.  We now discuss each contribution in detail.  To simplify the notaton, we will set $\mu^2 = \hat{Q}^2 = x_1^{'} x_2^{'} Q^2$, where $x_{1,2}^{'}$ are the arguments of the PDFs.
\begin{enumerate}

\item {\boldmath $H^{(1)} \mathcal{I}_n^{(0)} \mathcal{I}_{\bn}^{(0)}\mathcal{S}^{-1 (0)}$}: This contribution has the same structure as the leading-order expression; using the Wilson coefficient in 
Eq.~(\ref{NLOwilson}), we find the result
\begin{equation}
\frac{d^2 \hat{\sigma}^{(1,a)}}{d\hat{u} \,d\hat{t}} = \frac{\pi}{576 v^2} \left( \frac{\alpha_s}{\pi}\right)^3 C_A \left\{\frac{11}{2}+\frac{7\pi^2}{12} \right\}\delta(1-z) \delta (\hat{u})\delta (\hat{t}).
\end{equation}
After integration over the Mandelstam variables, the contribution of this term to the total partonic cross section is clearly
\begin{equation}
 \hat{\sigma}^{(1,a)} = \frac{\pi}{576 v^2} \left( \frac{\alpha_s}{\pi}\right)^3 C_A \left\{\frac{11}{2}+\frac{7\pi^2}{12} \right\} \delta(1-z).
\end{equation}

\item {\boldmath $H^{(0)} \mathcal{I}_n^{(0)} \mathcal{I}_{\bn}^{(0)} \mathcal{S}^{-1 (1)}$}: After implementing the simplifications described above, the only terms remaining from the iSF in Eq.~(\ref{softfunc}) that contribute to the cross section are given by
\begin{eqnarray}
\mathcal{S}^{-1(1,b)} &=& -\frac{N_c^2-1}{4} \frac{\alpha_s C_A}{\pi} Q^2 \left\{-\frac{\pi^2}{12} \delta(t_{\bar{n}}^{max}-t_{\bar{n}}^-)  \delta(t_n^{max}-t_n^+) \right. \nonumber \\ 
	&+& \left.
	   \frac{1}{\hat{Q}^4}\left[ \frac{\hat{Q}^2}{t_n^{max}-t_n^+}\right]_+  \left[ \frac{\hat{Q}^2}{t_{\bar{n}}^{max}-t_{\bar{n}}^-} \right]_+  {_{0}F_{1}}\left( 1; -\frac{b_{\perp}^2 (t_n^{max}-t_n^+)	(t_{\bar{n}}^{max}-t_{\bar{n}}^-)}{4Q^2}\right) \right\}. \nonumber \\
\end{eqnarray}
The first term gives a contribution proportional to the leading order cross section.  When computing the contribution of the second term to Eq.~(\ref{pT-fac-final}), the following identity is needed:
\begin{equation}
\int_{0}^{\infty} db_{\perp} \, b_{\perp}\, J_0(k_{h\perp}b_{\perp})  {_{0}F_{1}}\left( 1; -\frac{b_{\perp}^2 y}{4} \right) = \frac{1}{k_{h\perp}} \delta (k_{h\perp} - \sqrt{y} ).
\end{equation}
This follows immediately from writing the hypergeometric function in terms of a Bessel function using Eq.~(\ref{hypergeo}), and then applying the orthogonality relation between $J_0$ functions of different arguments.  After performing the $b_{\perp}$ integral using the identity above, and expressing $t_{n,\bn}^{max}$ in terms of Mandelstam invariants using their definitions in Section~\ref{fixedorder}, we find the following contribution to the cross section:
\begin{eqnarray}
\frac{d^2 \hat{\sigma}^{(1,b)}}{d\hat{u} \,d\hat{t}} &=& -\frac{\pi}{576 v^2} \left( \frac{\alpha_s}{\pi}\right)^3 C_A \left\{ -\frac{\pi^2}{12} \delta(1-z)\delta (\hat{u})\delta (\hat{t}) \right. \nonumber \\
	&+&\left. \frac{1}{\hat{Q}^2}\left[\frac{\hat{Q}^2}{\hat{t}}\right]_+ \left[\frac{\hat{Q}^2}{\hat{u}}\right]_+ \delta( \hat{Q}^2+\hat{t}+\hat{u}-m_h^2 ) \right\}.
\end{eqnarray}	
To find the total cross section resulting from this term, we note that the range of the $\hat{t}$ integration is $-\hat{Q}^2 (1-z) \leq \hat{t} \leq 0$.  The definition of the plus distribution $1/[x]_+$ as the $\mathcal{O}(\epsilon^0 )$ term in the expansion of $x^{-1-\epsilon}$ allows us to derive the integral
\begin{equation}
\int_{-\hat{Q}^2 (1-z)}^0 d\hat{u} \int_{-\hat{Q}^2 (1-z)}^0 d\hat{t} \frac{1}{\hat{Q}^2} \left[ \frac{\hat{Q}^2}{\hat{t}}\right]_+ \left[ \frac{\hat{Q}^2}{\hat{u}}\right]_+ 
	\delta( \hat{Q}^2+\hat{t}+\hat{u}-m_h^2 ) = 2\left[ \frac{\text{ln} (1-z)}{1-z}\right]_+ -\frac{\pi^2}{6} .
\end{equation}	
Using this result, it is straightforward to derive
\begin{equation}
\hat{\sigma}^{(1,b)} = -\frac{\pi}{576 v^2} \left( \frac{\alpha_s}{\pi}\right)^3 C_A \left\{2\left[\frac{\text{ln}\,(1-z)}{1-z} \right]_+ -\frac{\pi^2}{4}\delta (1-z) \right\}.
\end{equation}

\item {\boldmath $H^{(0)} \mathcal{I}_n^{(1)} \mathcal{I}_{\bn}^{(0)} \mathcal{S}^{-1 (0)}$}:

The terms from the next-to-leading order Wilson coefficient $\mathcal{I}_n$ in Eq.~(\ref{beamwilson}) that contribute to the cross section in Eq.~(\ref{pT-fac-final}) are 
\begin{eqnarray}
{\cal I}_{n,gg}^{(1,c)\alpha \beta}&&\hspace{-0.3cm}(\frac{x_1}{x_1^{'}},t_n^{+},b_\perp) = -g^2 C_A \frac{\alpha_s}{\pi} g_{\perp}^{\alpha\beta} \left\{ \frac{1}{\hat{Q}^2} 	\left[ \frac{\hat{Q}^2}{t_n^+} \right]_+ \left( 1-\frac{x_1}{x_1^{'}} +\frac{x_1^{'}(x_1^{'}-x_1)}{x_1^2} +\frac{1}{[ 1-x_1/x_1^{'}]_+} \right)  \right. \nonumber \\ 
	\times&& \hspace{-0.3cm} {_{0}F_{1}}\left( 1; -\frac{b_{\perp}^2 (t_n^{max}-t_n^+)(t_{\bar{n}}^{max}-t_{\bar{n}}^-)}{4Q^2}\right)
	+\delta (t_n^+) \frac{x_1}{x_1^{'}} \left[ \left( 1-\frac{x_1}{x_1^{'}} +\frac{x_1^{'}(x_1^{'}-x_1)}{x_1^2} \right) \right. \nonumber \\ &\times&  \left. \left. \text{ln} (1-x_1/x_1^{'}) 
	+ \left[ \frac{\text{ln}(1-x_1/x_1^{'})}{1-x_1/x_1^{'}}\right]_+  \right] -\frac{\pi^2}{12} \delta (t_n^+) \delta (1-x_1/x_1^{'} )\right\}.
\end{eqnarray}
The calculation of the first term in this expression proceeds identically to the previously discussed case.  Upon substitution of the second term into the cross section in 
 Eq.~(\ref{pT-fac-final}), the delta-functions constraints set $x_1 /x_1^{'}=z$.  Using the relation between hadronic and partonic Mandelstam invariants, it can also be shown that they set $\hat{u}=0$.  Since the Higgs boson transverse momentum can be written in terms of Mandelstam invariants as $p_T^2 = \hat{t}\hat{u}/\hat{Q}^2$, the second terms contributes only for zero transverse momentum.  The final contribution of the $ \mathcal{I}_n^{(1)}$ term to the differential cross section is
\begin{eqnarray}
\frac{d^2 \hat{\sigma}^{(1,c)}}{d\hat{u} \,d\hat{t}} &=& \frac{\pi}{576 v^2} \left( \frac{\alpha_s}{\pi}\right)^3 C_A \left\{ -\frac{\pi^2}{12} \delta(1-z)\delta (\hat{u})\delta (\hat{t}) \right. \nonumber \\
	&+& \frac{1}{\hat{Q}^2}\left[\frac{\hat{Q}^2}{\hat{t}}\right]_+ \left[\frac{\hat{Q}^2}{\hat{u}}\right]_+ \left( 1+\frac{\hat{t}}{\hat{Q}^2}+\frac{\hat{t}^2}				{\hat{Q}^4}\right)^2\delta( \hat{Q}^2+\hat{t}+\hat{u}-m_h^2 ) \nonumber \\
	&+& \left. \frac{1}{2}\left(1+z^4+(1-z)^4 \right) \left[ \frac{\text{ln} (1-z)}{1-z}\right]_+ \delta (\hat{u}) \,\delta( \hat{Q}^2+\hat{t}-m_h^2 )\right\}.
\end{eqnarray}
Upon integration over the Mandelstam invariants, we find the following contribution to the total cross section:
\begin{eqnarray}
\hat{\sigma}^{(1,c)} &=&  \frac{\pi}{576 v^2} \left( \frac{\alpha_s}{\pi}\right)^3 C_A \left\{ -\frac{\pi^2}{4} \delta(1-z)  -\frac{1-z}{6} \left( 11-4z+11z^2 \right) \right. \nonumber \\
	&+& \left. \left( 3-4z+6z^2-4z^3+2z^4 \right) \left[ \frac{\text{ln} (1-z)}{1-z}\right]_+ \right\}.
\end{eqnarray}

\item {\boldmath $H^{(0)} \mathcal{I}_n^{(0)} \mathcal{I}_{\bn}^{(1)} \mathcal{S}^{-1 (0)}$}:

An identical calculation as that outlined above yields the following results for the differential and total partonic cross sections contributions from $\mathcal{I}_{\bn}$:
\begin{eqnarray}
\frac{d^2 \hat{\sigma}^{(1,d)}}{d\hat{u} \,d\hat{t}} &=& \frac{\pi}{576 v^2} \left( \frac{\alpha_s}{\pi}\right)^3 C_A \left\{ -\frac{\pi^2}{12} \delta(1-z)\delta (\hat{u})\delta (\hat{t}) \right. \nonumber \\
	&+& \frac{1}{\hat{Q}^2}\left[\frac{\hat{Q}^2}{\hat{t}}\right]_+ \left[\frac{\hat{Q}^2}{\hat{u}}\right]_+ \left( 1+\frac{\hat{u}}{\hat{Q}^2}+\frac{\hat{u}^2}				{\hat{Q}^4}\right)^2\delta( \hat{Q}^2+\hat{t}+\hat{u}-m_h^2 ) \nonumber \\
	&+& \left. \frac{1}{2}\left(1+z^4+(1-z)^4 \right) \left[ \frac{\text{ln} (1-z)}{1-z}\right]_+ \delta (\hat{t}) \,\delta( \hat{Q}^2+\hat{u}-m_h^2 )\right\}, \nonumber \\
\hat{\sigma}^{(1,d)} &=&  \frac{\pi}{576 v^2} \left( \frac{\alpha_s}{\pi}\right)^3 C_A \left\{ -\frac{\pi^2}{4} \delta(1-z)  -\frac{1-z}{6} \left( 11-4z+11z^2 \right) \right. \nonumber \\ &+& \left. \left( 3-4z+6z^2-4z^3+2z^4 \right) \left[ \frac{\text{ln} (1-z)}{1-z}\right]_+ \right\},
\end{eqnarray}
where the differential cross-section term above differs from the corresponding contributions from $\mathcal{I}_n$  only by the interchange $\hat{u} \leftrightarrow \hat{t}$.

\end{enumerate}

The total next-to-leading order result is obtained by summing the four contributions described above:
\begin{eqnarray}
\frac{d^2 \hat{\sigma}^{(1,total)}}{d\hat{u} \,d\hat{t}} &=& \frac{d^2 \hat{\sigma}^{(1,a)}}{d\hat{u} \,d\hat{t}}+\frac{d^2 \hat{\sigma}^{(1,b)}}{d\hat{u} \,d\hat{t}}
	+\frac{d^2 \hat{\sigma}^{(1,c)}}{d\hat{u} \,d\hat{t}}+\frac{d^2 \hat{\sigma}^{(1,d)}}{d\hat{u} \,d\hat{t}}, \nonumber \\ \\
\hat{\sigma}^{(1,total)} &=& \hat{\sigma}^{(1,a)}+\hat{\sigma}^{(1,b)}+\hat{\sigma}^{(1,c)}+\hat{\sigma}^{(1,d)}.
\end{eqnarray}
We now compare the differential cross section for $p_T >0$ and the total cross section with the full QCD expressions given in Eqs.~(\ref{ptspectrum}) and ~(\ref{qcdcross}).  Summing the four contributions to the differential cross section, and keeping only terms which contribute for $p_T >0$, we obtain the following result:
\begin{eqnarray}
\frac{d^2 \hat{\sigma}^{(1,p_T>0)}}{d\hat{u} \,d\hat{t}} &=& \frac{\pi}{192 v^2}  \left( \frac{\alpha_s}{\pi}\right)^3\,\delta (\hat{Q}^2+\hat{t}+\hat{u} -m_h^2 )  	\frac{\hat{Q}^2}{\hat{t}\hat{u}} \left\{ -1  + \left[ 1+\frac{\hat{u}}{\hat{Q}^2} +\frac{\hat{u}^2}{\hat{Q}^4}\right]^2\right. \nonumber \\ &+& 
\left. \left[ 1+\frac{\hat{t}}{\hat{Q}^2} +\frac{\hat{t}^2}{\hat{Q}^4}\right]^2 \right\}.
\end{eqnarray}
We have dropped the plus prescription on the Mandelstam invariants since we are only considering the region $p_T >0$.  This expression agrees with the QCD result in Eq.~(\ref{ptspectrum}) up to terms which are finite as $\hat{u},\hat{t} \to 0$.  Such terms can arise from suppressed operators in the EFT, and not from emission of collinear or soft gluons.  Our result therefore correctly reproduces that of QCD in the limit of soft or collinear gluon emission, as expected.  

The total cross section obtained by combining the four contributions above is
\begin{eqnarray}
\hat{\sigma}^{(1)} &=& \frac{\pi}{576 v^2} \left( \frac{\alpha_s}{\pi}\right)^3\left\{ \left(\frac{11}{2}+\pi^2 \right) \delta (1-z) + 6\left( 1+z^4+(1-z)^4\right) \left[ \frac{\text{ln} (1-z)}{1-z}\right]_+ \right. \nonumber \\ &-& \left.(1-z)(11 -4z +11z^2) \right\}.
\end{eqnarray}
The terms with $\delta (1-z)$ and $\text{ln} (1-z)$ in this result agree with the QCD cross section in Eq.~(\ref{qcdcross}).  Terms non-singular in the limit $z \to 1$ can arise also from emission of hard jets with the Higgs, and therefore agreement in the non-singular polynomial in $1-z$ is not expected.  The description of these pieces requires power-suppressed operators in the EFT.  Our total cross section therefore properly reproduces the expected result from fixed order QCD.

\subsection{Top down vs bottom up running: impact-parameter space}

Section \ref{topdownrunning} was devoted to the RG running of the hard Wilson coefficient $H(x_1x_2Q^2,\mu)$. First $H(x_1x_2Q^2,\mu_Q)$ is obtained from a matching calculation at the hard scale $\mu_Q\sim m_h$. Next, after computing  the anomalous dimension of $H(x_1x_2Q^2,\mu)$, it is RG evolved down to the $\mu_T \sim p_T$ scale to obtain $H(x_1x_2Q^2,\mu_Q;\mu_T)$ which has the logarithms of $m_h/p_T$ resummed. We refer to this procedure as top-down running. Equivalently, one can compute the anomalous dimensions of the iBFs and the inverse soft function and perform an RG running from the $\mu_T \sim p_T$ scale, where the iBFs and inverse soft function live, up to the hard scale $\mu_Q$. We refer to this procedure as bottom-up running. Scale invariance of the cross-section implies that the total anomalous dimension of the convolution of the iBFs and inverse soft function in the $t_{n,\bn}^\pm$ variables must exactly cancel the anomalous dimension of $H(x_1x_2Q^2,\mu)$. In this section, we use this consistency condition~\cite{Fleming:2007qr,Fleming:2007xt} to obtain the anomalous dimension of the iBF which we will compare with what was obtained from the fixed order calculation. We will derive the anomalous dimension of the iBF in a third way in section \ref{runaftermatch} as yet another check.

We define the quantity ${\cal D}$  as
\bea
{\cal D}&\equiv& H(x_1x_2Q^2,\mu) X(\mu), 
\eea
where $X(\mu)$ is the convolution of the iBFs and the inverse soft function 
\bea
X(\mu) &=& \frac{1}{Q^2} \int dt_n \int dt_\bn \tilde{B}_n^{\alpha \beta}(x_1,t_n,b_\perp,\mu)\> \tilde{B}_{\bn \alpha \beta}(x_2,t_\bn,b_\perp,\mu) \nn \\
&\times&\>{\cal S}^{-1}(\frac{t_\bn^{max}- t_\bn^-}{Q}, \frac{t_n^{max}- t_n}{Q},b_\perp, \mu).
\eea
The factorization formula for the differential cross-section at the $p_T$ scale in Eq.~(\ref{kpkmspacefac-3}) contains the quantity ${\cal D}$. Furthermore, all the renormalization scale dependent quantities are contained in ${\cal D}$. Scale invariance of the cross-section then requires that ${\cal D}$ has zero anomalous dimension or equivalently that the $\mu$-dependence cancels between $H(x_1x_2Q^2,\mu)$ and $X(\mu)$.

We begin by defining the renormalization constants which relate the bare and renormalized quantities as 
\bea
\label{bare}
H(x_1x_2Q^2,\mu) &=& Z_H^{-1}(\mu) H_b(x_1x_2Q^2), \nn \\
\tilde{B}_n^{\alpha \beta}(x_1, t_n,b_\perp,\mu) &=& \int dt_n' \>Z^{-1}_n(t_n-t_n',\mu) \tilde{B}_{n, b}^{\alpha \beta}(x_1, t_n,b_\perp)\nn \\
\tilde{B}_\bn^{\alpha \beta}(x_2, t_\bn,b_\perp,\mu) &=& \int dt_\bn' \>Z^{-1}_\bn(t_\bn-t_n',\mu) \tilde{B}_{\bn, b}^{\alpha \beta}(x_2, t_\bn,b_\perp) \nn \\
{\cal S}^{-1}(\frac{t_\bn}{Q}, \frac{t_n}{Q},b_\perp, \mu) &=& \frac{1}{Q^2}\int dt_\bn' \int dt_n' \>Z_{{\cal S}^{-1}}^{-1} (\frac{t_\bn'}{Q}-\frac{t_\bn}{Q},\frac{t_n'}{Q}-\frac{t_n}{Q},\mu)\>{\cal S}_b^{-1}(\frac{t_\bn'}{Q}, \frac{t_n'}{Q},b_\perp),\nn \\
\eea
where the sunscript $b$ on the objects on the RHS  denotes bare quantities and the objects on the LHS  are the renormalized quantities. Note that the renormalization of the hard coefficient of $H$ is multiplicative. On the other hand, the renormalization of the the iBFs and the inverse soft function involve a convolution. This implies that the hard coefficient has \textit{local} running while the iBFs and the inverse soft function undergo \textit{convolution} running. The consistency of top-down and bottom-up running implies that the total convolution of the iBFs and the inverse soft function undergoes local running that  precisely cancels the running of $H(x_1x_2Q^2,\mu)$.

The scale invariance of ${\cal D}$ leads to a consistency condition that relates the various renormalization constants as
\bea
\label{consistency}
Z_H(\mu) \delta(s_n - t_n)\delta(s_\bn - t_\bn) &=& \frac{1}{Q^2}\int dt_n'\int dt_\bn' Z^{-1}_n(t_n'- s_n,\mu)Z^{-1}_n(t_\bn'- s_\bn,\mu)\nn \\
&\times& Z_{{\cal S}^{-1}}^{-1}(\frac{-t_\bn' + t_\bn}{Q},\frac{-t_n'+t_n}{Q},\mu). \nn \\
\eea
To derive this consistency equation, we first write ${\cal D}$ entirely in terms of  bare quantities and use the first equation in Eq.~(\ref{bare}) to replace $H_b$. Next we write ${\cal D}$ entirely in terms of renormalized quantities and then use the last three equations in Eq.~(\ref{bare}) to replace the renormalized iBFs and the iSF in terms of the corresponding bare quantities. Finally by equating these two ways of writing ${\cal D}$ we arrive at Eq.~(\ref{consistency}).

We can make the content of Eq.~(\ref{consistency}) more transparent by noting that we can write the renormalization constants as a sum of a leading order piece and perturbative corrections as
\bea
Z_H &=& 1 + \delta Z_H, \nn \\
Z_n(t) &=& \delta(t) + \delta Z_n(t), \nn \\
Z_\bn(t) &=& \delta(t) + \delta Z_n(t), \nn \\
Z_{{\cal S}^{-1}}(\frac{t_\bn}{Q},\frac{t_n}{Q}) &=&  \delta(\frac{t_\bn}{Q})Z_{{\cal S}^{-1}}^n(\frac{t_n}{Q}) + \delta(\frac{t_n}{Q})Z_{{\cal S}^{-1}}^\bn(\frac{t_\bn}{Q}), \nn \\
\eea
where
\bea
Z_{{\cal S}^{-1}}^n(\frac{t_n}{Q}) &=& \delta (\frac{t_n}{Q}) + \delta Z_{{\cal S}^{-1}}^\bn(\frac{t_n}{Q}), \nn \\
Z_{{\cal S}^{-1}}^\bn(\frac{t_\bn}{Q}) &=& \delta (\frac{t_\bn}{Q}) + \delta Z_{{\cal S}^{-1}}^\bn(\frac{t_\bn}{Q}), \nn \\
\eea
We have used the fact that the dependence on $t_{n}/Q$ and $t_\bn/Q$ of the soft renormalization constant is separable~\cite{Fleming:2007xt} as was seen in the explicit calculation in section \ref{fixedorder}. The consistency condition in Eq.~(\ref{consistency}) can now be written as
\bea
\delta Z_n(t)+\frac{1}{Q} \delta Z_{{\cal S}^{-1}}^n(\frac{t}{Q}) =  - \frac{\delta Z_H}{2} \delta(t) =  \frac{\delta Z_O + \delta Z^*_O}{2} \delta(t), \nn \\
\delta Z_\bn(t) + \frac{1}{Q} \delta Z_{{\cal S}^{-1}}^\bn(\frac{t}{Q})=  - \frac{\delta Z_H}{2} \delta(t) =\frac{\delta Z_O + \delta Z^*_O}{2} \delta(t). \nn \\
\eea
The leading order results for the soft and hard coefficient renormalization constants from sections \ref{fixedorder} and \ref{topdownrunning} respectively are
\bea
\frac{\delta Z_O^{(1)} + \delta Z_O^{*(1)}}{2} &=& \frac{\alpha_s C_A}{\pi} \Big [ \frac{1}{2\epsilon^2} + \frac{1}{2\epsilon} \ln \Big (\frac{\mu^2}{\hat{Q}^2} \Big ) \Big ], \nn \\
\frac{1}{Q}\delta Z_{{\cal S}^{-1}}^{n,\bn(1)}(t/Q) &=&  \frac{\alpha_s C_A}{\pi} \Big [- \frac{\delta (t)}{2\epsilon^2} +\frac{1}{\epsilon}\frac{1}{\hat{Q}^2} \Big [\frac{\hat{Q}^2}{t}\Big ]_+ + \frac{1}{2\epsilon}\delta(t) \ln \frac{\hat{Q}^2}{\mu^2} \Big ], \nn \\
\eea
from which we can solve for the one loop renormalization constant of the iBF as
\bea
\label{jetcons}
\delta Z_{n,\bn}^{(1)}(t) &=& \frac{\alpha_s C_A}{\pi} \Big [\frac{\delta (t)}{\epsilon^2} - \frac{1}{\epsilon}\frac{1}{\hat{Q}^2} \Big [\frac{\hat{Q}^2}{t}\Big ]_+  - \frac{1}{\epsilon} \delta(t) \ln \frac{\hat{Q}^2}{\mu^2}  \Big ], \nn \\
&=&  \frac{\alpha_s C_A}{\pi} \Big [\frac{\delta (t)}{\epsilon^2} - \frac{1}{\epsilon}\frac{1}{\mu^2} \Big [\frac{\mu^2}{t}\Big ]_+  \Big ]
\eea
which is in agreement with the ultraviolet pole terms for the iBF  derived in section \ref{fixedorder}. Thus, we have an explicit demonstration of the equivalence of the top-down and bottom-up
running at leading order. Eq.~(\ref{jetcons}) implies that the leading order anomalous dimension of the iBF is given by
\bea
\gamma_{\tilde{B}_n}^{(1)} (t,\mu) = \frac{2\alpha_s C_A}{\pi}\frac{1}{\mu^2} \Big [\frac{\mu^2}{t}\Big ]_+.
\eea
In section \ref{runaftermatch}, we provide another independent check by computing the anomalous dimension by taking the derivative with respect to $\mu$ of the RHS of Eq.~(\ref{iBFPDF}) which expresses the iBF in terms of a matching coefficient and the standard QCD PDF.

\subsection{Top-down vs bottom-up running: momentum space}

The consistency of the top-down and bottom-up running can be checked in yet another way by looking at the momentum space factorization. We wrote down the momentum space factorization earlier in Eqs.~(\ref{pT-fac-final}) and (\ref{momspace-1}). However, in order to demonstrate the equivalence of the top-down and bottom-up running above the $p_T$ scale, we need the momentum space analog of Eq.(\ref{kpkmspacefac-3}) which is given by
\bea
\label{kpkmspacefac-5}
\frac{d^2\sigma}{du \>dt} &=& \frac{(2\pi)}{8(N_c^2-1)^2 } \int dp_h^+ dp_h^- \int d^2k_h^\perp  \int d^2k_{n}^\perp \int d^2k_\bn^\perp \int d^2k_{s}^\perp\> \delta(\vec{k}_h^\perp +\vec{k}_n^\perp +\vec{k}_\bn^\perp +\vec{k}_{s}^\perp )  \nn \\
%%%%
&\times & \delta \left [  u - m_h^2 +Q p_h^-\right ] \delta \left [  t - m_h^2 +Q  p_h^+\right ]\delta \left [p_h^+p_h^- - \vec{k}_{h\perp}^2 - m_h^2 \right ]  \nn \\
&\times&\int_0^1 dx_1 \int_0^1 dx_2 \int  dt_n^+ \int dt_\bn^- H(x_1x_2Q^2,\mu_Q;\mu_T) \>B_n^{\alpha \beta}(x_1,t_n^+,k_n^\perp,\mu_T)\> B_{\bn \alpha \beta}(x_2,t_\bn^-,k_\bn^\perp,\mu_T)\nn \\
&\times&{\cal S}(x_1 Q-p_h^- - \frac{t_\bn^-}{Q}, x_2 Q-p_h^+ - \frac{t_n^+}{Q},k_{s}^\perp,\mu_T). \nn \\
\eea
The momentum space functions are given in terms of the impact-parameter space functions by the Fourier transforms
\bea
B_{n,\bn}^{\alpha \beta}(x,t,k_\perp,\mu) &=& \int \frac{d^2b_\perp}{4\pi^2} e^{-i\vec{k}_\perp \cdot \vec{b}_\perp} B_{n,\bn}^{\alpha \beta}(x,t,b_\perp,\mu) , \nn \\
{\cal S}(k^-,k^+,k_\perp)&=&  \int \frac{d^2b_\perp}{4\pi^2} e^{-i\vec{k}_\perp\cdot \vec{b}_\perp} {\cal S}(k^-,k^+,b_\perp). \nn \\
\eea
We have made use of Eq.~(\ref{zero-soft}) to recast the factorization theorem in terms of purely collinear iBFs, defined with a zero-bin subtraction as in Eq.~(\ref{zero-iBF}), and the soft function. In this momentum space version of the factorization formula at the $p_T$ scale, the running of the purely collinear iBFs and the soft function is determined entirely by scaleless virtual graphs as shown in Figs.~\ref{jetdiags} and \ref{SCETsoft}. The real gluon emission graphs for the momentum space purely collinear iBFs and the soft function are finite and do not contribute to the running. This is in contrast to the impact parameter space factorization, examined in the previous section, in which the real emission graphs have UV divergences and do contribute to the running. The consistency of top-down and bottom-up running should work out in both impact-parameter and momentum space. In this section we show how the virtual graphs for the purely collinear iBFs and the soft function cancel the running of $H(x_1x_2Q^2,\mu)$ to maintain the $\mu$-independence of the differential cross-section.

For one loop running, we show that
\bea
\label{product-run}
\mu \frac{d}{d\mu} {\cal F} = - \gamma_{{\cal F}} {\cal F} =- (\gamma_{\cal O}^{(1)} + \gamma_{\cal O}^{(1)*}) {\cal F} = \gamma_{H}{\cal F} ,
\eea
where ${\cal F} $ is defined to be the product of the purely collinear iBFs and the soft function
\bea
\label{defF}
{\cal F}&\equiv& B_n^{\alpha \beta}(x_1,t_n^+,k_n^\perp,\mu_T)\> B_{\bn \alpha \beta}(x_2,t_\bn^-,k_\bn^\perp,\mu_T){\cal S}(x_1 Q-p_h^- - \frac{t_\bn^-}{Q}, x_2 Q-p_h^+ - \frac{t_n^+}{Q},k_{s}^\perp,\mu_T).  \nn \\ 
\eea
Eq.~(\ref{product-run}) ensures that Eq.~(\ref{kpkmspacefac-5}) is scale invariant as required
and $\gamma_{{\cal O}}^{(1)}$ is the one loop cusp anomalous dimension, given in Eq.~(\ref{cusp-anom}), of the operator ${\cal O}$ defined in Eq.~(\ref{SCETop}) and
\bea
\gamma_{\cal O}^{(1)} + \gamma_{\cal O}^{(1)*} = \frac{2\alpha_s C_A}{\pi} \ln \Big (\frac{\hat{Q}^2}{\mu^2} \Big).
\eea 
The bare and renormalized ${\cal F}$ are  related as
\bea
{\cal F}^{ b} = Z_{{\cal F}} {\cal F},
\eea
and writing the left-hand side in terms of renormalized fields and couplings we get
\bea
\label{ZF}
{\cal F} &=& B_n^{\alpha \beta}\> B_{\bn \alpha \beta} \>S + (\frac{Z_g^4 Z_A^2}{Z_F} -1) J_n^{\alpha \beta}\> J_{\bn \alpha \beta} \>S \nn \\
&=& B_n^{\alpha \beta}\> B_{\bn \alpha \beta} \>S + \Big [ 2 \{ 2 \delta Z_g + \delta Z_A\} - \delta Z_{{\cal F}} \Big ] B_n^{\alpha \beta}\> B_{\bn \alpha \beta} \>S,
\eea
where the quantities $B_n^{\alpha \beta},\> B_{\bn \alpha \beta}, \>S$ are all written in terms of renormalized fields and couplings.  We suppress the arguments in Eq.~(\ref{defF}) for notational simplicity.

The zeroth order expression for ${\cal F}$ is
\bea
{\cal F}^{(0)} &=& B_n^{(0)\alpha \beta}\> B^{(0)}_{\bn \alpha \beta} \>S^{(0)},
\eea
where
\bea
B_{n,\bn}^{(0)\alpha \beta}(x,t,k_\perp,\mu_T) &=& -g^2 g_\perp^{\alpha \beta} \delta(t) \delta(1-x)\delta^{(2)}(\vec{k}_\perp), \nn \\
S^{(0)}(k^-,k^+,k_\perp,\mu_T) &=& \frac{(N_c^2-1)}{4} \delta(k^-) \delta(k^+) \delta^{(2)}(\vec{k}_\perp), 
\eea
and the contribution of the virtual graphs in Figs.~\ref{jetdiags}  and \ref{SCETsoft} to ${\cal F}$ is
\bea
\label{virtual}
{\cal F}^{V(1)} &=& B_n^{V(1)\alpha \beta}\> B^{(0)}_{\bn \alpha \beta} \>S^{(0)} +B_n^{(0)\alpha \beta}\> B^{V(1)}_{\bn \alpha \beta} \>S^{(0)} + B_n^{(0)\alpha \beta}\> B^{(0)}_{\bn \alpha \beta} \>S^{V(1)} \nn \\
&=& \Big [ \tilde{B}_n^{V(1)\alpha \beta}- B_{n0}^{V(1)\alpha \beta}\Big ] \> B^{(0)}_{\bn \alpha \beta}S^{(0)}+\Big [ \tilde{B}_\bn^{V(1)\alpha \beta}-B_{\bn0}^{V(1)\alpha \beta}\Big ] \> B^{(0)}_{n \alpha \beta}S^{(0)} \nn \\
&+& B_n^{(0)\alpha \beta}\> B^{(0)}_{\bn \alpha \beta} \>S^{V(1)}.
\eea
The one loop virtual graphs contributing to the purely collinear jet functions and the soft function are given by
 \bea
 \label{vir-1}
  B_n^{V(1)\alpha \beta}&=&  B_n^{(0)\alpha \beta}\Big [  V_b(\hat{p}_1) + V_b(-\hat{p}_1) \Big ], \nn \\
 B_{n0}^{V(1)\alpha \beta} &=&  B_n^{(0)\alpha \beta} \Big [   V_{b0}(\hat{p}_1) + V_{b0}(-\hat{p}_1)  \Big],   \nn \\
 S^{V(1)} &=& S^{(0)}  \Big [   V_{a}(\hat{p}_1,\hat{p}_2) + V_{a}(-\hat{p}_1,-\hat{p}_2)  \Big], \nn \\
 \eea
where expressions for $ V_{a}(\hat{p}_1,\hat{p}_2)$, $V_b(\hat{p}_1) $, and $ V_{b0}(\hat{p}_1) $ were given earlier in Eq.~(\ref{Vscaleless}) and in Eq.~(\ref{VaVb-1}) with off-shell momenta $\hat{p}_{1,2}$ to regulate infrared divergences.
One can rearrange the terms in Eq.~(\ref{virtual}), at the level of the integrand, using Eq.~(\ref{vir-1}) to get
\bea
\label{Fgraphs}
{\cal F}^{V(1)} &=& B_n^{(0)\alpha \beta}B^{(0)}_{\bn \alpha \beta}S^{(0)} \nn \\
&\times& \Big \{V_a(\hat{p}_1,\hat{p}_2) + [V_b(\hat{p}_1) -V_{b0}(\hat{p}_1) ]+ [V_b(\hat{p}_2) -V_{b0}(\hat{p}_2) ]    \nn \\
&+&V_a(-\hat{p}_1,-\hat{p}_2)+  [V_b(-\hat{p}_1) -V_{b0}(-\hat{p}_1) ]+ [V_b(-\hat{p}_2) -V_{b0}(-\hat{p}_2) ]  \Big \}.
\eea
The first line in curly brackets is precisely the sum of graphs in Fig.~\ref{QCDmatchSCETdiags} which determine the anomalous dimension at one loop for ${\cal O}(\bn \cdot \hat{p}_1,n\cdot \hat{p}_2)$ and the second line in curly brackets is just the sum of conjugate diagrams. From Eqs.~(\ref{ZF}), (\ref{Fgraphs}), (\ref{sum-ver}), and Eq.~(\ref{wf}) we get
\bea
Z_{{\cal F}} = 1 + \frac{\alpha_s C_A}{2\pi}\Big [ \frac{2}{\epsilon^2} + \frac{2}{\epsilon}\ln \Big (\frac{\mu^2}{\hat{Q}^2}\Big ) \Big ],
\eea
so that
\bea
\gamma_{{\cal F}} = \gamma_{{\cal O}} + \gamma_{{\cal O}}^* = \frac{2\alpha_s C_A}{\pi} \ln \Big (\frac{\hat{Q}^2}{\mu^2} \Big), 
\eea
as expected for the scale invariance of Eq.~(\ref{kpkmspacefac-5}). 

\subsection{Running after matching}
\label{runaftermatch}

For completeness, we will also verify that the running after matching our results to the standard PDFs cancels the running of the hard Wilson coefficient $H(Q^2,\mu^2)$.  To do so, we begin with the relevant quantity appearing in the factorization formula of Eq.~(\ref{below-pT-fac}):
\begin{eqnarray}
X &=& \int \frac{dx^{'}_1}{x^{'}_1} \frac{dx^{'}_2}{x^{'}_2}\int dt_n^+ dt_{\bar{n}}^- \,\mathcal{I}_n^{\alpha\beta}({x^{'}_1},t_n^+,b_{\perp},\mu) \mathcal{I}_{\bar{n} \alpha\beta}({x^{'}_2},t_{\bar{n}}^-,b_{\perp},\mu)\nonumber \\ &\times& S^{-1}(\frac{t_n^{max}-t_n}{Q}, \frac{t_{\bn}^{max}-t_\bn}{Q},b_{\perp},\mu) f_{g/P}(\frac{x_1}{x^{'}_1},\mu^2)f_{g/P}(\frac{x_2}{x^{'}_2},\mu).
\label{xdef}
\end{eqnarray}
We have rearranged the $x_i^{'}$ integrals appearing in Eq.~(\ref{below-pT-fac}) into a form convenient for this calculation.  The quantity $X$ must obey the renormalization group equation
\begin{equation}
\mu \frac{dX}{d\mu} = -\frac{2\alpha_s C_A}{\pi} \text{ln}\frac{\hat{Q}^2}{\mu^2} \,X,
\label{neededrunning}
\end{equation}
in order for the cross section to be scale invariant and for the operator $\mathcal{O}$ defined in Eq.~(\ref{SCETop}) to have the correct anomalous dimension.

It is simplest to derive the running of $X$ by differentiating directly the beam function defined in Eq.~(\ref{iBFPDF}).  This accounts for the running of the product 
$\mathcal{I}\times f_{g/p}$ in Eq.~(\ref{xdef}).  From Eqs.~(\ref{iBFPDF}), (\ref{iBFPDF1}), (\ref{Itree}), and (\ref{beamwilson}), we can check that the terms appearing in the $n$-collinear beam function that contribute to the $\mu$ dependence are
\begin{eqnarray}
\tilde{B}_{n[\mu]}^{\alpha\beta}(x_1, t_n,b_{\perp},\mu)&=& -g^2(\mu) g_{\perp}^{\alpha\beta}\left\{ f_{g/P}(x,\mu) +\frac{\alpha_sC_A}{\pi} \int_x^1 \frac{dx^{'}_1}{x^{'}} f_{g/P}\left( \frac{x_1}{x_1^{'}},\mu \right) \right. \nonumber \\ &\times &\left[ \delta(1-x_1^{'}) \frac{1}{\hat{Q}^2}\left[\frac{\hat{Q}^2}{t_n^{+}}\right]_+ \text{ln}\frac{\hat{Q}^2}{\mu^2}  +\delta(t_n^{+}) \left[x_1^{'}(1-x_1^{'}) +\frac{1-x_1^{'}}{x_1^{'}} +\frac{x_1^{'}}{[1-x_1^{'}]_+} \right] \text{ln}\frac{\hat{Q}^2}{\mu^2} \right. \nn \\ &+& \left.\left. \frac{1}{2} \delta(t_n^{+})\delta(1-x_1^{'})\text{ln}^2\frac{\hat{Q}^2}{\mu^2} \right] \right\}. \nn \\
\label{jetfunc}
\end{eqnarray}
The running of the coupling constant and PDF are given by the standard expressions
\begin{eqnarray}
\mu \frac{d g^2(\mu)}{d\mu} &=& -\frac{\alpha_s}{2\pi}\left\{ \frac{11}{3}C_A - \frac{2}{3} N_F\right\} \nonumber \\
\mu \frac{d f_{g/P}(x,\mu)}{d\mu} &=&2 \frac{\alpha_s}{\pi} \int_x^1 \frac{dx^{'}}{x^{'}} f_{g/P}\left( \frac{x}{x^{'}},\mu \right) P_{gg}(x^{'})= 2 \frac{\alpha_sC_A}{\pi}  \int_x^1 \frac{dx^{'}}{x^{'}} f_{g/P}\left( \frac{x}{x^{'}},\mu \right) \nonumber \\ &\times& \left[x^{'}(1-x^{'}) +\frac{1-x^{'}}{x^{'}} +\frac{x^{'}}{[1-x^{'}]_+} 	+\left( \frac{11}{12}C_A -\frac{N_F}{6}\right)\delta(1-x^{'})\right]. \nn \\
\end{eqnarray}
We now differentiate the beam function in Eq.~(\ref{jetfunc}) with respect to $\mu$, and arrive at the result
\begin{equation}
\mu \frac{d \tilde{B}_{n}^{\alpha\beta}}{d\mu} = 2g^2 \frac{\alpha_s C_A}{\pi} g_{\perp}^{\alpha\beta} \int_{x_{1}}^1 \frac{dx^{'}_1}{x_1^{'}} f_{g/P}\left( \frac{x_1}{x_1^{'}},\mu \right)
	\left\{\delta(1-x_1^{'}) \frac{1}{\hat{Q}^2}\left[\frac{\hat{Q}^2}{t_n^{+}}\right]_+ +\delta(t_n^{+})\delta(1-x_1^{'})\text{ln}\frac{\hat{Q}^2}{\mu^2} \right\}.
	\label{ndiff}
\end{equation}
We note that this result is in agreement with the anomalous dimension found in Eq.~(\ref{jetcons}).  An identical calculation for the other $\bn$-collinear beam function yields
\begin{equation}
\mu \frac{d \tilde{B}_{\bar{n}}^{\alpha\beta}}{d\mu} = 2g^2 \frac{\alpha_s C_A}{\pi} g_{\perp}^{\alpha\beta} \int_{x_{2}}^1 \frac{dx^{'}_2}{x_2^{'}} f_{g/P}\left( \frac{x_2}{x_2^{'}},\mu \right)
	\left\{\delta(1-x_2^{'}) \frac{1}{\hat{Q}^2}\left[\frac{\hat{Q}^2}{t_{\bar{n}}^{-}}\right]_+ +\delta(t_{\bar{n}}^{-})\delta(1-x_2^{'})\text{ln}\frac{\hat{Q}^2}{\mu^2} \right\},
	\label{bndiff}
\end{equation}
We next differentiate the finite part of the iSF defined in Eq.~(\ref{softfunc}), and find
\begin{eqnarray}
\mu \frac{d \mathcal{S}^{-1}}{d\mu} &=& \frac{N_c^2-1}{4} 2\frac{\alpha_s C_A}{\pi}Q^2 \left\{ \delta(t_{\bar{n}}^{max}-t_{\bar{n}}^-)\delta(t_n^{max}-t_n^+) \text{ln} \frac{\hat{Q}^2}{\mu^2}
	+\frac{1}{\hat{Q}^2}\delta(t_{\bar{n}}^{max}-t_{\bar{n}}^-) \left[ \frac{\hat{Q}^2}{t_n^{max}-t_n^+}\right]_+ \right. \nonumber \\ &+& \left. \frac{1}{\hat{Q}^2}\delta(t_n^{max}-t_n^+)  \left[ \frac{\hat{Q}^2}{t_{\bar{n}}^{max}-	t_{\bar{n}}^-} \right]_+ \right\}.
	\label{sdiff}
\end{eqnarray}
We note that it is possible to manipulate the result of Eq.~(\ref{softfunc}) such that $\hat{Q}$ rather than $Q$ appears in the logarithms and distributions, and we have done so in writing this expression. 

We are now ready to plug all of this into $X$ and determine its running.  The leading order expression for $X$ can be easily determined forom the results of Section~\ref{fixedorder} to be
\begin{eqnarray}
X^{(0)} &=&   \int_{x_{1}}^1\frac{dx^{'}_1}{x_1^{'}} f_{g/P}\left( \frac{x_1}{x_1^{'}},\mu \right)  \int_{x_{2}}^1 \frac{dx^{'}_2}{x_2^{'}} f_{g/P}\left( \frac{x_2}{x_2^{'}},\mu \right) 
 \nonumber \\ &\times&
 g^4 (d-2) \frac{N_c^2-1}{4} Q^2\,\delta(1-x_1^{'})\delta(1-x_2^{'})\delta(t_n^{max}) \delta(t_{\bar{n}}^{max}).
\end{eqnarray}
Upon substitution of the derivatives computed above in Eqs.~(\ref{ndiff}), (\ref{bndiff}), and~(\ref{sdiff}) into the derivative of $X$, we find the following result through 
next-to-leading order:
\begin{eqnarray}
\mu \frac{ d X}{d\mu} &=&-2\frac{\alpha_s}{\pi}C_A \,\text{ln}\frac{\hat{Q}^2}{\mu^2} \,\int_{x_{1}}^1\frac{dx^{'}_1}{x_1^{'}} f_{g/P}\left( \frac{x_1}{x_1^{'}},\mu \right)  \int_{x_{2}}^1 \frac{dx^{'}_2}{x_2^{'}} f_{g/P}\left( \frac{x_2}{x_2^{'}},\mu \right) 
 \nonumber \\ &\times&
 g^4 (d-2) \frac{N_c^2-1}{4} Q^2\,\delta(1-x_1^{'})\delta(1-x_2^{'})\delta(t_n^{max}) \delta(t_{\bar{n}}^{max})\nonumber \\
 &=& -2\frac{\alpha_s}{\pi}C_A \,\text{ln}\frac{\hat{Q}^2}{\mu^2} \,X^{(0)}.
\end{eqnarray}
The computation is simplified by noting that all plus distribution terms appearing in this expression cancel due to the following identity:
\begin{equation}
 \int_0^{t_n^{max}} dt_n^+ \frac{1}{\hat{Q}^2}\left\{ \delta(t_n^+)\left[\frac{\hat{Q}^2}{t_n^{max}-t_n^+}\right]_+ - \delta(t_n^{max}-t_n^+)\left[\frac{\hat{Q}^2}{t_n^+}\right]_+ \right\}=0,
\end{equation}
such that only the terms in Eqs.~(\ref{ndiff}), (\ref{bndiff}), and~(\ref{sdiff}) with an explicit $\text{ln}(Q^2/\mu^2)$ contribute to the running.  This verifies that the 
running of $X$ matches the required form in Eq.~(\ref{neededrunning}).

%%%%%%%%%
%%%%%%%%%
%%%%%%%%

\section{Conclusion}
\label{sec:conc}

We have derived a factorization theorem for the transverse momentum and rapidity distributions  of the Higgs boson at hadronic colliders, in the region  $m_h \gg p_T\gg \Lambda_{QCD}$, employing the methods of effective field theory. The factorization theorem naturally allows for a resummation of the low $p_T$ region and can be straightforwardly generalized to other processes that involve the production of one or more color neutral particles, such as the Drell-Yan process. The problem of resummation  in the low $p_T$ region has been extensively studied in the QCD literature~\cite{Dokshitzer:1978yd,Parisi:1979se,Curci:1979bg,Collins:1984kg,Bozzi:2003jy} and more recently with effective field theory methods~\cite{Idilbi:2005er,Gao:2005iu}. However, the form of our factorization theorem and the method of our analysis contains several new and interesting elements that are worth pursuing further.

The derived factorization theorem takes the form shown in Eq.~(\ref{intro-fac}) which is a convolution between the perturbative coefficients $H(x_1x_2Q^2,\mu_Q;\mu_T)$ and ${\cal G}^{ij}(x_1,x_1',x_2,x_2',p_T,Y,\mu_T)$  and   standard QCD PDFs. This form was obtained by matching onto a sequence of
effective field theories. First the top quark is integrated out to obtain the standard effective coupling of the Higgs to gluons. The hard
coefficient $H(x_1x_2Q^2,\mu_Q;\mu_T)$ is obtained by matching the effective Higgs operator onto an effective SCET$_{p_T}$ operator, integrating out the hard scale $m_h$ in the process, followed by  RG evolution between the scales $\mu_Q \sim m_h$ and $\mu_T \sim p_T$. The RG evolution of the hard coefficient, determined in the effective theory SCET$_{p_T}$, sums up  logarithms of $m_h/p_T$.  The perturbative coefficient ${\cal G}^{ij}$, evaluated at the $\mu_T \sim p_T$ scale, takes the form in Eq.~(\ref{intro-G}) which involves a convolution over the $n$ and $\bn$ collinear functions ${\cal I}_{n,\bn;g,i}^{\alpha \beta}$ and the inverse Soft Function ${\cal S}^{-1}$. This form of ${\cal G}^{ij}$ into two collinear and one soft sectors is  the result of the soft-collinear decoupling property of the leading order SCET$_{p_T}$ Lagrangian.  The collinear functions ${\cal I}_{n,\bn}^{\alpha \beta}$ are obtained from an OPE in $\Lambda_{QCD}/p_T$ of the impact-parameter Beam Functions (iBFs) $\tilde{B}_{n,\bn}^{\alpha \beta}$ which appear in the factorization theorem at the $p_T$ scale as in Eq.~(\ref{kpkmspacefac-3}). At leading order in $\Lambda_{QCD}/p_T$ the iBFs match onto the standard QCD PDFs with Wilson coefficients ${\cal I}_{n,\bn;g,i}^{\alpha \beta}$ as in Eq.~(\ref{iBFPDF}). The form of the factorization theorem is pictorially summarized in Fig.~\ref{cartoon}.

The iBFs are universal objects in that they will appear in the $p_T$ distributions for any gluon initiated process of color neutral particles. The iBFs defined here will have analogs for quark-initiated scattering as well, and the generalization to such processes is straightforward. Similarly, the iSF is also  universal and again has a straightforward analog for quark-initiated processes. The iBFs are defined without a soft zero-bin subtraction, so that in addition to describing the emissions and virtual effects  of collinear gluons they also do the same for soft gluons. The presence of two iBFs in the factorization theorem leads to this soft region being double-counted. The iSF plays the role of \textit{subtracting} a soft region so that there is only one soft region in the full factorization theorem as required. Each iBF is still defined implicitly with an ultrasoft zero-bin subtraction to avoid double counting with ultrasoft modes which are unaffected by $p_T$ constraints. Each iBF and correspondingly the collinear functions ${\cal I}_{n,\bn;g,i}^{\alpha \beta}$ are defined in the class of covariant or non-singular gauges. The iBF is not invariant under singular gauge transformations where the gauge potential is non-vanishing at infinity and must be modified. However,  the quantity ${\cal G}^{ij}$ in the factorization theorem which is a convolution over the two iBFs and an iSF is fully gauge invariant. The factorization of ${\cal G}^{ij}$ into two iBFs and an iSF can be quite useful for the computation of higher order radiative corrections at the $p_T$ scale since the computation of each of the factored objects is much simpler.

The factorization theorem can also be formulated entirely in momentum space with no reference to an impact-parameter. This form of the factorization theorem is given in Eq.~(\ref{intro-fac}) with the quantity ${\cal G}^{ij}$ given in Eq.~(\ref{momspace-1}) entirely in terms of momentum-space  objects. In this effective field theory formulation one avoids Landau poles associated with impact-parameter integrations in the usual CSS formulation.  The use of SCET allows power-suppressed corrections of the form $\Lambda_{QCD}/p_T$  and $p_T/m_h$ to be systematically included.  When $p_T \sim \Lambda_{QCD}$, a model of the non-perturbative region can be included as in the usual approach.  

The structure of our factorization theorem differs also from previous SCET analyses.  We are led to gluon iBFs that have an index structure, which at higher orders is needed for the separate calculation of effects from each collinear sector. 

We believe our study provides a new framework for the investigation of the low-$p_T$ region in hadron collider processes.  Several interesting differences from previous approaches arise naturally in our derivation within SCET that require further investigation.  We look forward to future research investigating the implications of the factorization theorem derived here.

\appendix

\section{Details of deriving factorization in SCET$_{p_T}$}\label{fac-derv}

After matching $n_f=5$ QCD onto SCET$_{p_T}$ as in Eq.~(\ref{QCDtoSCETmatch}), the differential cross-section in the Mandelstam variables $u$ and $t$ is given by 

\bea
\frac{d^2\sigma}{du \>dt} &=& \frac{1}{2Q^2} \Big [\frac{1}{4} \Big ]\int \frac{d^2 p_{h_\perp}}{(2\pi)^2} \int \frac{dn\cdot p_h d\bn\cdot p_h}{2(2\pi)^2} (2\pi)\theta (n\cdot p_h + \bn \cdot p_h)\delta (n\cdot p_h \bn \cdot p_h - \vec{p}_{h_\perp}^{\> 2} -m_h^2 ) \nn \\
%%%
&\times& \delta (u-(p_2-p_h)^2) \delta (t-(p_1 -p_h)^2)  
\sum_{\text{initial pols.}}\sum_{X} \big |C(\omega_1,\omega_2) \otimes \> \langle h X_n X_\bn X_{s} |  {\cal O}(\omega_1,\omega_2) | pp \rangle \>\big |^2 \nn \\
&\times& (2\pi)^4\delta^{(4)} (p_1+p_2-P_{X_n}-P_{X_\bn}-P_{X_s}-p_h) ,\nn \\
\eea
where the delta functions in $u$ and $t$ pick out the appropriate final states. The overall factor of 1/4 in square brackets comes from the average over initial proton polarizations. The final state hadrons have been broken up into n-collinear, $\bn$-collinear, and soft states as
\bea
|X \rangle = | X_n, X_\bn, X_{s} \rangle,
\eea
since these are the states with non-zero overlap with the SCET$_{p_T}$ operator ${\cal O}(\omega_1,\omega_2)$
\bea
{\cal O}(\omega_1,\omega_2) = g_{\mu \nu}h \> \text{Tr} \Big [ T\{ S_n (g B_{n\perp}^{(0)\mu})_{\omega_n}S_n^\dagger S_\bn (g B_{\bn \perp}^{(0)\nu})_{\omega_\bn}S_\bn^\dagger \} \Big ],
\eea
where
\bea
 S_n (x^\mu) = P {\rm exp} \left(i g_s \int_{\infty}^0 ds \, n \cdot A^a_{s}(x^\mu + s n^{\mu}) \right)
\eea
with an analogous expression for $S_n$. Since the collinear and soft fields are decoupled from each other, we can factorize the matrix element of ${\cal O}(\omega_1,\omega_2)$ into collinear and soft sectors as
\bea
&&\langle h X_n X_\bn X_{s} |  {\cal O}(\omega_1,\omega_2) | pp \rangle \nn \\
%%%%
&=& \langle h X_n X_\bn X_{s} |  g_{\mu \nu}h \> \text{Tr} \Big [T\{  S_n (g B_{n\perp}^\mu)_{\omega_n}S_n^\dagger S_\bn (g B_{\bn \perp}^\nu)_{\omega_\bn}S_\bn^\dagger \} \Big ] | p_1 p_2 \rangle  \nn \\
%%%%
&=& \langle X_{s} | \text{Tr}  \big ( T\{ S_nT^A S_n^\dagger  S_\bn T^B S_\bn^\dagger  \}\big ) | 0 \rangle  \langle X_n |  \left (B_{n \perp}^{\alpha A} \right )_{\omega_{1}} | p_1 \rangle  \langle X_\bn |  \left (B_{\bn \perp \alpha }^B \right )_{\omega_{2}} | p_2 \rangle .
\eea
The absolute value squared of the matrix element now takes the form
\bea
&& | \langle h X_n X_\bn X_s | C\otimes {\cal O} | pp \rangle |^2 \nn \\
%%%%%
&=& C^\dagger \otimes C \otimes \langle 0 |  \text{Tr}  \big (\bar{T}\{  S_\bn T^D S_\bn^\dagger  S_n T^C S_n^\dagger \}\big ) | X_{s} \rangle  \langle X_{s} | \text{Tr}  \big ( T\{ S_nT^A S_n^\dagger  S_\bn T^B S_\bn^\dagger \} \big )  | 0 \rangle \nn \\
%%%%%%
&\times & \langle p_1 |  \left (B_{1n \perp \beta}^C \right )_{\omega_{1}'} | X_n \rangle \langle X_n | \left (B_{1n \perp \alpha}^A \right )_{\omega_{1}} | p_1 \rangle \nn \\
%%%%%
&\times & \langle p_2 |  \left (B_{2\bn \perp \beta}^D \right )_{\omega_{2}'} | X_\bn \rangle \langle X_\bn | \left (B_{2\bn \perp \alpha}^B \right )_{\omega_{2}} | p_2 \rangle, \nn \\
&=& C^\dagger \otimes C \otimes \langle 0 |  \text{Tr}  \big (  \bar{T} \{S_\bn T^D S_\bn^\dagger  S_n T^C S_n^\dagger \} \big ) | X_{s} \rangle  \langle X_{s} | \text{Tr}  \big ( T\{ S_nT^A S_n^\dagger  S_\bn T^B S_\bn^\dagger \}\big )  | 0 \rangle \nn \\
%%%%%%
&\times & \frac{\delta^{CA}}{N_c^2-1}\langle p_1 |  \left (B_{1n \perp \beta}^A\right )_{\omega_{1}'} | X_n \rangle \langle X_n |\left (B_{1n \perp \alpha}^A\right)_{\omega_1} | p_1 \rangle \nn \\
%%%%%
&\times & \frac{\delta^{DB}}{N_c^2-1}\langle p_2 |  \left (B_{2\bn \perp \beta}^B\right )_{\omega_{2}'} | X_\bn \rangle \langle X_\bn | \left (B_{2\bn \perp \alpha}^B \right )_{\omega_2}  | p_2 \rangle \nn \\
%%%%
&=& C^\dagger \otimes C \otimes \langle 0 |  \text{Tr}  \big ( \bar{T}\{ S_\bn T^D S_\bn^\dagger  S_n T^C S_n^\dagger \} \big ) | X_{s} \rangle  \langle X_{s} | \text{Tr}  \big (  T \{ S_nT^C S_n^\dagger  S_\bn T^D S_\bn^\dagger \} \big )  | 0 \rangle \nn \\
%%%%%%
&\times & \frac{1}{N_c^2-1}\langle p_1 |  \left (B_{1n \perp \beta}^A\right )_{\omega_{1}'} | X_n \rangle \langle X_n |\left (B_{1n \perp \alpha}^A\right )_{\omega_1} | p_1 \rangle \nn \\
%%%%%
&\times & \frac{1}{N_c^2-1}\langle p_2 |  \left (B_{2\bn \perp \beta}^B\right )_{\omega_{2}'} | X_\bn \rangle \langle X_\bn |\left (B_{2\bn \perp \alpha}^B\right)_{\omega_2}  | p_2 \rangle ,\nn \\
\eea
where we have simplified the contraction of color indices in the last two equalities. Next we insert the identity operator
\bea
\label{identityop}
1 &=& \int d^4p_n \int d^4p_\bn \int d^4p_s \delta^{(4)} (p_n - P_{X_n})\delta^{(4)} (p_\bn - P_{X_\bn}) \delta^{(4)}(p_{s}-K_{X_{s}}),\nn \\
%%%%%
\eea
and we decompose the momentum components of order $m_h$ for the final state collinear particles into label and residual parts so that 
\bea
P_{X_n}^- &=& \tilde{P}_{X_n}^- + K_{X_n}^-, \qquad P_{X_\bn}^+ = \tilde{P}_{X_\bn}^+ + K_{X_\bn}^+, \nn \\
\eea
where $\tilde{P}_{X_n}^-,\tilde{P}_{X_\bn}^+ \sim m_h$ and $K_{X_n}^-,K_{X_\bn}^+ \ll m_h$ and we write the remaining momentum components as
\bea
P_{X_n}^{+,\perp} &=& K_{X_n}^{+,\perp}, \qquad P_{X_\bn}^{-,\perp} = K_{X_\bn}^{-,\perp}, \qquad P_{X_{s}}^\mu =  K_{X_s}^\mu.\nn \\
\eea
Similarly, we write
\bea
p_n^- &=& \tilde{p}_n^- + k_n^-, \qquad p_\bn^+ = \tilde{p}_\bn^+ + k_\bn^+,  \nn \\
p_n^{+,\perp}&=& k_n^{+,\perp}, \qquad p_\bn^{-,\perp}= k_\bn^{-,\perp}, \qquad p_{s}^\mu= k_{s}^\mu, \nn \\
\eea
where again  $\tilde{p}_n^{-}, \tilde{p}_\bn^+ \sim m_h $ and  $\tilde{k}_n^{-}, \tilde{k}_\bn^+ \ll m_h $. The delta functions in Eq.~(\ref{identityop}) for the large collinear momentum components  break up into Kronecker delta functions over the hard collinear label momenta and residual delta functions which we can write using the integral representation to get
\bea
1 &=& \sum_{\tilde{p}_n,\tilde{p}_\bn} \delta_{\tilde{p}_n^-,\tilde{P}_{X_n}^-}  \delta_{\tilde{p}_\bn^+,\tilde{P}_{X_\bn}^+}\int d^4k_n d^4k_\bn d^4k_s \delta^{(4)} (k_n - K_{X_n})\delta^{(4)} (k_\bn - K_{X_\bn}) \delta^{(4)}(k_{s}-K_{X_{s}})\nn \\
%%%%
&=&\sum_{\tilde{p}_n,\tilde{p}_\bn} \delta_{\tilde{p}_n^-,\tilde{P}_{X_n}^-}  \delta_{\tilde{p}_\bn^+,\tilde{P}_{X_\bn}^+} \int d^4k_n d^4k_\bn d^4k_s \int \frac{d^4x}{(2\pi)^4} \frac{d^4 y}{(2\pi)^4} \frac{d^4z}{(2\pi)^4} \> e^{i(k_n-K_{X_n})\cdot x}\> e^{i(k_\bn-K_{X_\bn})\cdot y}\> e^{i(k_{s}-K_{X_{s}})\cdot z}. \nn \\
\eea
The exponential factors above involving the momenta of the final state collinear and soft states  allow us to perform a shift in coordinates and perform a sum over the states $X_{n,\bn,s}$ states.
Furthermore we break up the Higgs momentum as
\bea
n\cdot p_h &=& n\cdot \tilde{p}_h + n\cdot k_h, \nn \\
\bn \cdot p_h &=& \bn\cdot \tilde{p}_h + \bn\cdot k_h, \nn \\
\vec{p}_h^\perp &=& k^\perp_h,
\eea
where $ n\cdot \tilde{p}_h,  \bn\cdot \tilde{p}_h  \sim m_h$ and 
$n\cdot \tilde{k}_h,  \bn\cdot \tilde{k}_h, k_h^\perp  \ll m_h$.
The Higgs phase space integrals can now be written  as a sum over label momenta and integrals over the residual momenta as 
\bea
 &&\sum_{ \tilde{p}_h^+, \tilde{p}_h^-}  \int d^2 k_{h_\perp}  \int \frac{dk_h^+ dk_h^-}{2} \theta (n\cdot \tilde{p}_h + \bn \cdot \tilde{p}_h)\nn \\
&\times&\delta ( \tilde{p}_h^+ \tilde{p}_h^- + \tilde{p}_h^+k_h^- + \tilde{p}_h^-k_h^+ +k_h^+k_h^- - \vec{k}_{h_\perp}^{\> 2}  -m_h^2 ). \nn \\
\eea
We also break up the momentum conserving delta function into Kronecker deltas over label momenta and delta functions over the residual residual momenta  as
\bea
\delta^{(4)} (p_1+p_2-P_{X_n}-P_{X_\bn}-P_{X_{s}}-p_h) &=&  \delta_{\omega_1,\tilde{p}_h^-} \delta_{\omega_2, \tilde{p}_h^+ }  \delta^{(2)} (K_{X_{s \perp}} + K_{X_{n\perp}} + K_{X_{\bn\perp}} + k_{h\perp}) \nn\\
%%%%%
 &\times & \delta (K_{X_n}^+  + K_{X_\bn}^+ + K_{X_{s}}^+ + k_h^+) \delta (K_{X_n}^-  + K_{X_\bn}^- + k_{X_{s}}^- + k_h^-), \nn \\
 &=&  \delta_{\omega_1,\tilde{p}_h^-} \delta_{\omega_2, \tilde{p}_h^+ }  \delta^{(2)} (k_{s\perp} + k_{n\perp} + k_{\bn \perp} + k_{h\perp}) \nn\\
%%%%%
 &\times & \delta (k_n^+  + k_\bn^+ + k_{s}^+ + k_h^+) \delta (k_n^-  + k_\bn^- + k_{s}^- + k_h^-). \nn \\
\eea
After all of the above manipulations, the SCET$_{p_T}$ double differential cross-section can be brought into the form
\bea
\frac{d^2\sigma}{du dt} &=& \frac{(2\pi)^4}{16 Q^2 (2\pi)^3 (N_c^2-1)^2 }  \sum_{\tilde{p}_h^+, \tilde{p}_h^-}  \int dk_h^+ dk_h^- \int d^2 k_{h_\perp}  \int d^4k_n d^4k_\bn d^4k_{s}  \int \frac{d^4x}{(2\pi)^4}  \, \frac{d^4 y}{(2\pi)^4} \, \frac{d^4z}{(2\pi)^4}  \nn \\
%%%%
&\times& \theta (\tilde{p}_h^+ + \tilde{p}_h^-)\delta (  \tilde{p}_h^+ \tilde{p}_h^- + \tilde{p}_h^+k_h^- + \tilde{p}_h^-k_h^+ +k_h^+k_h^- - \vec{k}_{h_\perp}^{\> 2}  -m_h^2 )  \nn \\
%%%
&\times& \delta (u-m_h^2 + Q\tilde{p}_h^- + Q k_h^-) \delta (t-m_h^2 + Q\tilde{p}_h^+ + Q k_h^+) \nn \\
%%%
&\times& \sum_{\text{initial pols.}} \sum_{X_n,X_\bn,X_{s}}  e^{i k_n\cdot x}\> e^{ik_\bn\cdot y}\>e^{ik_s\cdot z}  \nn \\
%%%%%
%%%%%
&\times & \int d\omega_1 d\omega_2 \int d\omega_1'd\omega_2' C^*(\omega_1', \omega_2') C(\omega_1,\omega_2)\nn \\
&\times& \langle p_1 |  \left (B_{1n \perp \beta}^A\right )_{\omega_{1}'} (x) \left (B_{1n \perp \alpha}^A \right )_{\omega_{1}} (0)| p_1 \rangle \nn \\
&\times& \langle p_2 |  \left (B_{2\bn \perp \beta}^B\right )_{\omega_{2}'}(y) \left (B_{2\bn \perp \alpha}^B \right )_{\omega_{2}}(0) | p_2 \rangle \nn \\
&\times& \langle 0 |  \text{Tr}  \big ( \bar{T}\{ S_\bn T^D S_\bn^\dagger  S_n T^C S_n^\dagger \} \big ) (z) \text{Tr}  \big ( T\{ S_nT^C S_n^\dagger  S_\bn T^D S_\bn^\dagger \} \big )(0)  | 0 \rangle \nn \\
%%%%
&\times& \delta_{\omega_1,  \tilde{p}_h^-} \delta_{\omega_2, \tilde{p}_h^+ } \delta^{(2)} (k_{s\perp} + k_{n\perp} + k_{\bn\perp} + k_{h\perp}) \nn\\
%%%%%
 &\times & \delta (k_n^+  + k_\bn^+ + k_{s}^+ + k_h^+) \delta (k_n^-  + k_\bn^- + k_{s}^- + k_h^-).
\eea
Next we perform the integral over $x^+$ to get the fourier transform of the n-collinear field $B_{1n \perp \beta}^A(x)$ with respect to $x^+$ as
\bea
&&\int \frac{dx^+}{2(2\pi)} \>e^{\frac{i}{2}k_n^-x^+}\left (B_{1n \perp \beta}^A\right )_{\omega_{1}'} (x^+,x^-,x_\perp) =\left (B_{1n \perp \beta}^A\right )_{\omega_{1}'} (k_n^-, x^-,x_\perp), \nn \\
\eea
with an analogous equation for the $\bn$-collinear field $B_{2\bn \perp \beta}^B(y)$. Next we combine the sum over labels and integral over residual  as usual so that
\bea
\int d\omega_1 dk_n^- \to \int d\omega_1, \qquad \int d\omega_2 dk_\bn^+ \to \int d\omega_2,
\eea
and absorb the residual momenta $k_n^-$ and $k_\bn^+$ into $\omega_1^{(')}$ and $\omega_2^{(')}$ respectively.  We arrive at
\bea
\frac{d^2\sigma}{du dt} &=& \frac{(2\pi)}{16 Q^2 }  \int dp_h^+ dp_h^- \int d^2 k_{h_\perp}  \frac{1}{2}\int dk_n^+ d^2k_n^\perp \frac{1}{2} \int dk_\bn^- d^2k_{\bn}^\perp d^4k_s \int \frac{dx^- d^2x_\perp}{(2\pi)^3}  \, \frac{dy^+ d^2y_\perp}{(2\pi)^3} \, \frac{d^4z}{(2\pi)^4}  \nn \\
%%%%
&\times& \theta (p_h^+ +p_h^-)\delta ( p_h^+p_h^- - \vec{k}_{h_\perp}^{\> 2}  -m_h^2 )   \delta (u-m_h^2 + Q p_h^-) \delta (t-m_h^2 + Q p_h^+ ) \nn \\
%%%
&\times& \sum_{\text{initial pols.}} \sum_{X_n,X_\bn,X_{s}}  e^{\frac{i}{2}k_n^+ x^- - i \vec{k}_{n\perp} \cdot \vec{x}_\perp}\> e^{\frac{i}{2}k_\bn^- y^+-i\vec{k}_{\bn\perp} \cdot \vec{y}_\perp}\>e^{ik_s\cdot z}  \nn \\
%%%%%
%%%%%
&\times & \int d\omega_1 d\omega_2 \int d\omega_1'd\omega_2' C^*(\omega_1', \omega_2') C(\omega_1,\omega_2)\nn \\
&\times& \langle 0 |  \text{Tr}  \big (  \bar{T} \{ S_\bn T^D S_\bn^\dagger  S_n T^C S_n^\dagger \} \big ) (z) \text{Tr}  \big ( T\{ S_nT^C S_n^\dagger  S_\bn T^D S_\bn^\dagger \} \big )(0)  | 0 \rangle \nn \\
%%%%
&\times& \langle p_1 |  \left (B_{1n \perp \beta}^A\right )_{\omega_{1}'} (x^-,x_\perp) \left (B_{1n \perp \alpha}^A \right )_{\omega_{1}} (0)| p_1 \rangle \nn \\
%%%%
&\times&\langle p_2 |  \left (B_{2\bn \perp \beta}^B\right )_{\omega_{2}'}(y^+,y_\perp) \left (B_{2\bn \perp \alpha}^B \right )_{\omega_{2}}(0) | p_2 \rangle \nn \\
%%%%
&\times& \delta^{(2)} (k_{s} + k_{n\perp} + k_{\bn\perp} + k_{h\perp}) \nn\\
%%%%%
 &\times & \delta (\omega_2 -p_h^+ - k_n^+   - k_{s}^+ ) \delta (\omega_1 - p_h^-  - k_\bn^- - k_{s}^- )\nn \\
%%%
&\times &  \frac{1}{(N_c^2-1)^2}. \nn \\
\eea
Next we note that the n-collinear matrix element is proportional to $\delta(\omega_1-\omega_1')$ which can be seen as follows:
\bea
&&\langle p_1 |  \left (B_{1n \perp \beta}^A\right )_{\omega_{1}'} (x^-,x_\perp) \left (B_{1n \perp \alpha}^A \right )_{\omega_{1}} (0)| p_1 \rangle \nn \\
&=&\langle p_1 | \big [  B_{1n \perp \beta}^A  (x^-,x_\perp) \delta({\cal P}^\dagger- \omega'_1)\big ]\big [\delta({\cal P} -\omega_1 ) B_{1n \perp \alpha}^A (0)\big ]| p_1 \rangle  \nn \\
&=&\sum_{X_n}\langle p_1 | \big [  B_{1n \perp \beta}^A  (x^-,x_\perp) \delta({\cal P}^\dagger- \omega'_1) \big ]|X_n \rangle  \langle X_n | \delta({\cal P} -\omega_1 ) B_{1n \perp \alpha}^A (0)| p_1 \rangle  \nn \\
%%%%
&=&  \sum_{X_n}\ \langle p_1 | \big [  B_{1n \perp \beta}^A  (x^-,x_\perp) \delta(\omega_1- \omega'_1) \big ]|X_n \rangle  \langle X_n |\big [ \delta({\cal P} -\omega_1 ) B_{1n \perp \alpha}^A  (0) \big ]| p_1 \rangle , \nn \\
&=&   \langle p_1 | \big [  B_{1n \perp \beta}^A  (x^-,x_\perp) \delta(\omega_1- \omega'_1) \big ] \big [ \delta({\cal P} -\omega_1 )B_{1n \perp \alpha}^A  (0) \big ]| p_1 \rangle , \nn \\
\eea
where we have inserted a complete set of states and used momentum conservation and then performed the sum overs states again. Analogously, the $\bn$-collinear matrix element is proportional to $\delta(\omega_2-\omega_2')$ so that one can perform the integral over $\omega_{1,2}'$ to get

\bea
\frac{d^2\sigma}{du dt} &=& \frac{2\pi}{16 Q^2 (N_c^2-1)^2}  \int dp_h^+ dp_h^- \int d^2 k_{h_\perp}  \frac{1}{2}\int dk_n^+ d^2k_n^\perp \frac{1}{2} \int dk_\bn^- d^2k_{\bn}^\perp d^4k_{s} \nn \\
&\times& \int \frac{dx^- d^2x_\perp}{(2\pi)^3}  \, \frac{dy^+ d^2y_\perp}{(2\pi)^3} \, \frac{d^4z}{(2\pi)^4} \int \frac{db^+db^-}{2(2\pi)^2}\frac{d^2b_\perp}{(2\pi)^2} \nn \\
%%%%
&\times& \theta (p_h^+ +p_h^-)\delta ( p_h^+p_h^- - \vec{k}_{h_\perp}^{\> 2}  -m_h^2 )   \delta (u-m_h^2 + Q p_h^-) \delta (t-m_h^2 + Q p_h^+ ) \nn \\
%%%
&\times&   e^{\frac{i}{2}k_n^+ (x^--b^-)} e^{-i \vec{k}_{n\perp} \cdot (\vec{x}_\perp-\vec{b}_\perp)}\> e^{\frac{i}{2}k_\bn^- (y^+-b^+)}e^{-i\vec{k}_{\bn\perp} \cdot (\vec{y}_\perp-\vec{b}_\perp)}\nn \\
&\times& e^{\frac{i}{2}k_{s}^+(z^- -b^-)} e^{\frac{i}{2}k_{s}^-(z^+-b^+)} e^{-i \vec{k}_{us\perp}\cdot (\vec{z}_\perp-\vec{b}_\perp)}e^{i\vec{k}_{h\perp}\cdot \vec{b}_\perp} \nn \\
%%%%%
%%%%%
&\times & \int d\omega_1 d\omega_2  e^{\frac{i}{2}(\omega_1 - p_h^-)b^+}e^{\frac{i}{2}(\omega_2-p_h^+)b^-} |C(\omega_1,\omega_2)|^2 J_n^{\alpha \beta}(\omega_1,x^-,x_\perp)J_\bn^{\alpha \beta}(\omega_2,y^+,y_\perp)S(z)\nn \\
\eea
where we have defined the jet and soft functions
\bea
 J_n^{\alpha \beta}(\omega_1,x^-,x_\perp) &=&\sum_{\text{initial pols.}}\langle p_1 |  \big [ g B_{1n \perp \beta}^A  (x^-,x_\perp)\delta(\bar{{\cal P}} -\omega_1 )g B_{1n \perp \alpha}^A  (0)  \big ] | p_1 \rangle \nn \\
 %%%
J_\bn^{\alpha \beta}(\omega_1,y^+,y_\perp) &=&\sum_{\text{initial pols.}}\langle p_2 |  \big [ g B_{1n \perp \beta}^A  (y^+,y_\perp)\delta(\bar{{\cal P}} -\omega_2 )g B_{1n \perp \alpha}^A  (0)  \big ] | p_2 \rangle \nn \\
%%%%
S(z) &=& \langle 0 | \text{Tr} \big ( \bar{T} \{S_\bn T^D S_\bn^\dagger S_n T^C S_n^\dagger \} \big ) (z) \text{Tr} \big( T\{ S_n T^C S_n^\dagger S_\bn T^D S_\bn^\dagger \} \big ) (0) | 0 \rangle .\nn \\
\eea
After performing the integrals over the residual momenta $k_n^+, k_n^\perp, k_\bn^-, k_{\bn}^\perp, k_{s}^\mu$ and the $x,y,z$ coordinates we arrive at the simpler form
\bea
\frac{d^2\sigma}{du dt} &=& \frac{2\pi}{8Q^2 (N_c^2-1)^2}  \int dp_h^+ dp_h^- \int d^2 k_{h_\perp}  \int d\omega_1 d\omega_2  \int \frac{d^4b}{4(2\pi)^4}  e^{\frac{i}{2}(\omega_1 - p_h^-)b^+}e^{\frac{i}{2}(\omega_2-p_h^+)b^-} e^{i\vec{k}_{h\perp}\cdot \vec{b}_\perp}\nn \\
%%%%
&\times& \theta (p_h^+ +p_h^-)\delta ( p_h^+p_h^- - \vec{k}_{h_\perp}^{\> 2}  -m_h^2 )   \delta (u-m_h^2 + Q p_h^-) \delta (t-m_h^2 + Q p_h^+ ) \nn \\
%%%
%%%%%
%%%%%
&\times &  |C(\omega_1,\omega_2)|^2 J_n^{\alpha \beta}(\omega_1,b^-,b_\perp)J_\bn^{\alpha \beta}(\omega_2,b^+,b_\perp)S(b^+,b^-,b_\perp),\nn \\
\eea
which appears in Eq.~(\ref{bspacefac}).

%%%%%%%%%%%%%

%%%%%
%%%%

%%%%%%%
%%%%%%%%

%%%%%%%%%%%
%%%%%%%%%%%%
%%%%%%%%
\section{Equivalence of zero-bin and soft subtractions}\label{equiv-zero}

In this section we demonstrate the validity of Eq.~(\ref{zero-soft}), which we write here again for convenience 
\bea
\label{zero-soft-appex}
 E&\equiv&\int d\omega_1 d\omega_2|C(\omega_1, \omega_2,\mu)|^2 \int dk_n^+ dk_\bn^- B_n^{\alpha \beta}(\omega_1,k_n^+,b_\perp,\mu)\> B_{\bn \alpha \beta}(\omega_2,k_\bn^-,b_\perp,\mu) \nn \\
&\times& \>{\cal S}(\omega_1-p_h^- - k_\bn^-, \omega_2-p_h^+ - k_n^+,b_\perp,\mu) \nn \\
&=&\int d\omega_1 d\omega_2|C(\omega_1, \omega_2,\mu)|^2 \int dk_n^+ dk_\bn^- \tilde{B}_n^{\alpha \beta}(\omega_1,k_n^+,b_\perp,\mu)\> \tilde{B}_{\bn \alpha \beta}(\omega_2,k_\bn^-,b_\perp,\mu) \nn \\
&\times& \>{\cal S}^{-1}(\omega_1-p_h^- - k_\bn^-, \omega_2-p_h^+ - k_n^+,b_\perp,\mu). \nn \\
\eea
The statement of Eq.~(\ref{zero-soft-appex}) is that the factorization theorem of SCET$_{p_T}$ in terms of the purely collinear iBFs $B_{n,\bn}^{\alpha \beta}$, defined with a zero-bin subtraction, and the soft function ${\cal S}$ can be equivalently formulated in terms of the iBFs $\tilde{B}_{n,\bn}^{ \alpha \beta}$, defined without a zero-bin subtraction, and an iSF ${\cal S}^{-1}$. 
At zeroth order the purely collinear iBFs $B_{n,\bn}^{(0)\alpha \beta}$ and the soft function are given by
\bea
\label{zeroth}
B_n^{(0)\alpha \beta}(\omega_1,k_n^+,b_\perp) &=& -g^2g_\perp^{\alpha \beta} \delta(k_n^+) \delta(\omega_1 - \bn \cdot p_1), \nn \\
B_\bn^{(0)\alpha \beta}(\omega_2,k_\bn^-,b_\perp) &=&-g^2 g_\perp^{\alpha \beta} \delta(k_\bn^-) \delta(\omega_2 - n \cdot p_2), \nn \\
 {\cal S}^{(0)}(\omega_1-p_h^- - k_\bn^-, \omega_2-p_h^+ -k_n^+,b_\perp) &=& \frac{(N_c^2-1)}{4}\delta(\omega_1-p_h^- - k_\bn^-)\delta( \omega_2-p_h^+ - k_n^+),\nn \\
\eea
and Eq.~(\ref{zero-soft-appex}) is trivially satisfied since 
\bea
\label{zeroth1}
B_n^{(0)\alpha \beta}(\omega_1,k_n^+,b_\perp) &=&\tilde{B}_n^{(0)\alpha \beta}(\omega_1,k_n^+,b_\perp)\nn \\
B_\bn^{(0)\alpha \beta}(\omega_2,k_\bn^-,b_\perp) &=&\tilde{B}_\bn^{(0)\alpha \beta}(\omega_2,k_\bn^-,b_\perp) \nn \\
 {\cal S}^{(0)}(\omega_1-p_h^- - k_\bn^-, \omega_2-p_h^+ - k_n^+,b_\perp) &=&  {\cal S}^{(0)-1}(\omega_1-p_h^- - k_\bn^-, \omega_2-p_h^+ - k_n^+,b_\perp). \nn \\
\eea
We now demonstrate the validity of Eq.~(\ref{zero-soft-appex}) at the next order in perturbation theory. The the purely collinear iBFs and the soft function up to the first order in perturbation theory take the form
\bea
B_n^{\alpha \beta}(\omega_1,k_n^+,b_\perp) &=& B_n^{(0)\alpha \beta}(\omega_1,k_n^+,b_\perp)  + \big [ \tilde{B}_n^{(1);\alpha \beta}(\omega_1,k_n^+,b_\perp) - B_{n0}^{(1);\alpha \beta}(\omega_1,k_n^+,b_\perp)  \big ] \nn \\
&+& \cdots \nn \\
B_\bn^{\alpha \beta}(\omega_2,k_\bn^-,b_\perp) &=& B_\bn^{(0)\alpha \beta}(\omega_2,k_\bn^-,b_\perp)  + \big [ \tilde{B}_\bn^{(1);\alpha \beta}(\omega_2,k_\bn^-,b_\perp) - B_{\bn0}^{(1);\alpha \beta}(\omega_2,k_\bn^+,b_\perp)  \big ] \nn \\
&+& \cdots \nn \\
{\cal S}(\omega_1-p_h^- - k_\bn^-, \omega_2-p_h^+ - k_n^+,b_\perp) &=& {\cal S}^{(0)}(\omega_1-p_h^- - k_\bn^-, \omega_2-p_h^+ - k_n^+,b_\perp) \nn \\
&+& {\cal S}^{(1)}(\omega_1-p_h^- - k_\bn^-, \omega_2-p_h^+ - k_n^+,b_\perp) + \cdots , \nn \\
\eea
where the dots denote terms that are higher order in perturbation theory and $B_{n0,\bn0}^{(1);\alpha \beta}$ denote the zero-bin subtraction terms. There are two types types of terms at first order in perturbation theory for the convolution of the left-hand side of Eq.~(\ref{zero-soft-appex}) so that we can write
\bea
E= E^{(0)} + E^{(1)}+\cdots, \qquad E^{(1)} = E^{(1)}_A + E^{(1)}_B,
\eea
where $E^{(0)}$ is the zeroth order term, $E^{(1)}$ is the first order term, and $E^{(1)}_A $ and $E^{(1)}_B$ denote  the two types of terms at first order in perturbation theory and  are defined as
\bea
E^{(1)}_A &\equiv &\int d\omega_1 d\omega_2 |C(\omega_1,\omega_2)|^2\int dk_n^+ dk_\bn^- \nn \\
&\times& \Big [ \tilde{B}_n^{(1);\alpha \beta}(\omega_1,k_n^+,b_\perp) \tilde{B}_{\bn;\alpha \beta}^{(0)}(\omega_2,k_\bn^-,b_\perp){\cal S}^{(0)}(\omega_1-p_h^- + k_\bn^-, \omega_2-p_h^+ + k_n^+,b_\perp) \nn \\
&+& \tilde{B}_n^{(0);\alpha \beta}(\omega_1,k_n^+,b_\perp) \tilde{B}_{\bn;\alpha \beta}^{(1)}(\omega_2,k_\bn^-,b_\perp){\cal S}^{(0)}(\omega_1-p_h^- + k_\bn^-, \omega_2-p_h^+ + k_n^+,b_\perp) \Big ], \nn \\
\eea
and 
\bea
\label{soft-zero-cancel}
E^{(1)}_B&\equiv &\int d\omega_1 d\omega_2 |C(\omega_1,\omega_2)|^2\int dk_n^+ dk_\bn^- \nn \\
&\times&\Big [ \tilde{B}_n^{(0);\alpha \beta}(\omega_1,k_n^+,b_\perp) \tilde{B}_{\bn;\alpha \beta}^{(0)}(\omega_2,k_\bn^-,b_\perp){\cal S}^{(1)}(\omega_1-p_h^- - k_\bn^-, \omega_2-p_h^+ -k_n^+,b_\perp) \nn \\
&-&B_{n0}^{(1);\alpha \beta}(\omega_1,k_n^+,b_\perp) \tilde{B}_{\bn;\alpha \beta}^{(0)}(\omega_2,k_\bn^-,b_\perp){\cal S}^{(0)}(\omega_1-p_h^- - k_\bn^-, \omega_2-p_h^+ - k_n^+,b_\perp) \nn \\
&-&  \tilde{B}_{n}^{(0);\alpha \beta}(\omega_1,k_n^+,b_\perp)B_{\bn 0;\alpha \beta}^{(1)}(\omega_2,k_\bn^-,b_\perp){\cal S}^{(0)}(\omega_1-p_h^- - k_\bn^-, \omega_2-p_h^+ - k_n^+,b_\perp) \Big ].\nn \\
\eea
At the first order in perturbation theory, Eq.~(\ref{zero-soft-appex}) is equivalent to demonstrating that $E^{(1)}_B$ in Eq.~(\ref{soft-zero-cancel}) can be rewritten as
\bea
\label{x1B}
E^{(1)}_B &=& - \int d\omega_1 d\omega_2 |C(\omega_1,\omega_2)|^2\int dk_n^+ dk_\bn^- \nn \\
&\times& \tilde{B}_n^{(0);\alpha \beta}(\omega_1,k_n^+,b_\perp) \tilde{B}_{\bn;\alpha \beta}^{(0)}(\omega_2,k_\bn^-,b_\perp) {\cal S}^{(1)}(\omega_1-p_h^--k_\bn^- , \omega_2-p_h^+ -k_n^+,b_\perp),\nn \\
\eea
which is the statement that each of the zero-bin subtraction terms in the last two lines of Eq.~(\ref{soft-zero-cancel}) is equivalent to a soft subtraction. Using the zeroth order expressions for the iBFs and the soft function from  Eq.~(\ref{zeroth}) in Eq.~(\ref{soft-zero-cancel}) we can write $E^{(1)}_B$ as 
\bea
E^{(1)}_B&=&\int d\omega_1 d\omega_2 |C(\omega_1,\omega_2)|^2 \nn \\
&\times&\Big [ 2 g^4 \delta(\omega_1 - \bn\cdot p_1)\delta(\omega_2 - \bn \cdot p_2){\cal S}^{(1)}(\omega_1-p_h^- , \omega_2-p_h^+ ,b_\perp)\nn \\
&+& \frac{g^2(N_c^2-1)}{4}\delta(\omega_2 - n\cdot p_2)\delta(\omega_1 - p_h^-) g_\perp^{\alpha \beta}B_{n 0;\alpha \beta}^{(1)}(\omega_1,\omega_2-p_h^+,b_\perp)\nn \\
&+& \frac{g^2(N_c^2-1)}{4}\delta(\omega_1 - \bn\cdot p_1)\delta(\omega_2 - p_h^+) g_\perp^{\alpha \beta}B_{\bn 0;\alpha \beta}^{(1)}(\omega_2,\omega_1-p_h^-,b_\perp)\Big ].\nn \\
\eea
The  iBFs, the zero-bin subtraction terms,  and the soft function at first order in perturbation theory receive contributions from virtual (V) and real (R) emission graphs so that
\bea
\label{RplusV}
 \tilde{B}_{n,\bn}^{(1);\alpha \beta} &=&  \tilde{B}_{n,\bn}^{V(1);\alpha \beta} + \tilde{B}_{n,\bn}^{R(1);\alpha \beta}, \nn \\
 \tilde{B}_{n0,\bn0}^{(1);\alpha \beta} &=&  \tilde{B}_{n0,\bn0}^{V(1);\alpha \beta} + \tilde{B}_{n0,\bn0}^{R(1);\alpha \beta}, \nn \\
 {\cal S}^{(1)}&=&  {\cal S}^{V(1)} +  {\cal S}^{R(1)},\nn \\
\eea
and correspondingly we write
\bea
\label{x1BRplusV}
E^{(1)}_B &=& E^{V(1)}_B + E^{R(1)}_B,
\eea
where $E^{V(1)}_B$ and $E^{R(1)}_B$ denote the contributions from virtual and real graphs respectively to $E^{(1)}_B$ and are given by
\bea
\label{x1BVR}
E^{V,R(1)}_B&=&\int d\omega_1 d\omega_2 |C(\omega_1,\omega_2)|^2 \nn \\
&\times&\Big [ 2 g^4 \delta(\omega_1 - \bn\cdot p_1)\delta(\omega_2 - \bn \cdot p_2){\cal S}^{V,R(1)}(\omega_1-p_h^- , \omega_2-p_h^+ ,b_\perp)\nn \\
&+& \frac{g^2(N_c^2-1)}{4}\delta(\omega_2 - n\cdot p_2)\delta(\omega_1 - p_h^-) g_\perp^{\alpha \beta}B_{n 0;\alpha \beta}^{V,R(1)}(\omega_1,\omega_2-p_h^+,b_\perp)\nn \\
&+& \frac{g^2(N_c^2-1)}{4}\delta(\omega_1 - \bn\cdot p_1)\delta(\omega_2 - p_h^+) g_\perp^{\alpha \beta}B_{\bn 0;\alpha \beta}^{V,R(1)}(\omega_2,\omega_1-p_h^-,b_\perp)\Big ].\nn \\
\eea
We demonstrate the validity of Eq.~(\ref{zero-soft-appex}) at first order in perturbation theory, or equivalently Eq.~(\ref{x1B}), separately for the virtual and real contributions. 
We first look at the virtual contributions for which the soft function and the zero-bin subtraction terms of the purely collinear  iBFs take the form
 \bea
 \label{jvsv}
 B_{n0,}^{V(1)\alpha \beta} (\omega_{1},\>\omega_{2}-p_h^+ ,b_\perp) &=&  B_{n}^{(0)\alpha \beta} (\omega_{1},\>\omega_{2}-p_h^+ ,b_\perp)(-2i g^2 C_A) I_s   \nn \\
B_{\bn0}^{V(1)\alpha \beta} (\omega_{2},\>\omega_{1}-p_h^- ,b_\perp) &=& {\cal  J}_{\bn}^{(0)\alpha \beta} (\omega_{2},\>\omega_{1}-p_h^- ,b_\perp)(-2i g^2 C_A) I_s   \nn \\
 {\cal S}^{V(1)}(\omega_1-p_h^- , \omega_2-p_h^+ ,b_\perp) &=& {\cal S}^{(0)}(\omega_1-p_h^- , \omega_2-p_h^+ ,b_\perp) (-2i g^2 C_A) I_s\nn \\
 \eea
 where $I_s$ is the scaleless integral
 \bea
 \label{Is}
 I_s &=& 2\int \frac{d^d \ell}{(2\pi)^d}\frac{1}{(\ell^2 +i0) \>(\bn \cdot \ell - i0) \> (n\cdot \ell +i0)} .\nn \\
 \eea
 Using Eqs.~(\ref{jvsv}), (\ref{Is}),  (\ref{zeroth}), and (\ref{zeroth1}) in Eq.~(\ref{x1BVR}) for $E^{V(1)}_B$ we find that
 \bea
 \label{Vequiv}
 E^{V(1)}_B&=& -2 g^4\int d\omega_1 d\omega_2 |C(\omega_1,\omega_2)|^2 \nn \\
&\times& \delta(\omega_1 - \bn\cdot p_1)\delta(\omega_2 - \bn \cdot p_2){\cal S}^{V(1)}(\omega_1-p_h^- , \omega_2-p_h^+ ,b_\perp)\nn \\
&=& - \int d\omega_1 d\omega_2 |C(\omega_1,\omega_2)|^2\int dk_n^+ dk_\bn^- \nn \\
&\times& \tilde{B}_n^{(0);\alpha \beta}(\omega_1,k_n^+,b_\perp) \tilde{B}_{\bn;\alpha \beta}^{(0)}(\omega_2,k_\bn^-,b_\perp) {\cal S}^{V(1)}(\omega_1-p_h^--k_\bn^- , \omega_2-p_h^+ -k_n^+,b_\perp),\nn \\
 \eea
 which is in agreement with Eq.~(\ref{x1B}) for the virtual graph contributions at first order in perturbation theory.

Next we look at the real contribution $E_B^{R(1)}$ for which we need expressions for the single gluon emission contribution to the iBFs and the soft function which are given by
\bea
\label{x1BRjs}
{\cal S}^{R(1)}(\omega_1-p_h^-, \omega_2-p_h^+ ,b_\perp) &=& g^2 C_A(N_c^2-1)\int \frac{d^4k_\perp}{(2\pi)^3} \theta(k^0) \delta(k^2) \nn \\
&\times& \delta(\omega_1-p_h^--k^-)\delta(\omega_2-p_h^+-k^+) \frac{e^{i \vec{b}_\perp \cdot \vec{k}_\perp}}{k^+ k^-}\nn \\
B_{n0}^{R(1);\alpha \beta}(\omega_1,\omega_2-p_h^+,b_\perp) &=& -g_\perp^{\alpha \beta} 4g^4 C_A\int \frac{d^4k}{(2\pi)^3}  \theta(k^0) \delta(k^2) \delta(\bn \cdot p_1 - \omega_1 - k^-)\nn \\
&\times&\delta(\omega_2- p_h^+ -k^+ )\frac{e^{i\vec{k}_\perp\cdot \vec{b}_\perp}}{k^+k^-}\nn \\
B_{\bn0}^{R(1);\alpha \beta}(\omega_2,\omega_1-p_h^-,b_\perp) &=& -g_\perp^{\alpha \beta} 4g^4 C_A \int \frac{d^4k}{(2\pi)^4} (2\pi) \theta(k^0) \delta(k^2) \delta(\bn \cdot p_2 - \omega_2 - k^+)\nn \\
&\times&\delta(\omega_1-p_h^- -k^- )\frac{e^{i\vec{k}_\perp\cdot \vec{b}_\perp}}{k^+k^-}.\nn \\
\eea
Using the expressions in Eq.~(\ref{x1BRjs}) in Eq.~(\ref{x1BVR}) for $E^{R(1)}_B$ and using the freedom to perform residual shifts in the label momenta $\omega_{1,2}$, but ignoring these shifts in the Wilson coefficient $C(\omega_1,\omega_2)$ as they are power suppressed, we find
\bea
\label{Requiv}
E_B^{R(1)}&=& - 2 g^4\int d\omega_1 d\omega_2 |C(\omega_1,\omega_2)|^2  \delta(\omega_1 - \bn\cdot p_1)\delta(\omega_2 - \bn \cdot p_2){\cal S}^{(1)}(\omega_1-p_h^- , \omega_2-p_h^+ ,b_\perp)\nn \\
&=& - \int d\omega_1 d\omega_2 |C(\omega_1,\omega_2)|^2\int dk_n^+ dk_\bn^- \nn \\
&\times& \tilde{B}_n^{(0);\alpha \beta}(\omega_1,k_n^+,b_\perp) \tilde{B}_{\bn;\alpha \beta}^{(0)}(\omega_2,k_\bn^-,b_\perp) {\cal S}^{R(1)}(\omega_1-p_h^-k_\bn^- , \omega_2-p_h^+ -k_n^+,b_\perp),\nn \\
\eea
which is in agreement with Eq.~(\ref{x1B}) for the real graph contributions. Thus, together with Eqs.~(\ref{Vequiv}), (\ref{Requiv}), (\ref{x1B}), (\ref{RplusV}), and (\ref{x1BRplusV}) we have explicitly shown the validity of Eq.~(\ref{zero-soft-appex}) up to first order in perturbation theory.

\section{Consistency of top-down and bottom-up running above $p_T$}

In this section we give details of the derivation of the consistency relation in Eq.~(\ref{consistency}). After performing a shift in the integration variables, the scale invariant quantity ${\cal D}$
\bea
\label{kpkmspacefac-8}
{\cal D} &\equiv& \int  dt_n \int dt_\bn H(x_1x_2Q^2,\mu) \>\tilde{B}_n^{\alpha \beta}(x_1,t_n,b_\perp,\mu)\> \tilde{B}_{\bn \alpha \beta}(x_2,t_\bn,b_\perp,\mu )\nn \\
&\times&{\cal S}^{-1}( \frac{t_{\bn}^{max} -t_\bn}{Q}, \frac{t_{n}^{max} -t_n}{Q},b_\perp,\mu_T), \nn \\
\eea
can be written as
\bea
\label{kpkmspacefac-9}
{\cal D} &=& \int  dt_n \int dt_\bn H(x_1x_2Q^2,\mu) \>\tilde{B}_n^{\alpha \beta}(x_1,t_n+t_n^{max},b_\perp,\mu)\> \tilde{B}_{\bn \alpha \beta}(x_2,t_\bn+t_\bn^{max},b_\perp,\mu )\nn \\
&\times&{\cal S}^{-1}( -\frac{t_\bn}{Q}, -\frac{t_n}{Q},b_\perp,\mu_T). \nn \\
\eea
We now write the above expression in two equivalent ways. First we write ${\cal D}$ entirely in terms of bare quantities and then use the first relation in Eq.~(\ref{bare}) to write the bare hard coefficient $H_b$ in terms of the renormalized  hard coefficient $H$ to get
\bea
\label{D-11}
{\cal D}&=& \int dt_n''\int dt_\bn''\int  dt_n' \int dt_\bn'  \>H(x_1x_2Q^2,\mu) Z_H \>\delta(t_n''-t_n'-t_n^{max})\delta(t_\bn''-t_\bn'-t_\bn^{max})\nn \\&\times&\tilde{B}_{n,b}^{\alpha \beta}(x_1,t_n'',b_\perp,\mu)\> \tilde{B}_{\bn,b; \alpha \beta}(x_2,t_\bn'',b_\perp,\mu ){\cal S}^{-1}_b( -\frac{t_\bn'}{Q}, -\frac{t_n'}{Q},b_\perp,\mu_T). \nn \\
\eea
Next we write ${\cal D}$ entirely in terms of renormalized quantities and then use the last three relations in Eq.~(\ref{bare}) to write the bare iBFs and the bare  iSF in terms of renormalized quantities to get
\bea
\label{D-12}
{\cal D} &=& \frac{1}{Q^2}\int dt_n''\int dt_\bn''\int  dt_n' \int dt_\bn'  \Bigg \{\int ds_n\int ds_\bn \int  dt_n\int dt_\bn\> H(x_1x_2Q^2,\mu) \>\nn \\
&\times& \delta(s_n-t_n-t_n^{max})\delta(s_\bn-t_\bn-t_\bn^{max})\nn \\&\times&Z^{-1}_n(s_n-t_n'',\mu)Z^{-1}_\bn(s_\bn'-t_\bn'',\mu)Z^{-1}_{{\cal S}^{-1}}(\frac{-t_\bn+t_\bn'}{Q},\frac{-t_n+t_n'}{Q},\mu)\nn \\
&\times& \tilde{B}_{n,b}^{\alpha \beta}(x_1,t_n'',b_\perp,\mu)\> \tilde{B}_{\bn,b; \alpha \beta}(x_2,t_\bn'',b_\perp,\mu ){\cal S}^{-1}_b( -\frac{t_\bn'}{Q}, -\frac{t_n'}{Q},b_\perp,\mu_T)\Bigg \}. \nn \\
\eea
Since the expressions on the RHS of Eqs.~(\ref{D-11}) and (\ref{D-12}) must be equal, by comparing the quantity in curly brackets in Eq.~(\ref{D-12}) with the integrand of Eq.~(\ref{D-11}) one can derive the relation
 \bea
\label{consistency-appex}
Z_H(\mu) \delta(s_n - t_n)\delta(s_\bn - t_\bn) &=& \frac{1}{Q^2}\int dt_n'\int dt_\bn' Z^{-1}_n(t_n'- s_n,\mu)Z^{-1}_n(t_\bn'- s_\bn,\mu)\nn \\
&\times& Z_{{\cal S}^{-1}}^{-1}(\frac{-t_\bn' + t_\bn}{Q},\frac{-t_n'+t_n}{Q},\mu). \nn \\
\eea
as quoted in Eq.~(\ref{consistency}).

%%%%%%%%%%%
%%%%%%%%%

\section*{Acknowledgments}

We thank Michael Trott for collaboration during the early stages of this work. We also thank Adam Leibovich, Gavin Salam, Iain Stewart, and Frank Tackmann for useful conversations. We thank Markus Diehl and Carola Berger for questions that led to a modification of the formulation in section \ref{sec:eft}.  F. P. thanks the Institute for Theoretical Physics at ETH-Zurich for their kind hospitality during the completion of this work. This work was supported by the Alfred P. Sloan foundation and by the DOE grant DE-FG02-95ER40896.
\OMIT{
\bibliographystyle{h-physrev3.bst}
\bibliography{Higgs}
}
%%%%%%%%%

\bibliographystyle{h-physrev3.bst}
\bibliography{HiggspT}

\end{document}